# Turbulent density fluctuations in the solar wind

A thesis
Submitted in partial fulfillment of the requirements
Of the degree of
Doctor of Philosophy

By

## Madhusudan Ingale
20083019

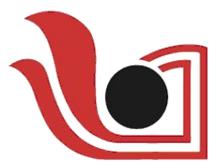

INDIAN INSTITUTE OF SCIENCE EDUCATION AND RESEARCH
PUNE

May, 2015

# Certificate

Certified that the work incorporated in the thesis entitled "*Turbulent density fluctuations in the solar wind*", submitted by *Madhusudan Ingale* was carried out by the candidate, under my supervision. The work presented here or any part of it has not been included in any other thesis submitted previously for the award of any degree or diploma from any other University or institution.

*Date*                                                                       **Dr. Prasad Subramanian**

# Declaration

I declare that this written submission represents my ideas in my own words and where others' ideas have been included, I have adequately cited and referenced the original sources. I also declare that I have adhered to all principles of academic honesty and integrity and have not misrepresented or fabricated or falsified any idea/data/fact/source in my submission. I understand that violation of the above will be cause for disciplinary action by the Institute and can also evoke penal action from the sources which have thus not been properly cited or from whom proper permission has not been taken when needed.

*Date* **Madhusudan Ingale**

Roll No.- 20083019

# Contents

# List of Figures

# List of Tables

# Acknowledgements

*About 8 light minutes away from the known living planet lies the inspiration of this work. The sun constantly fueling life and providing enduring stimulus with all its beauty and challenges. In my journey of PhD I not only learn how to approach few of these challenges but also to appreciate the beauty of the nature. During this period many lives touched upon and influenced me with their support and encouragement. It is a great pleasure to acknowledge them, who contributed directly and indirectly to the success of this thesis.*

*There is no doubt that the most influential person throughout this journey is my mentor, supervisor and what I cherish most is a great human being, Dr. Prasad Subramanian. I am always be grateful to his support and encouragement. Several times when I think about challenges, may it be personnel or others, I find myself admiring the way he approach to just any problem. Right from the PhD interview, through some difficult period at the beginning he was and is there with me.*

*I would like to thank Dr. K. N. Ganesh, Director, IISER-Pune, for the inspiring research facilities and academic environment. I acknowledge the financial support from IISER-Pune in the form of research fellowship. I am thankful to the IISER administration, security, library, and house-keeping staff.*

*I acknowledge the support by CAWSES II program administered by Indian Space Research Organization (ISRO). I acknowledge the financial support by Asian Office of Aerospace Research and Development (AOARD). I am thankful for the support by Center for Excellence in Space Sciences (CESSI),*

*India, Kolkata and prof. Nandy.*

*I am grateful to Iver Cairns, University of Sydney for his support and helpful discussions. During my short stay at CESSI, Kolkata discussions with Soumitra and Aveek da were very helpful. I am thankful to Prof. Mahendra Verma and Dr. M. S. Santhanam, my research advisory committee members for their valuable feedback and useful suggestions regarding my study. I have enjoyed monthly meetings of the Solar-Physics Journal club and I am thankful to Dr. Durgesh Tripathi for an initiative in this regard.*

*Life at IISER-Pune and now as a whole, cannot be thought of without my friends Abhishek, Arun, Arthur, Kajari, Kanika, Mandar, Murthy, Nitin, Padmashri, Snehal, Resmi, Ramya, Sesi and Vimal. At the beginning phase of my PhD I have gone through some difficut situations, of which I only manage to come out due to an active support from Arun and Kanika. Debates with Arun, Kanika, Resmi and Vimal are unforgettable. I can see quiet a change in my thinking before joining IISER and today at the time of submitting this thesis and I owe this to the company of Kanika, Resmi and Vimal. Cooking sessions at HR2, trekking and chit chatters spiced up my journey. With Nishtha and Tomin we share a nice and light discussions may it be a technical or just anything. Kanika and Vimal by and large helped me to go through and overcome "a bit" my own imperfect perfections during PhD and I know they won't hesitate to do so in the future also.*

*My family, which include my parent as well as my guardians at Pune believed me, kept hope in me throughout this period and admired my every little success and equally stood by my failures. I am grateful to them for being in my life and making it so much worth to live.*

**Madhusudan Ingale**

# Abstract

Measurements of density, electric field, magnetic field, in the solar wind have revealed fluctuations in these quantities spanning a large range of scales, indicative of turbulence. The nature of turbulence in the solar wind has been the subject of intense research as it plays an essential role in several aspects of plasma behavior such as, solar wind acceleration and heating of the extended solar corona and solar wind. While considerable progress has been made, the nature of turbulent dissipation, especially in the extended solar corona, and the role of density turbulence therein remains a significant unsolved problem. This thesis is concerned with the nature of density turbulence in the extended solar corona, especially near the inner/dissipation scale.

Electromagnetic waves traversing the solar wind experience scattering due to turbulent density fluctuations, which leads to a wide variety of observed phenomena such as intensity scintillations, angular broadening, pulse smearing, etc. These observations provide useful constraints on the quantities characterizing density turbulence. Chapter 2 provides an overview of the phenomenon of angular broadening.

Treatments of the radio scattering due to density turbulence in the solar wind typically employ asymptotic approximations to the phase structure function. In chapter 3 we use a general structure function (GSF) that straddles the asymptotic limits and quantify the relative error introduced by the approximations. We show that the regimes where GSF predictions are accurate than those of its asymptotic approximations is not only of practical relevance, but are where inner scale effects influence the estimate of the scatter-broadening. Thus we propose that GSF should henceforth be used for scatter broadening calculations and estimates of quantities characterizing density turbulence in the solar corona and solar wind.

In the next part of this thesis we use measurements of density turbulence in the solar wind from observations of radio wave scattering and interplanetary scintillations. Density fluctuations are inferred using the GSF for radio scattering data and existing analysis methods for IPS. Assuming that the density fluctuations below proton scales are due to kinetic Alfvèn waves, we constrain the rate at which the extended solar wind is heated due to turbulent dissipation. These results, elaborated in chapter 4, provide the first estimates of the solar wind heating rate all the way from the Sun to the Earth.

# Chapter 1

# Introduction

*In this chapter we will introduce the basic theme of this thesis. We briefly review the history and important findings in the solar wind physics in the last century. This is followed by the solar wind observations with main emphasis on the discussion of the turbulence in the solar wind.*

The correlation between the events on the Sun and the disturbances in the geomagnetic activities on the Earth has been known from long. The beautiful Auroras (Figure 1.1a), a natural light display in the sky predominantly observed in the high altitude regions, and geomagnetic storms giving rise to strong currents are typical examples of such correlations. The idea of connection between the Sun and the terrestrial magnetic disturbances was taken seriously by some physicists near the end of the nineteenth century. An important step came by the work of Kristian Birkeland (Birkeland, 1908), who based on his extensive geomagnetic survey concluded that, the Earth was bombarded with the continuum of 'rays of electric corpuscles emitted by the Sun'. In other words Birkeland was suggesting a continuous outflow of charged particles from the Sun feeling up the interplanetary space - much closer to our modern concept of the solar wind. These ideas lay in obscurity for many years only to resurface in entirely different context, around the middle of the twentieth century. The concept of a continuous outflow from the Sun re-emerged as an answer to an intriguing question posed by the gaseous tails of comets.

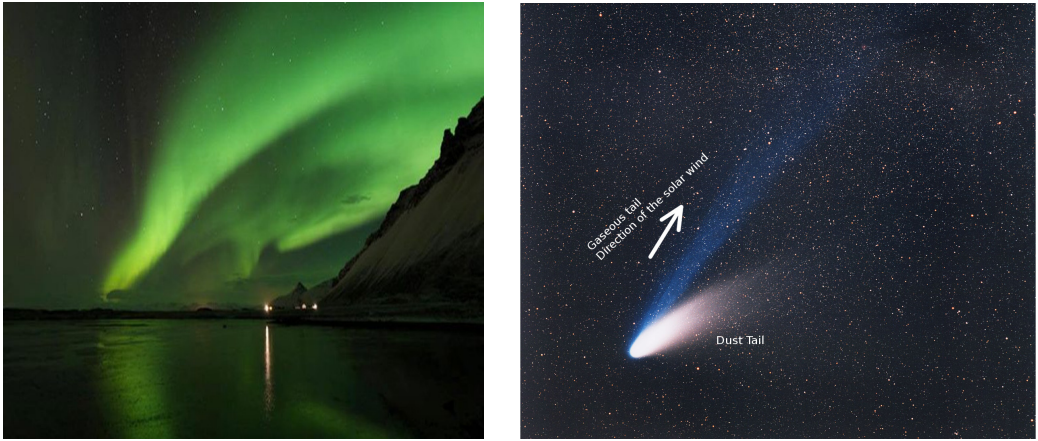

(a) Aurora : an example of correlation between the activity on the Sun and geomagnetic disturbances (image courtesy : https://en.wikipedia.org/wiki/Aurora)

(b) Comet Hale-Bopp observed in 1997, credit (ESO). (image courtesy : http://www.space.com/19931-hale-bopp.html)

Figure 1.1: correlation between geomagnetic disturbances and solar activity

Comets have two classes of tails (i) one made of dust and curved away, with curvature explained by solar radiation pressure and the gravity and (ii) gaseous tail that always points straight away, in the opposite direction of the Sun (Figure 1.1b). Ludwig Biermann in the early 1950s successfully applied the ideas due to Birkeland to the challenge of explaining the observed properties of the gaseous tails (Biermann, 1953). The gaseous tails of comets can therefore be neatly explained if they were subjected to a permanent flux of charged particles coming from the Sun. Around the same time Sydney Chapman proposed the existence of static atmosphere of the Sun extended from the corona with the Earth and other planets immersed in it (Chapman, 1957).

## 1.1 The solar wind

It was difficult to reconcile the two seemingly different results, where one suggests continuous flux of charge particles from the Sun and other talks about the static solar atmosphere. However, Eugene Parker in 1958 realized that, 'however unlikely it seemed, the only possibility is that the Biermann and Chapmann were talking about the same thing' (Parker, 1958). So Biermanns continuous flux of solar particles was just Chapman's extended solar atmosphere, with the temperature so high that neither the solar gravity nor the pressure of the tenuous interstellar medium can confine it. Thus was born the modern concept of the solar wind.

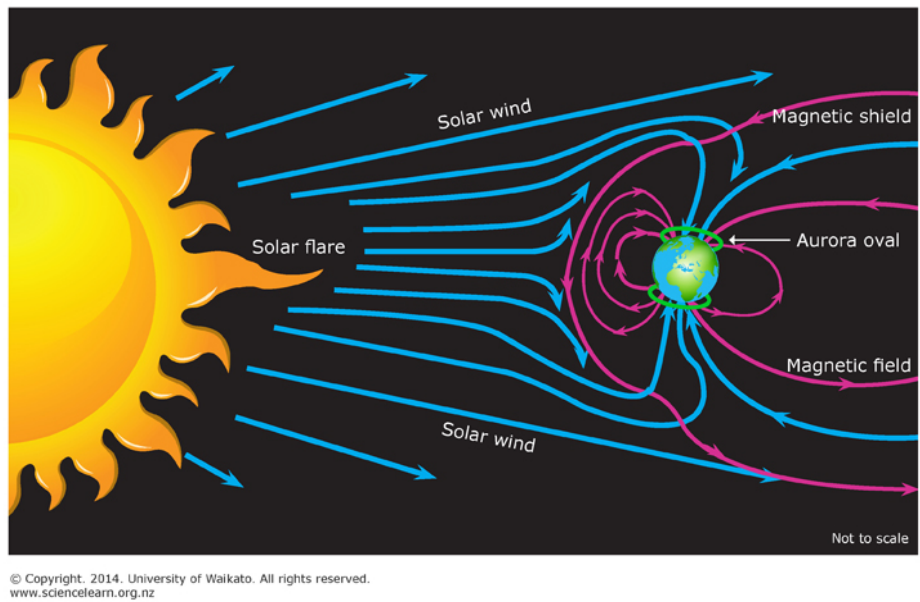

Figure 1.2: Schematic of the solar wind and its effect on the Earth's magnetosphere. Copyright. 2014. University of Waikato (www.sciencelearn.org.nz)

Parker supported his arguments with the comprehensive theory of solar wind, verified and confirmed later by the various space explorations e.g., Mariner 2, Voyager and Helios (Neugebauer, 1962, 1997). So the picture we have now of the solar wind is as a huge bubble of supersonic plasma blown away by the matter ejection from the Sun. The region it fills out is known as heliosphere. The solar wind engulfs Earth and other planets and shape their environments. It is therefore important to understand the physics of the solar wind (Meyer-Vernet, 2007).

## 1.2   Turbulence in the solar wind

The solar wind emerging from the solar corona is found to be inhomogeneous and generates complicated three dimensional structure of the plasma heliosphere. Outflowing streams of different speeds and solar transients introduce further complications and enriches the solar wind with enormous variability in its basic properties. In situ as well as remote sensing observations of magnetic field, velocity and density have revealed fluctuations spanning a broad range of timescales.

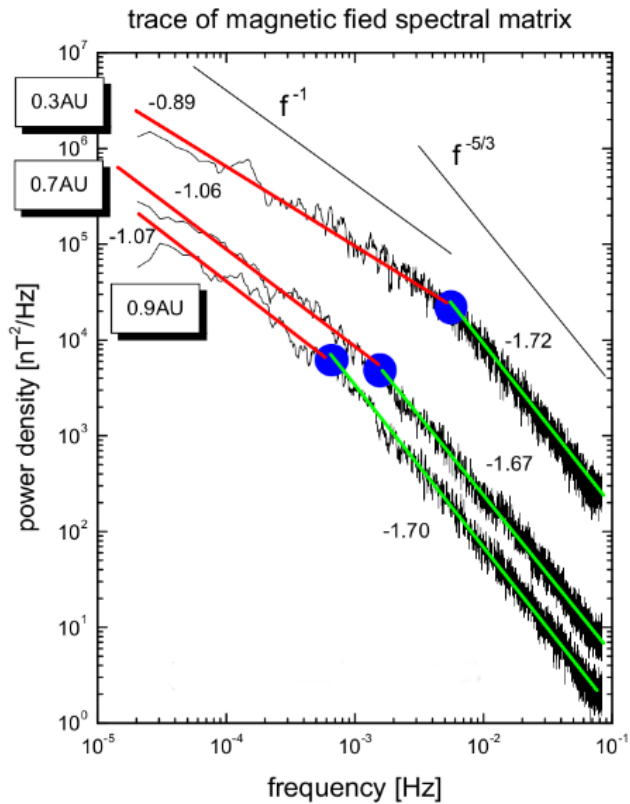

Figure 1.3: Power spectrum of magnetic field fluctuations observed by Helios 2 and Ulysses between 0.3 AU and 1AU, (Adapted from (Bruno & Carbone, 2013))

In situ observations measure quantities like magnetic and electric field, flow speeds, densities and temperatures of plasma species. Representative measurements of the power spectrum of magnetic field fluctuations are presented in figures 1.3 and 1.4. Figure (1.3) shows that the spectrum has a power law shape with index $-5/3$ above a frequency that decreases with increasing heliocentric distance. The spectrum at 1 AU also exhibits cut-offs indicative of dissipation, (figure 1.4). Similar spectral characteristics are observed at all locations explored by spacecraft in the solar wind. Observations and subsequent measurements therefore indicate that the solar wind develops a strong turbulent character towards a state that resembles well known Kolmogorov-like scaling.

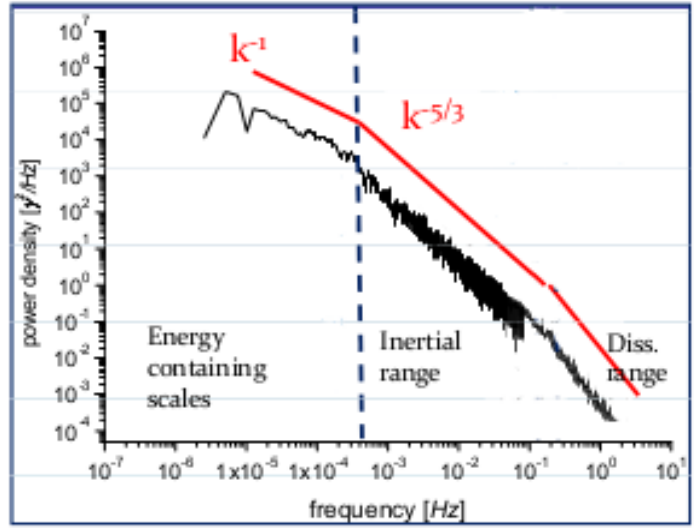

Figure 1.4: Power spectrum of a typical interplanetary magnetic filed fluctuations observed by Helios 2 (low frequency range, Bruno et al. (2009) and WIND (high frequency range, Leamon et al. (1998)). (Adapted from (Bruno & Carbone, 2013))

The spectrum of solar wind turbulence consists of large scales, also referred as outer scales ($l_{\text{out}}$) at which energy injection takes place and marks the beginning of the non-linear cascade. This is followed by the "inertial" range characterized by an omni-directional power law shape with index -5/3, ($-11/3$ in three dimensions) and holds upto very large frequencies or wavenumber i.e., smaller scales (Zank & Matthaeus, 1992; Dastgeer & Zank, 2009). At the smallest scales the spectrum exhibits sharp steepening indicative of the presence of the inner scale ($l_{\text{i}}$) corresponding roughly to the proton gyroradius where dissipation is presumed to take place (Coles & Harmon, 1989; Harmon, 1989; Verma, 1996; Yamauchi et al., 1998; Leamon et

al., 1999, 2000; Smith et al., 2001; Bruno & Trenchi, 2014).

Detailed measurements also indicate that the solar wind turbulence is strongly intermittent and anisotropic depending on the angle between the wave vector and the background magnetic field. Several questions arise : what produces the energy injection and therefore determines the outer scale. The solar wind is known to be weakly compressible; however what give rise to a spectrum with Kolmogorov power law index, which corresponds to incompressible, isotropic fluid turbulence. What is the mechanism behind turbulent dissipation, which should involve kinetic effects. Magnetohydrodynamics, which is valid for the low frequency regime, is out of the question at such small scales.

These questions have large in common with the existing major problems in the solar wind and in general space plasma physics. Turbulence plays a relevant role in several aspect of plasma behavior such as solar wind generation, acceleration of high energy particles, plasma heating and cosmic ray propagation. Ultimately the problem of extended heating in the solar corona and solar wind are those of the storage, transfer and dissipation of the abundant energy present in the photospheric motions.

Magnetohydrodynamic (MHD) turbulence in the solar wind is a subject of intense research. While considerable progress has been made, the nature of turbulence specially near the inner scale where dissipation sets in has remained a significant unsolved problem. Recent observations of perpendicular ion heating by the UVCS instrument aboard SOHO (Kohl et al. 1997, 1998, 1999; Noci et al. 1997; Cranmer, Field & Kohl 1999) have posed questions about whether there is enough power at the dissipation scales to enable direct perpendicular heating of the ions (Cranmer & Van Ballegooijen 2003) In this regard the damping of Alfvén wave turbulence on ions has attracted considerable interest. Near the dissipation scale the Alfvén wave turbulence is found to be increasingly compressive, which suggests that the density fluctuations in the solar wind can constrain the power in high wave-number kinetic Alvén wave turbulence. This in turn can explain some of the heating of the extended solar corona and solar wind (Chandran et al., 2009) as well as acceleration of the solar wind.

One way of addressing this issue is via observations of radio scintillations which includes scattering of radio waves and interplanetary scintillations. The density turbulence in the solar wind introduces both systematic and random variations in the refractive index which effects the propagation of radio waves. The radio waves experiences scattering as a result of interaction with the irregularities in the plasma they traverse. This leads to a wide variety of propagation phenomena, which offers a variety of tools for studying the medium responsible for the scattering. Such tools have been widely used for understanding of interstellar medium, planetary atmosphere, ionosphere and of particular interest here - the solar wind.

It is worth mentioning the upcoming space missions - Solar probe plus and solar orbiter which will be of great importance in probing the solar wind properties and its behavior across the inner heliosphere. These future observations will not only provide the much needed test bed for the existing theories of the solar wind but will also be helpful to improve our understanding of the physics of processes in the solar corona and solar wind. The planned trajectory of the solar probe plus has several passes within 10 $R_\odot$, which in general believed to be the region of strong scattering and solar wind acceleration. On-board experiments like FIELDS and SWEAP will provide in-situ measurements of electric and magnetic field, densities, flow speeds including velocity fluctuations and temperature of electrons, alpha particles and protons. This will greatly clarify the physics responsible for solar wind origin.

## 1.3 Organization of thesis

The rest of this thesis is organized as follows.

In the chapter 2 we review the relevant results from the theory of wave propagation in random medium. Starting from the parabolic wave equation we obtain general solution for the first moment of the wave field. The properties of the medium are described by the dielectric function $\epsilon(r)$. The solution yields phase structure function which defines wide class of observed phenomena produced by scattering of electromagnetic waves.

We know that a useful statistic of the scattering medium is described by the phase structure function which is the subject of the chapter 3 of this thesis. Generally observations of angular broadening due to radio wave scattering in turbulent medium employ asymptotic approximations valid either when interferometric baseline is $\gg$ or $\ll$ the inner scale ($l_i$) of the turbulence. We consider general structure function (GSF) that does not use these approximations. We demonstrate that the regimes where GSF predictions are more accurate than those of the asymptotic approximations is of practical relevance and more importantly it is the regime where inner scale effects influence the estimate of angular broadening.

In the chapter 4 we use previously published observations of radio wave scattering and interplanetary scintillations (IPS) in the solar wind. Taken together these observations provide measurements of density turbulence over wide range of spatial scales including an important dissipation range. The density fluctuations are inferred using a recently developed tool (i.e. GSF) for radio wave scattering and existing methods for IPS. Assuming that the entire contribution to density fluctuations comes by Kinetic Alfvén waves we obtain constraints on the rate of turbulent heating.

We then conclude with a brief discussion of future extension of our work and scope in chapter 5.

In the Appendix we review the salient features of the Alfvén waves as they non-linearly interact and propagate to higher wavenumber through turbulent cascade. At high wavenumbers the solution to Alfvén branch is known as the kinetic Alfvén waves (KAW), which unlike Alfvén waves, exhibit compressibility. We outline the KAW dispersion relation which relates velocity fluctuations to density turbulence.

# Chapter 2

# Electromagnetic wave propagation in random media

*We review relevant results from the theory of wave propagation in a random medium. The aim is to describe statistical properties of waves in terms of statistical properties of the random medium. The starting point is the parabolic wave equation (PWE), which describes small angle wave propagation in a random medium. The PWE is manipulated to obtain equations for the moment of a wave field in terms of two point correlations of the dielectric constant $\epsilon(\mathbf{r})$. A solution to the first order moment equation of a wave field yields the structure function. We can quantitatively define a wide range of observed phenomena produced by scattering of electromagnetic waves using the structure function.*

## 2.1 Introduction

Electromagnetic wave propagation through a random medium leads to a wide variety of observed phenomena such as pulse smearing, spectral broadening, intensity scintillations, angular broadening etc. Observations of these phe-nomena can be used to study the nature of the turbulence in the intervening random medium (e.g. the solar wind). The first step in this regard is to establish the physical foundations. We reproduce important results in this field, relevant to our work.

We discuss the problem of wave propagation and scattering in a random continuum, defined as a medium whose dielectric constant $\epsilon(\mathbf{r}, t)$ is a continuous random function of position and time. Examples of random media are the solar wind, the solar corona and the Earths atmosphere.

The approach presented here is known as analytical or multiple scattering theory, which starts with the fundamental differential equations for field quantities (e.g. Maxwell's equations for electromagnetic field) and then introduces statistical considerations.

Since we are dealing with a random medium we need to use the statistical approach to describe the medium and its effects on the electromagnetic wave propagating through it. Specifically, we concentrate on the correlation of fluctuations at two points separated in space and time. There are several ways to do so but describing these correlations in terms of the structure function is the most elegant way and of great practical value. We use turbulence theory to express the structure function in terms of the spectrum of fluctuations in a random medium. We will see that this treatment requires to specify certain physical parameters (such as the amplitude of turbulence $C_N^2(\mathbf{r})$ and its inner scale ($l_i(\mathbf{r})$)). These parameters characterizes the turbulent fluctuations ambient in the medium and need to be determined using observations. We will see that the observations of radio scattering and interplanetary scintillations (IPS) provide useful constraints on these parameters.

From now on we use the term turbulent medium to denote a random medium having fluctuations in $\epsilon(r, t)$, spanning a wide range of scales.

## 2.2 Parabolic wave equation

Consider an electromagnetic wave, $\mathbf{E}(\mathbf{r},t) = \mathbf{E}_0 e^{i(\kappa z - \omega t)}$ propagating in $+z$ direction. Assume source free region and the magnetic permeability, $\mu_m = 1$, ($\mathbf{B} = \mathbf{H}$). Maxwell's equations relate electric and magnetic fields by following equations :

$$\nabla \times \mathbf{E}(\mathbf{r},t) = \frac{1}{c}\frac{\partial \mathbf{H}(\mathbf{r},t)}{\partial t};$$
$$\nabla \times \mathbf{H}(\mathbf{r},t) = \frac{1}{c}\frac{\partial}{\partial t}(\epsilon(\mathbf{r},t)\mathbf{E}(\mathbf{r},t)). \qquad (2.1)$$

Here $\epsilon(\vec{r},t)$ is the dielectric constant that describes the propagation properties of a medium. Consider this wave incident upon the turbulent medium extending over the range $0 < z < L$. Turbulent fluctuations in the dielectric constant scatter the wave. Assuming a stime dependence of $e^{-i\omega t}$ we can write :

$$\mathbf{E}(\mathbf{r},t) = \mathbf{E}(\mathbf{r})e^{-i\omega t} \qquad (2.2)$$
$$\mathbf{H}(\mathbf{r},t) = \mathbf{H}(\mathbf{r})e^{-i\omega t} \qquad (2.3)$$

Using this in (2.1) we get :

$$\nabla \times \mathbf{E}(\mathbf{r}) = \frac{i\omega}{c}\mathbf{H}(\mathbf{r});$$
$$\nabla \times \mathbf{H}(\mathbf{r}) = \frac{\mathbf{E}(\mathbf{r})}{c}\left[\frac{\partial}{\partial t}\epsilon(\mathbf{r},t) - i\omega\epsilon(\mathbf{r},t)\right]$$

For the sake of brevity, we will henceforth drop the explicit dependence

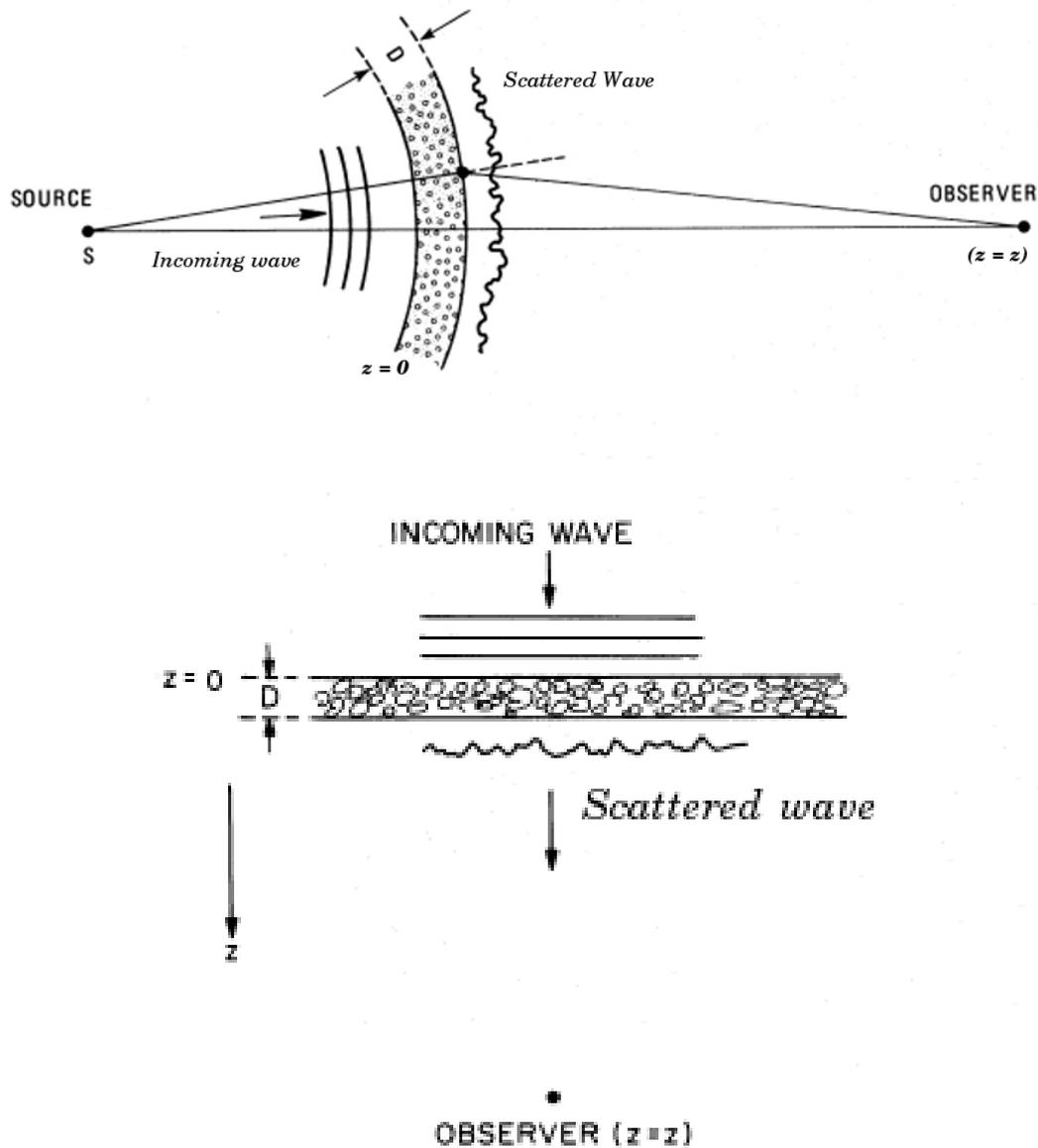

Figure 2.1: A schematic diagram showing the geometry for a thin screen of irregularities. The electromagnetic wave propagates in the $+z$ direction and is incident on the turbulent medium at $z = 0$. The top panel shows spherical wave propagation whereas the bottom panel is for plane wave propagation (chapter 3)

on space (e.g. $\mathbf{E}(\mathbf{r}) \mapsto \mathbf{E}$). Combining these two equations :

$$\nabla \times \nabla \times \mathbf{E} = \frac{i\omega}{c^2}\mathbf{E}\frac{\partial \epsilon}{\partial t} + \frac{\omega^2}{c^2}\epsilon\mathbf{E} \qquad (2.4)$$

Consider the magnitudes of the terms :

$$\frac{\omega}{c^2}|\mathbf{E}|\frac{\partial \epsilon}{\partial t} \quad \text{and} \quad \frac{\omega^2}{c^2}\epsilon|\mathbf{E}| \qquad (2.5)$$

To estimate the time dependence of $\epsilon$, we note that the first derivative with respect to time is related with the average speed of irregularities and their size $l$ in the turbulent medium i.e., $\partial/\partial t \mapsto v/l$. Here $v$ is the typical velocity scale; in case of the slow solar wind, $v = 400$km/s. Since $v \ll c$ :

$$\frac{\omega}{c^2}\frac{\partial \epsilon}{\partial t} \mapsto \frac{v}{l}\frac{\omega^2}{c^2}\epsilon \ll \frac{\omega^2}{c^2}\epsilon \qquad (2.6)$$

Thus we can neglect the first term on the right hand side of 2.4 and write :

$$\vec{\nabla} \times \vec{\nabla} \times \vec{E} = \kappa^2 \epsilon \vec{E} \qquad (2.7)$$

where $\kappa = \omega/c$. Noting that :

$$\nabla \times \nabla \times \mathbf{E} = -\nabla^2\mathbf{E} + \nabla(\nabla \cdot \mathbf{E}) \quad \text{and} \quad \nabla \cdot (\epsilon\mathbf{E}) = 0, \text{ we get} \qquad (2.8)$$

$$\nabla^2 \mathbf{E} + \nabla\left\{\mathbf{E} \cdot \frac{\nabla \epsilon}{\epsilon}\right\} + \kappa^2 \epsilon \mathbf{E} = 0 \qquad (2.9)$$

At this point it is customary to decompose $\epsilon$ in terms of its average value and a fluctuating part, which is a stochastic function of space and time : It is this part that gives rise to the scattering of electromagnetic radiation. Defining :

$$\epsilon = \epsilon_0 + \delta\epsilon, \text{ where} \langle \epsilon(\mathbf{r})\rangle = \epsilon_0 \text{ and } \epsilon_1 \equiv \frac{\delta\epsilon}{\epsilon_0} \ll 1. \qquad (2.10)$$

We can write (2.9) as :

$$\nabla^2 \mathbf{E} + \nabla\left\{\mathbf{E} \cdot \frac{\nabla \epsilon_1}{1+\epsilon_1}\right\} + \kappa^2(1+\epsilon_1)\mathbf{E} = 0 \qquad (2.11)$$

The term in the curly braces describes the effects of polarization. We define $l_\epsilon$ as the smallest length scale over which $\epsilon$ changes only by a small amount and assume :

$$\kappa l_\epsilon \gg 1 \qquad (2.12)$$

$$\text{Then}, \ \left|\nabla\left\{\mathbf{E} \cdot \frac{\nabla \epsilon_1}{1+\epsilon_1}\right\}\right| \ \ll \ |\mathbf{E}|\frac{\kappa \epsilon_1}{l_\epsilon} \qquad (2.13)$$

Since the smallest scale of $|\mathbf{E}|$ being $1/\kappa$, it is clear that when compared with $|\kappa \epsilon \vec{E}|$, the polarization term can be neglected, since $\kappa l_\epsilon \gg 1$.
These arguments suggests that when $\kappa l_\epsilon \gg 1$ and $\epsilon_1 \ll 1$ the scalar wave equation is a good approximation to equation (2.9) and we can write (Tatarskii, 1969) :

$$\nabla^2 E + \kappa^2(1+\epsilon_1) = 0 \qquad (2.14)$$

This equation describes effects of diffraction as well as refraction but neglects polarization. In deriving 2.14 we ignored contribution from the Faraday rotation due to ambient magnetic field. This is justified when the difference in the refractive indices for the right and left circular polarized wave is negligible.

As the wave (**E**) propagates in the $z$ direction its phase $\propto e^{i\kappa z}$. Thus it is useful to define $u$ as follows (Lee & Jokipii, 1975I) :

$$E = u e^{i\kappa z} \qquad (2.15)$$

Using this with 2.14 we get :

$$2i\kappa\frac{\partial u}{\partial z} + \nabla^2 u + \kappa^2 \epsilon_1 u = 0 \qquad (2.16)$$

As mentioned above we are interested in a situation where $\kappa l_\epsilon \gg 1$. In other words the geometrical dimensions of the irregularities are much greater than the wavelength $\lambda$ of radiation, i.e., :

$$\left|\kappa\,\frac{\partial u}{\partial z}\right| \gg \left|\frac{\partial^2 u}{\partial z^2}\right| \qquad (2.17)$$

This is equivalent to neglecting reflection and considers only small angle propagation (Figure 2.2). The condition 2.17 is often called the quasi-optical or parabolic wave approximation. Using this we can replace $\nabla^2$ by the transverse Laplacian yielding :

$$2i\kappa\frac{\partial u}{\partial z} + \nabla_\perp^2 u + \kappa^2 \epsilon_1 u = 0 \qquad (2.18)$$

This is known as the parabolic wave equation (PWE) and is the starting point for the theory of strong fluctuations.

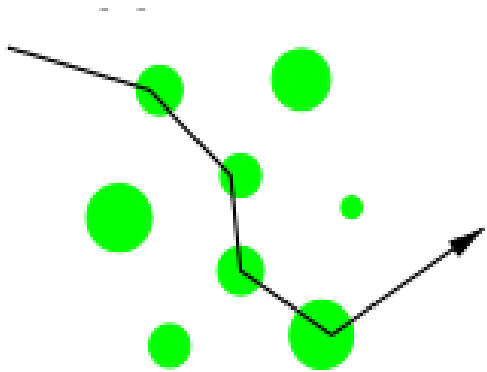

Figure 2.2: An example of small angle scattering. The parabolic wave equation accounts for multiple episodes of forward scattering and neglects backscattering.

## 2.3 Parabolic equation for moment

For further analysis it is important to describe statistical properties of u in terms of the statistical properties of the medium. We assume that the fluctuations in the dielectric constant $\epsilon(\mathbf{r}, t)$ are time-stationary. We therefore omit the time dependence from the arguments of $\epsilon_1$. Define $\mathbf{s} = (x, y)$ as transverse coordinates and z as the direction of the propagation. We can express statistical properties of $u$ in terms of an infinite set of correlation function (Lee & Jokipii, 1975I) :

$$\Gamma_{m,n}(z,\ s_1,\ s_2,\ldots,\ s_m,\ s_1^*,\ s_2^*,\ldots,\ s_n^*)$$
$$= \langle u(z, s_1)...u(z, s_m)u^*(z, s_1)...u^*(z, s_n)\rangle \quad (2.19)$$

Where $\langle \ \rangle$ defines ensemble average defined over all the configurations of the medium. $m$ and $n$ are the integers.

Next we need to find a statistical description for the turbulent medium. Since a medium is described by the fluctuations in dielectric constant ($\epsilon_1$) we need to learn how $\epsilon_1$ can be expressed in terms measured properties of a turbulent medium. Phase fluctuations $\delta\phi(z, \mathbf{s})$ can be considered as a reliable surrogate for the dielectric fluctuations. The question to be asked is how $\delta\phi$ varies from one point to another in the turbulent medium. Consider two sensors at positions $r_1 = (z, \mathbf{s}_1)$ and $r_2 = (z, \mathbf{s}_2)$. We are interested in the spatial dependence of the sensor readings averaged over all configurations of the medium. A natural way to achieve this is to define a spatial correlation function :

$$B_\phi(z, \mathbf{s}_1, \mathbf{s}_2) = \langle \delta\phi(z, \mathbf{s}_1)\delta\phi(z, \mathbf{s}_2)\rangle \quad (2.20)$$

Assuming homogeneity further simplifies the calculations. Assume that the spatial correlation function defined above remains invariant under the spa-tial translation; in other words, $B_\phi$ depends only on the difference in the arguments :

$$\text{Homogeneity implies } B_\phi(z, \mathbf{s}) = \langle \delta\phi(z,0)\delta\phi(z,\mathbf{s})\rangle, \text{ where } \mathbf{s} = \mathbf{s}_1 - \mathbf{s}_2 \quad (2.21)$$

Similarly for fluctuations in dielectric constant, we define :

$$B_\epsilon(z, \mathbf{s}) = \langle \epsilon_1(z,0)\epsilon_1(z,\mathbf{s})\rangle \quad (2.22)$$

The assumption of homogeneity can be relaxed without affecting the results but this complicates the algebra.

The problem now is to manipulate (2.18) to obtain equation for $\Gamma_{m,n}$ in terms of the statistical properties of $\epsilon_1$, specified by two point correlation function, $B_\epsilon(z, \mathbf{s})$. We stick to the dielectric constant description for now. In the next chapter, we introduce phase fluctuations as it is relevant from the point of view of measurements. The problem is considerably simpler as we need to consider only lowest values of $m$ and $n$. For example, $m = 1$ and $n = 1$ defines a second order moment, $\Gamma_{1,1}(z, \mathbf{s}_1, \mathbf{s}_2)$ which gives angular spectrum and hence angular broadening. The quantity $\Gamma_{1,1}(z, \mathbf{s}_1, \mathbf{s}_2)$ is defined as :

$$\Gamma_{1,1}(z, \mathbf{s}_1, \mathbf{s}_2) = \langle u(z, \mathbf{s}_1)u^*(z, \mathbf{s}_2)\rangle \quad (2.23)$$

To obtain a differential equation for $\Gamma_{1,1} = \Gamma_{1,1}(z, \mathbf{s}_1, \mathbf{s}_2)$, consider (2.18), for $u(z, \mathbf{s}_1)$, and multiply it with $u^*(z, \mathbf{s}_2)$ :

$$2i\kappa \frac{\partial u_1}{\partial z} u_2^* + \nabla_{\perp 1}^2 u_1 u_2^* + \kappa^2 \epsilon_1(z, \mathbf{s}_1) u_1 u_2^* = 0 \quad (2.24)$$

Here $u_1 = u(z, \mathbf{s}_1)$ and $u_2^* = u(z, \mathbf{s}_2)$, and $\nabla_{\perp 1}^2$ is the transverse laplacian with respect to $\mathbf{s}_1 = (x_1, y_1)$. Similarily we can write the conjugate of (2.24)

as follows :

$$2i\kappa \frac{\partial u_2^*}{\partial z}u_1 + \nabla_{\perp 2}^2 u_2^* u_1 + \kappa^2 \epsilon_1(z, \mathbf{s}_2)u_2^* u_1 = 0 \qquad (2.25)$$

Here $\nabla_{\perp 2}^2$ is the transverse laplacian with respect to $\mathbf{s}_2 = (x_2, y_2)$. Subtracting (2.25) from (2.24) and averaging we get :

$$2i\kappa \frac{\partial}{\partial}\langle u_1 u_2^*\rangle + (\nabla_{\perp 1}^2 - \nabla_{\perp 2}^2)\langle u_1 u_2^*\rangle + \kappa^2 \langle[\epsilon_1(z, \mathbf{s}_1) - \epsilon_1(z, \mathbf{s}_2)]u_1 u_2^*\rangle = 0 \quad (2.26)$$

If we can express the third term in (2.26) in the form of an average field, we can construct the differential equation for $\Gamma_{1,1} = \langle u_1 u_2^*\rangle$. We are thus looking for an equation of the form :

$$\langle \epsilon_1(z, \mathbf{s}_1) u_1 u_2^*\rangle = A(z, \mathbf{s}_1, \mathbf{s}_2)\langle u_1 u_2^*\rangle \qquad (2.27)$$

Where $\mathbf{A}(z, \mathbf{s}_1, \mathbf{s}_2)$ is proportional to $\langle \epsilon_1(z, \mathbf{s}_1)\epsilon_1(z, \mathbf{s}_2)\rangle$. The turbulent medium is described by the fluctuations in dielectric constant ($\epsilon_1(z, \mathbf{s})$), and the last term in (2.26), which couples $\epsilon_1(z, \mathbf{s})$ to the wave field, $u$, is therefore crucial to explain how the wave field is modified due to irregularities. One way to deal with this term is to use the Rytov approximation (Tatarskii, 1969; Ishimaru, 1978). The Rytov approximation is characterized by the Rytov condition : The logarithmic amplitude variations $\langle\chi\rangle$ in the turbulent media are small :

$$\langle\chi\rangle \ll 1 \qquad (2.28)$$

In other words this approximation assumes the intensity fluctuations to be very small as compared to the mean intensity. The Rytov approximation works well for interplanetary scintillations but for the situations where fluctuations can be comparable to the mean (e.g. interstellar scintillations, solar corona and solar wind $< 30R_\odot$), it is inapplicable. We use here the more general, Markov approximation (Lee & Jokipii, 1975I), which assumes the wave field $u$ changes only by a small amount over the correlation length ($l_\epsilon$) of $\epsilon_1(z, \mathbf{s})$. This approximation helps us deal with the third term in (2.26)

We consider following three important assumptions

- Parabolic wave approximation

- Turbulent medium is homogeneous and is described by the correlations of dielectric constant.

$$B_\epsilon(z, \mathbf{s}) = \langle \epsilon_1(z, 0)\epsilon_1(z, \mathbf{s}) \rangle$$

- Markov approximation

As noted earlier, the correlations of the dielectric constant has a direct bearing on the wave field (last term in 2.26) in the transverse direction, but it has little effect on the wave field along the direction of the propagation ($z$). This is the essence of the Markov approximation. We can quantify this approximation in a following way : fluctuations in dielectric constant ($\epsilon_1$) are delta correlated in the direction of propagation :

$$\langle \epsilon_1(z, 0)\epsilon_1(z', \mathbf{s}) \rangle = \delta(z - z')\mathbf{A}(z, \mathbf{s}) \tag{2.29}$$

The term, $\mathbf{A}(z, \mathbf{s})$ will be used later, but the link between the Markov approximation and the third term of (2.26) may now be seen if we compare (2.29) with (2.27).

The Markov approximation amounts to introducing an intermediate length scale (say, $\Delta z$ in the direction of propagation), such that $\Delta z > l_\epsilon$ but small enough so that $u(z - \Delta z, \mathbf{s}) \sim u(z, \mathbf{s})$, (Lee & Jokipii, 1975I). Given the above three assumptions with (2.29) and since $u(z, \mathbf{s})$ is a function of $\epsilon_1(z, \mathbf{s})$ we can obtain an expression for the third term in (2.26) using the Furutsu - Navikov formula (Furutsu (1963); Navikov 1965) :

$$\langle \epsilon(z, \mathbf{s}_1) u_1 u_2^* \rangle = \int d\mathbf{s}_1' \mathbf{A}(\mathbf{s}_1 - \mathbf{s}_1') \left\langle \frac{\delta(u_1 u_2^*)}{\delta \epsilon_1(z, \mathbf{s}_1)} \right\rangle \tag{2.30}$$

Here $\delta/\delta\epsilon_1(z, \mathbf{s}_1)$ denotes a functional derivative. To solve this we need to find an expression for $\delta(u_1 u_2^*)/\delta\epsilon_1(z, \mathbf{s}_1)$. Using the Markov approximation we can express the functional derivative in terms of the delta function in transverse coordinates and second order moment, $\Gamma_{1,1}$, defined in (2.26). Following Tatarskii (1969); Ishimaru (1978) we can write :

$$\frac{\delta u_1 u_2^*}{\epsilon_1(z, \mathbf{s}_1')} = \frac{i\kappa}{4}[\delta(\mathbf{s}_1 - \mathbf{s}_1')u_1 u_2^* - \delta(\mathbf{s}_2 - \mathbf{s}_1')u_1 u_2^*] \qquad (2.31)$$

Taking the ensemble average of (2.31) and substituting in (2.30), we get :

$$\langle \epsilon(z, \mathbf{s}_1) u_1 u_2^* \rangle = \frac{i\kappa}{4}[A(z, 0) - A(z, \mathbf{s}_1 - \mathbf{s}_2)]\Gamma_{1,1} \qquad (2.32)$$

We can write the conjugate of (2.32) as :

$$\langle \epsilon(z, \mathbf{s}_2) u_2^* u_1 \rangle = -\frac{i\kappa}{4}[A(z, 0) - A(z, \mathbf{s}_1 - \mathbf{s}_2)]\Gamma_{1,1} \qquad (2.33)$$

Substituting (2.32) and (2.33) in (2.26) we get the differential equation for the second order moment :

$$\left\{ 2i\kappa\frac{\partial}{\partial z} + (\nabla_{\perp 1}^2 - \nabla_{\perp 2}^2) + \frac{i\kappa^3}{2}[\mathbf{A}(z, 0) - \mathbf{A}(z, \mathbf{s}_1 - \mathbf{s}_2)] \right\} \Gamma_{1,1} = 0 \qquad (2.34)$$

This is the parabolic equation for the second order monent, $\Gamma_{1,1}$. We can further simplify it using the assumption of homogeneity, which says that :

$$\langle u_(z, \mathbf{s}_1) u^*(z, \mathbf{s}_2) \rangle = \Gamma_{1,1}(z, \mathbf{s}) \qquad (2.35)$$

where $\mathbf{s} = \mathbf{s}_1 - \mathbf{s}_2$. This implies $\nabla \langle u(z, \mathbf{s}) \rangle = 0$, and hence $(\nabla_{\perp 1}^2 -$

$\nabla_{\perp 2}^2)\Gamma_{1,1}(z,\mathbf{s}) = 0$. Thus we can write (2.34) as :

$$\left\{\frac{\partial}{\partial z} + \frac{\kappa^2}{4}[\mathbf{A}(z,0) - \mathbf{A}(z,\mathbf{s})]\right\}\Gamma_{1,1} = 0 \tag{2.36}$$

Integrating (2.36) along the direction of propagation and imposing the initial condition, $\Gamma(z,0) = 1$, the general solution can be written as :

$$\Gamma_{1,1}(z,\mathbf{s}) = \exp\left\{-\frac{\kappa^2}{4} z \left[\mathbf{A}(z,0) - \mathbf{A}(z,\mathbf{s})\right]\right\} \tag{2.37}$$

The term in the curly braces of Eq. (2.37) is called the structure function.

$$D_\epsilon(z,\mathbf{s}) = \frac{\kappa^2}{4}\left[\mathbf{A}(z,0) - \mathbf{A}(z,\mathbf{s})\right] z \tag{2.38}$$

We will be concerned primarily with the structure function in (chapter 3) and it will also play an important role as a tool in the calculations of rate of turbulent heating (chapter 4). The $D_\epsilon(z,\mathbf{s})$ defined in (2.38) is a structure function for $\epsilon_1(z,\mathbf{s})$, which are fluctuations in the dielectric constant of a turbulent medium. In chapter 3 when we introduce the structure function for the phase fluctuations $(D_\phi(z,\mathbf{s}))$, we will demonstrate that both these defenitions are identical.

To find a general expression for the structure function $(D_\epsilon(z,\mathbf{s}))$ we need to simplify the term in square bracket of (2.38). We now take a detour to see how the correlations of $\epsilon_1$ in a turbulent medium can be related to $\mathbf{A}(z,\mathbf{s})$.

## 2.4  The wavenumber spectrum

Consider the Fourier transform of the correlation of $\epsilon_1$ at points $\mathbf{r}_1$ and $\mathbf{r}_2$ in a turbulent medium, (Ishimaru, 1978) :

$$\langle\epsilon_1(\mathbf{r}_1)\epsilon_1(\mathbf{r}_2)\rangle = \iiint d\boldsymbol{\kappa}\ \Phi_\epsilon(\boldsymbol{\kappa})\exp[i\boldsymbol{\kappa}\cdot(\mathbf{r}_1 - \mathbf{r}_2)] \tag{2.39}$$

Here $d\boldsymbol{\kappa} = d\kappa_x d\kappa_y d\kappa_z$. Taking the inverse Fourier transform of (2.39) and using (2.29) :

$$\Phi_\epsilon(\boldsymbol{\kappa}) = \frac{1}{2\pi} \int dz' \iint d\mathbf{s}\ \delta(z-z')\, \mathbf{A}(z',\mathbf{s}) \exp[-i\boldsymbol{\kappa}_\perp \cdot \mathbf{s}]$$

Here $d\mathbf{s} = dxdy$ and $\mathbf{s} = \mathbf{r}_\perp = (x,y)$, is a transverse coordinate. Thus with $d\boldsymbol{\kappa}_\perp = d\kappa_x d\kappa_y$ we get :

$$\mathbf{A}(z,\mathbf{s}) = 2\pi \int \left\{ \iint d\boldsymbol{\kappa}_\perp\ \Phi_\epsilon(\boldsymbol{\kappa}) \exp[i\boldsymbol{\kappa}_\perp \cdot \mathbf{s}] \right\} dz \qquad (2.40)$$

In deriving this, the assumption of homogeneity has been used to simplify the problem. We can further simplify it by assuming isotropy :

$$\Phi_\epsilon(\boldsymbol{\kappa}) \mapsto \Phi_\epsilon(\kappa) \qquad (2.41)$$
$$B_\epsilon(z,\mathbf{s}) \mapsto B_\epsilon(z,s) \qquad (2.42)$$

In other words the correlation depends only on the magnitude of the difference between the transverse coordinates. Though the problem in general is anisotropic we will learn in chapter 3 isotropy is a good assumption in several situations.

The isotropic form of (2.40) is given by :

$$\mathbf{A}(z,s) = 4\pi^2 \int \left\{ \int \kappa d\kappa\ \Phi_\epsilon(\kappa) \left(\frac{\sin(\kappa s)}{\kappa s}\right) \right\} dz \qquad (2.43)$$

The irregularities of a turbulent medium are completely described by $\Phi_{\epsilon(\boldsymbol{\kappa})}$, the spatial power spectrum of the correlations of $\epsilon_1(z,\mathbf{s})$. The expression for

**A**(z, 0) can be obtained easily using **s** = 0 in (2.40). From (2.38) we get :

$$D_\epsilon(z, \mathbf{s}) = \frac{\pi \kappa^2}{2} \int_0^z \mathrm{d}z \left\{ \iiint_{-\infty}^{\infty} \mathrm{d}\boldsymbol{\kappa}\ \Phi_\epsilon(\boldsymbol{\kappa})[1 - \exp(i\boldsymbol{\kappa}_\perp \cdot \mathbf{s})] \right\} \quad (2.44)$$

Assuming isotropy :

$$D_\epsilon(z, \mathbf{s}) = 2\pi^2 \kappa^2 \int_0^z \mathrm{d}z \left\{ \int_0^{\infty} \kappa \mathrm{d}\kappa\ \Phi_\epsilon(\kappa) \left[1 - \frac{\sin(\kappa \mathrm{s})}{\kappa \mathrm{s}}\right] \right\} \quad (2.45)$$

Equation (2.44) forms the basis for our work presented in chapter 3.

### 2.4.1 Interpretation of $\kappa$

The physical interpretation of the wavenumber can be understood as follows. Instabilities in the ambient field produces large scale irregularities, often reffered to as eddies. The large scale eddies breaks into smaller ones producing hierarchy of scales ($l$), which is the hallmark of turbulence. Eddies of different sizes affect measurements of spatial covariance in different way. Data obtained from an interferometer of baseline **s** (vector seperation between the sensors/antennas) shows that the covariance decreases with increasing **s**. This gives an important clue suggesting that only those eddies which satisfy (Wheelon, 2001) :

$$|\mathbf{s}| < l \quad (2.46)$$

are going to contribute significantly to the measurements of spatial covariance. This condition can be expressed in terms of the wavenumber as follows. For simplicity we choose the isotropic case. The contribution from the sine term (known as weighting function) in (2.43) is significant when the argument ($\kappa s$) satisfies :

$$\kappa s \ll 1 \qquad (2.47)$$

Comparing the conditions expressed by (2.46) and (2.47) we conclude :

$$\kappa \sim 1/l \qquad (2.48)$$

This suggests that large eddies corrosponds to small wavenumber and vice versa. Although we introduced the spectral representation as a mathematical convinience we see that the sizes of the different irregularities can be identified with inverse wavenumbers. Therefore the spectrum describes the relative ability of different eddy sizes to influence the measurements of correlations of the dielectric constant in turbulent medium.

The second important advantage of this approach is the simplification of the propagation integral (2.44). Transforming the integral into space and wavenumber representation allows one to express a measured quantity in the following way (Wheelon, 2001) :

$$\text{Measured Quantity} = \int d\boldsymbol{\kappa} \; \Phi_\epsilon(\boldsymbol{\kappa}) \left[\text{Weighting function of } \kappa\right] \qquad (2.49)$$

The problem can now be seen in terms of two modules coupled by the wavenumber integral (e.g. Eq. 2.45 in the light of Eq. 2.49). The modules are,

- Wavenumber spectrum $\Phi_\epsilon(\boldsymbol{\kappa})$, which describes the turbulent medium.
- Weighting function, which characterizes the propagation features of the electromagnetic wave.

Together they generate a complete description of a measured quantity. In the discussion till now we have analysed almost all the electromagnetic features of propagation without an explicit mention of a model for the turbulent medium. The approach of representing turbulent medium in terms of a wavenumber spectrum makes this possible. The choice of the model for a turbulent medium is postponed to the next chapter.

## 2.5 The structure function

We now make a few remarks regarding the structure function which demonstrates its utility.

Consider the correlation of refractive index fluctuations at two points $n(A)$ at A and $n(B)$ at B in the turbulent medium. The points A and B are seperated by a distance $l$ as shown in figure (2.3). Let $l_1$ be the scale of the irregularities smaller than that of the distance between A and B. $l_2$, on the other hand, is the scale of irregularities that is $\gg$ the distance between A and B. The distance between A and B is denoted by $l$. We can see from figure that $l_2 \gg l$ and its effect on both points is almost identical. However it does not contribute to the difference $n(A) - n(B)$. Similarly irregularities with $l_1 \ll l$ contributes only a small amount to the difference $n(A) - n(B)$. Therefore the main contribution is caused by the irregularities having scale comparable to the distance between A and B. This suggests that the value of the structure function is a measure of the intensity of turbulent irregularities having scale $l$. We can readily connect $l$ with the interferrometric baseline $|\mathbf{s}|$. This not only allows us to produce a quantitative description of the observed phenomenon due to scattering, but also helps in constraining parameters that characterize turbulence in the intervening medium.

## 2.6 Summary

We have provided a brief description of the theory of scattering of electromagnetic waves in a turbulent media. Starting with the Maxwell equations in a source free region and using the small angle scattering we reproduced an equation for the wave propagation in turbulent medium Eq. (2.18). Fluctuations in the dielectric constant necessitate a statistical approach to deduce the effects of turbulent fluctuations on the wave propagation. Using the Markov approximation on the variability of the wave field in the direction of propagation, we can obtain the parabolic equation for the second moment of the wave field Eq. (2.34). A solution of this equation to first order yields angular broadening and defines structure function via two point correlations of fluctuations in the dielectric constant Eq. (2.44). The structure function described in terms of the wavenumber spectrum helps us simplify the propagation integral. The physical insight this picture offers is the simple identification of the wavenumber with the size of irregularities (eddies) in a turbulent medium. This connects the theoratical description with measured properties of the intervening medium like the solar wind.

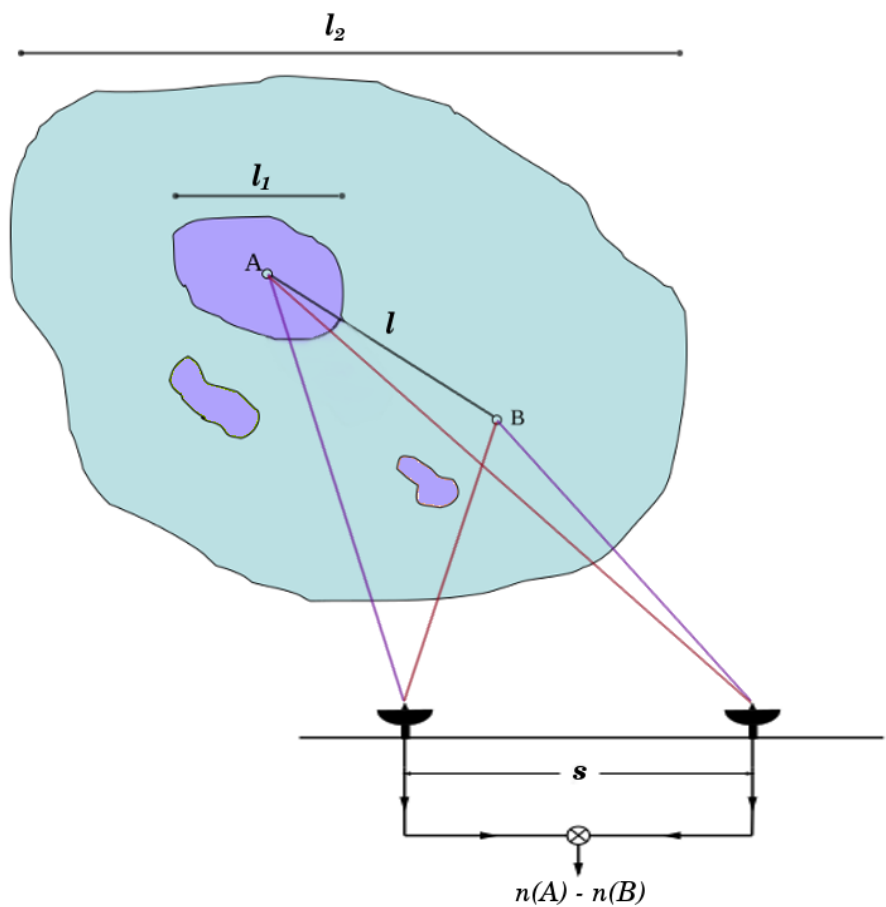

Figure 2.3: Schematic diagram to illustrate effects of irregularities on the measured properties of a turbulent medium. The structure function is a measure of the intensity of irregularities having scales comparable to the distance ($l$) between points A and B in a turbulent medium. Small scale ($l_1 \ll l$) irregularities are shown in purple. Irregularities with $l_2 \gg l$ are shown in blue. The structure function can be defined in terms of difference in correlations of refractive index fluctuations, at two different points in a turbulent medium, $n(A) - n(B)$.

In the next chapter we continue bulding on these concepts and explore the general form of the structure function and its role in the radio wave scattering. We will see how the interplay between the scales of irregularities and interferometric baselines provides constraints on the form of the structure function to be used while predicting phenomenon due to the scattering of electromagnetic waves.

# Chapter 3

# The general structure function (GSF)

*Calculations of angular broadening of radio sources due to scattering off turbulent density inhomogeneities in the solar wind usually employ asymptotic approximations to the phase structure function. We use a general structure function (GSF) that does not use these approximations. We show that the regimes where the GSF predictions are more accurate than those of its asymptotic approximations is of practical relevance, and are also the regimes where inner scale effects influence the estimate of angular broadening. The findings in chapter (3) have been published in Ingale et al. (2015)*

## 3.1 Introduction

We learned from chapter 2 that the interaction between plasma inhomogeneities and electromagnetic waves give rise to the scattering of electromagnetic waves in the solar wind. This interaction is determined by the dielectric constant, $\epsilon$ or equivalently by refractive index ($\epsilon = \mu^2$). If $\omega$ is the angular frequency of the propagating wave and $\omega_p = 4\pi n_e^2 e^2/m_e$ is theÂăelectron plasma frequency, the dielectric constant, $\epsilon$ is given by :

$$\epsilon = 1 - \frac{\omega_p^2}{\omega^2} = 1 - \frac{\lambda^2 r_e n_e}{\pi} \tag{3.1}$$

where $\lambda$ is the wavelength of the electromagnetic wave, $r_e = e_2/m_e c^2$ is the classical electron radius. Plasma irregularities are described by the fluctuations in the dielectric constant, ($\epsilon_1$) given by :

$$\epsilon_1 = \frac{\lambda^2 r_e}{\pi} \delta N_e = 2\delta\mu \tag{3.2}$$

Here $\delta N_e$ denotes the fluctuations in the background electron density. Interaction between the electromagnetic wave and plasma irregularity, over a characteristic length scale $\delta z$ in the direction of the propagation give rise to fluctuations in the phase ($\delta\phi$) :

$$\delta\phi = \kappa \Delta z \delta\mu = \lambda r_e \Delta z \delta N_e \tag{3.3}$$

Therefore the phase front of an electromagnetic wave passing through the turbulent medium gets corrugated and an observer receives radiation from a range of angles; in other words the source has suffered angular broadening with a characteristic angular width $\theta_s$. Since the source is now broadened and not point-like there is a geometrical time delay resulting in temporal broadening of the source. If the turbulent medium is moving with the velocity $v_s$ in the direction transverse to the observer, a monochromatic source, in addition to angular / temporal broadening, experiences spectral broadening. Thus phase fluctuations of the electromagnetic wave traversing a turbulent medium gives rise to a range of scattering phenomena.

Phase fluctuations can be measured by using interferometric observations of a spatially coherent source with a pair of antennas separated by a distance **s** known as the baseline. Usually a phase deviation $\Delta\phi = \delta\phi(\mathbf{r}) - \delta\phi(\mathbf{r} + \mathbf{s})$ is measured over time scales much smaller than the scale over which solar wind irregularities causes a drift in the scale of the baseline **s** of the interferometer.

The structure function is then readily estimated using $D_\phi(\mathbf{s}) = \langle [\Delta\phi(\mathbf{s})]^2 \rangle$, known as phase structure function (Coles & Harmon, 1989; Spangler & Sakurai, 1995). Scattering phenomena can be quantitatively described by Eq. (2.44).

With the propagation integral given by Eq. (2.44) we can think of the problem in two ways. First if we know the form of the spatial power spectrum of density fluctuations we can integrate Eq. (2.44) and obtain the structure function; conversely, given measurements of the structure function we can determine the spatial power spectrum and thus directly get handle on the parameters that characterize density turbulence in the solar corona. We will explore both of these approaches. This chapter concerns with the first approach where we use the empirical information of the spatial power spectrum of density turbulence to compute the structure function. Chapter 4 deals with the later approach where we can constrain the parameters characterizing the density turbulence, using measurements of the structure function obtained by interferometric experiments.

## 3.2 Phase structure function

To illustrate various propagation phenomena, quantitatively described by $D_\phi$, consider a Cartesian coordinate system with $z$ being the direction of propagation. The observer is in the $x, y$ plane, located at a distance $L$ along the $z$ axis.

Suppose the turbulent medium fills the half space $z > 0$. The role of the turbulent medium is to change the phase of the incident wave randomly as a function of position and time (Figure 3.1). A coherent wave incident on the medium at $z = 0$, experiences loss of spatial as well temporal coherence due to random phase fluctuations. We have seen that this gives rise to angular broadening of the source with a characteristic angular width $\theta_c$, also known as the scattering angle. To obtain a quantitative description of angular broadening therefore we need to connect $\theta_s$ with the structure function $D_\phi$.

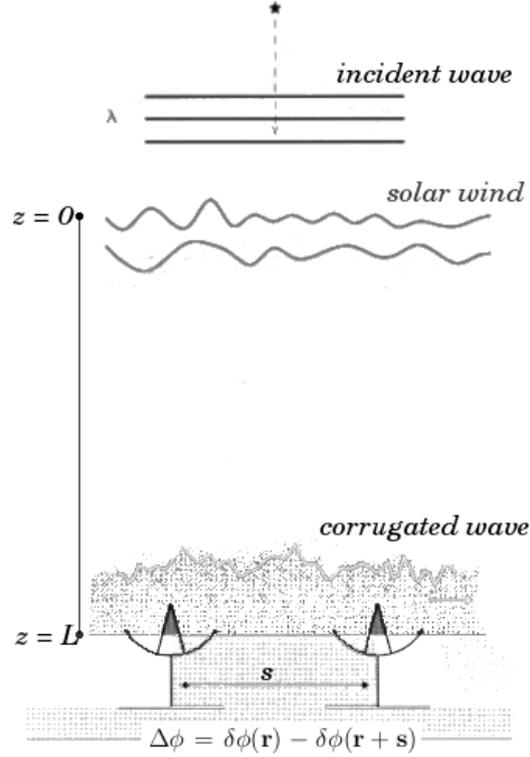

Figure 3.1: Schematic of an experiment to measure phase deviation of the incident plane wave due to turbulent solar wind.

We know that the statistical properties of a turbulent medium at $z = L$ can be described by the spatial correlation functions, $B_\epsilon, B_\phi$, or $B_{N_e}$ of $\epsilon_1, \delta\phi$ or $\delta N_e$ (Bastian, 2000), using (3.2) and (3.3) :

$$\begin{aligned} B_\epsilon(\mathbf{s}) &= \langle \epsilon_1(\mathbf{r})\epsilon_1(\mathbf{r}+\mathbf{s}) \rangle \\ &= \left(\frac{\pi^2}{\lambda^2 \Delta z}\right) B_\phi(\mathbf{s}) \\ &= (r_e^2 \lambda^4/\pi) B_{N_e}(\mathbf{s}) \end{aligned} \qquad (3.4)$$

Where $\mathbf{r} = (x, y, z = L)$ and $\mathbf{r_1} = \mathbf{r} + \mathbf{s}$, with $\mathbf{s} = (x', y', 0)$ are the

transverse coordinates. In the context of an interferometer on the Earth measuring the correlation, given by (3.4), the quantity $|\mathbf{s}|$ can be interpreted as the interferometer baseline. Following equations (2.29), (2.37) and (2.38) we can alternately define the structure function for $\epsilon_1$ as follows :

$$D_\epsilon(\mathbf{s}) = 2[\langle \epsilon_1(\mathbf{r})^2 \rangle - \langle \epsilon_1(\mathbf{r})\epsilon_1(\mathbf{r}+\mathbf{s}) \rangle] \quad (3.5)$$

It is evident from (3.2) and (3.3) that the treatment of the phase fluctuations is equivalent to the one described in previous chapter for the fluctuations in the dielectric constant (also see §2.3). We can therefore define the phase structure function as :

$$D_\phi(z, \mathbf{s}) = 2[\langle \delta\phi(\mathbf{r})^2 \rangle - \langle \delta\phi(\mathbf{r})\delta\phi(\mathbf{r}+\mathbf{s}) \rangle] \quad (3.6)$$

which is related to (3.5) via (3.4). The relation between the phase fluctuations $\delta\phi$ experienced by the wave propagating through the turbulent medium and the density fluctuations $\delta N_e$ depends on the geometry of the scattering. In general we can write (Spangler, 1996) :

$$\delta\phi(\mathbf{r}) = r_e \lambda \int_0^L \delta N_e(\mathbf{r})\,\mathrm{d}z \quad (3.7)$$

Therefore the phase structure function $D_\phi(\mathbf{s})$ can be written as :

$$D_\phi(z, \mathbf{s}) = 2r_e^2 \lambda^2 \int_0^L [\langle \delta N_e(\mathbf{r})^2 \rangle - \langle \delta N_e(\mathbf{r})\delta N_e(\mathbf{r}+\mathbf{s}) \rangle]\,\mathrm{d}z \quad (3.8)$$

Following a similar line of investigation as in Chapter 2 we can use the wavenumber representation. The Fourier transform of the correlations of density fluctuations is the spatial power spectrum :

$$\langle \delta N_e(\mathbf{r})\delta N_e(\mathbf{r}+\mathbf{s}) \rangle = \int_{-\infty}^{\infty} S_n(z, \boldsymbol{\kappa})\exp[i\boldsymbol{\kappa}\cdot\mathbf{s}]\,\mathrm{d}\boldsymbol{\kappa} \quad (3.9)$$

Here $\boldsymbol{\kappa}_\perp = (\kappa_x, \kappa_y, 0)$ is the transverse wave number and the $S_n(z, \boldsymbol{\kappa})$ is the spatial power spectrum of the density fluctuations. Thus using (3.8) and (3.9) we can write $D_\phi$ in terms of the wavenumber spectrum representation (equivalent to Eq. 2.49),

$$D_\phi(z, \mathbf{s}) = 8\pi^2 r_e^2 \lambda^2 \int_0^L dz \iint_{-\infty}^{\infty} d\boldsymbol{\kappa}\, S_n(z, \boldsymbol{\kappa})\left[1 - \exp(i\boldsymbol{\kappa} \cdot \mathbf{s})\right] \quad (3.10)$$

The value of $\mathbf{s}$ where $D_\phi(z, \mathbf{s}) = 1$ gives a measure of the extent of broadening of an ideal point source due to the scattering caused by the turbulent density fluctuations. This value of $\mathbf{s}$ is denoted by $\mathbf{s}_0$ and is called the coherence scale or diffraction scale. It is related to $\theta_s$ by :

$$\boldsymbol{\theta}_s = (|\boldsymbol{\kappa}|\, \mathbf{s}_0)^{-1} \quad (3.11)$$

Thus the phase structure gives direct information on the angular broadening of a source when observed against the background of a turbulent medium like the solar wind. Since the phase structure function contains the spatial power spectrum of the density fluctuations, information about the observed angular extent of the sources gives a good handle on the parameters characterizing the turbulence spectrum in the solar wind. We now briefly examine different geometries involved in the scattering problem.

### 3.2.1 Thin screen geometry

The irregularities of a turbulent medium can be assumed to be concentrated in a thin, two-dimensional screen. This assumption is well justified in many problems of interest (e.g. radio wave scattering from a distant celestial source) and also greatly simplifies the analysis.

If we consider the screen to be located at $z = 0$, the role of the screen is to introduce a position dependent phase change $\delta\phi(\mathbf{r})$ to the incident wave.

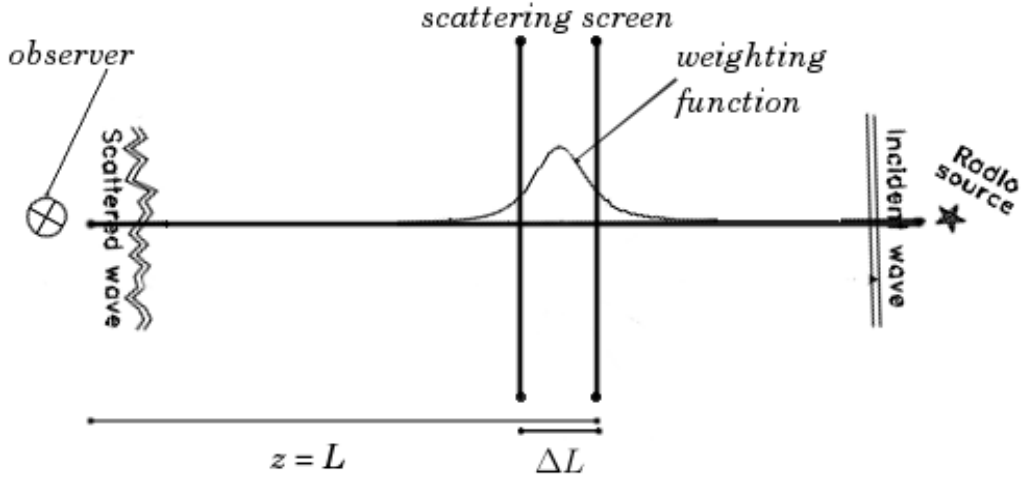

Figure 3.2: Illustration depicting the effect on a wave traversing a thin screen of turbulent irregularities. Scattering of electromagnetic wave is quantitatively described by the weighting function (2.49)

Equation (3.10) with a scattering screen of thickness $\Delta L$ yields :

$$D_\phi(z, \mathbf{s}) = 8\pi^2 r_e^2 \lambda^2 \Delta L \iint_{-\infty}^{\infty} d\boldsymbol{\kappa}\, S_n(z, \boldsymbol{\kappa}) \left[1 - \exp(i\boldsymbol{\kappa} \cdot \mathbf{s})\right] \quad (3.12)$$

### 3.2.2 Spherical and plane wave propagation

The extent of scatter broadening depends on whether the wavefront is planar (1-D) or spherical (3-D). When a source is embedded in the scattering medium, as is the case for sources in the solar corona (Bastian, 1994; Subramanian & Cairns, 2011) it is appropriate to adopt a formalism that includes the spherically diverging nature of the wavefront. We will later see that the assumption of isotropy in the turbulent spectrum is also justified in this situation.

For the spherically diverging wavefront, the observer is sensitive to a range of eddy sizes (scales of irregularities) given by $sa/b$, where $s$ is the interferometer baseline, $a$ is the distance of the scattering screen from the source and $b$ is the distance of the observer from the source; see Subramanian & Cairns (2011) for details. In other words, the effective baseline for spherical wave propagation at a given heliocentric distance $R$ is (Ishimaru, 1978) :

$$s_{\text{eff}} = sR/(R_1 - R_0),  \qquad (3.13)$$

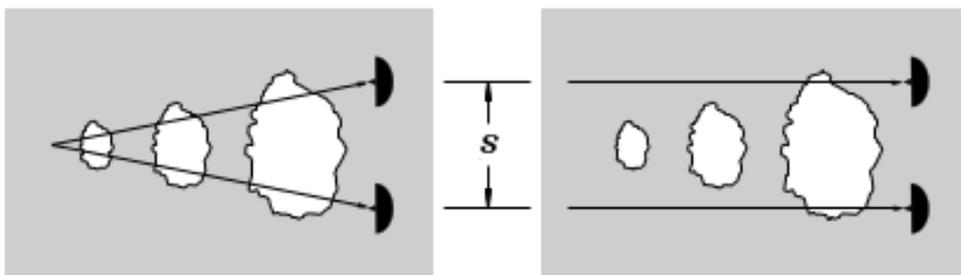

Figure 3.3: (Adapted from Wheelon (2001)) Schematic showing range of eddy sizes, observer is sensitive to, in case of spherical as well as plane wave propagation.

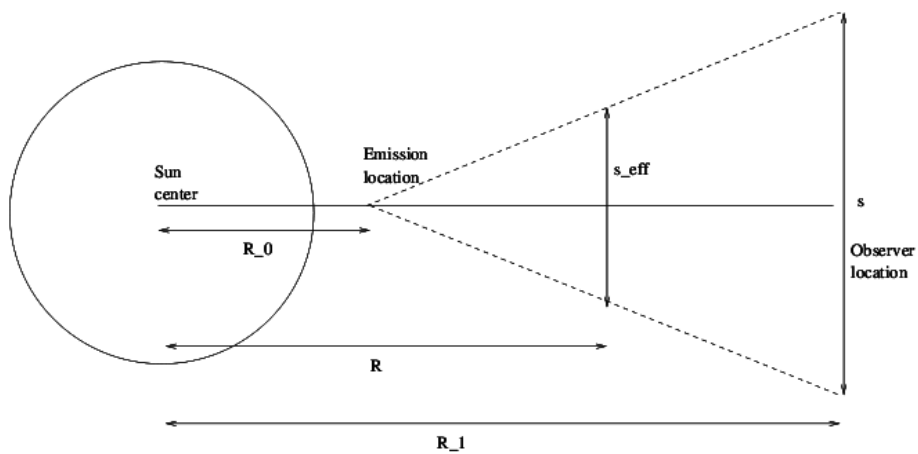

Figure 3.4: (Adapted from Subramanian & Cairns (2011)) Geometry for the spherical wave propagation

where $R_1$ is the heliocentric distance at which the observation of angular broadening is made and $R_0$ is the heliocentric distance at which the source is situated (figure 3.4).

The planar (1D) formalism, on the other hand, is appropriate when the source, scattering region(s), and observer are all far apart from each other, as assumed in calculations for distant celestial sources viewed through the scattering screen of the solar wind. We will see later that a treatment of anisotropic scattering is essential for this situation. In this case an observer is typically sensitive only to scattering regions (eddies) with sizes of the order of the interferometer baseline $s$ (Figure 3.3). In other words, $s_{\text{eff}} = s$ for plane wave propagation.

### 3.2.3 Turbulence spectrum

We know from the propagation integral (2.27) that the turbulent medium is described by the spatial power spectrum of density fluctuations $S_n(z, \boldsymbol{\kappa})$. To proceed further it is necessary to consider an actual model for the turbulence spectrum. Measurements of solar wind density fluctuations by various spacecraft have provided important inputs regarding the form of the turbulence spectrum. Observations in the low speed solar wind between 0.3 and 1 AU at frequencies $< 0.1$ Hz reveal that the spatial power spectrum of electron density fluctuations in the solar corona largely follow the Kolmogorov scaling (Marsch & Tu, 1990; Tu & Marsch, 1995; Coles & Harmon, 1989). The turbulent density spectrum is commonly modeled as a power law in wavenumber space. It is known that at the smallest scales the density spectrum displays an abrupt steepening (Coles & Harmon, 1989) indicating the existence of the inner scale. In this region an exponential cut-off is often a good approximation to a steeper power law (Bale et al., 2005; Alexandrova et al., 2012). Models also suggest that the dissipation range is an exponential cutoff, implying that observations of steeper power laws might arise from instrumental limitations (Howes et al., 2008). The turbulent density spectrum is often written as :

$$S_n(z, \boldsymbol{\kappa}) = C_N(z)^2 \, |\boldsymbol{\kappa}|^{-\alpha} \exp[-\kappa^2/\kappa_{\text{i}}^2] \qquad (3.14)$$

Here the quantity $C_N^2(z)$ is the amplitude of density turbulence and $\alpha$ is the power law index. The quantity $l_i = 2\pi/|\boldsymbol{\kappa_i}|$ is generally referred to as the inner scale, where dissipation sets in (exponential turnover). The steepening of the spectrum beyond the inner scale is often attributed to the dissipation of the turbulent eddies and associated waves (Coles, 1978; Woo & Armstrong, 1979).

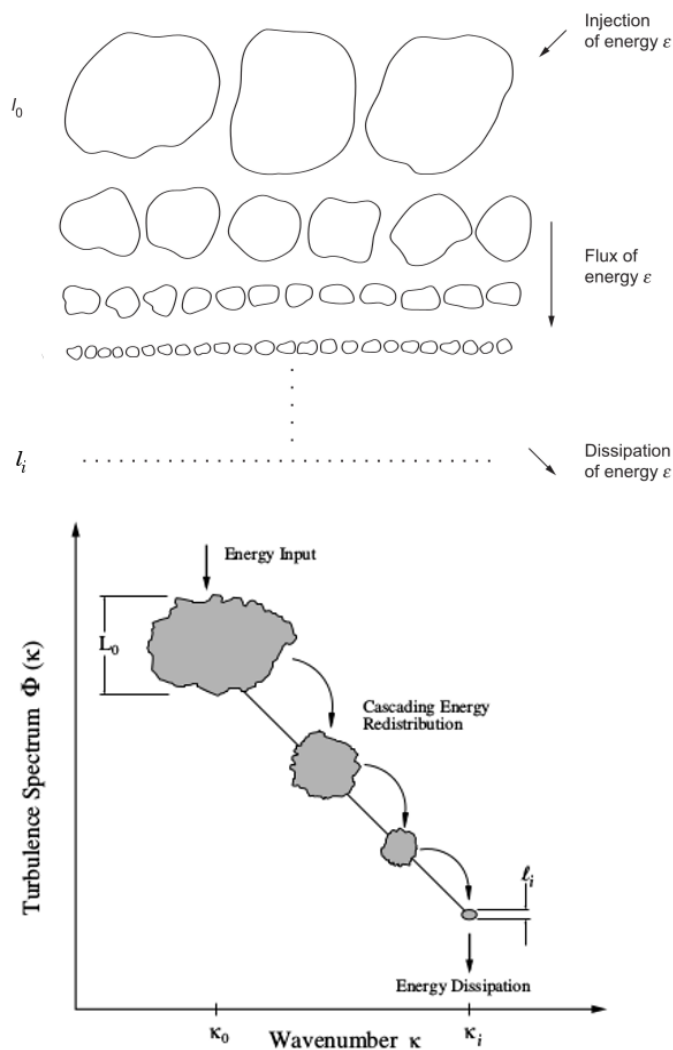

In-situ observations near the Earth at higher frequencies ($> 0.1$Hz) find no evidence for any deviation from the Kolmogorov scaling as a function of the angle with respect to the local magnetic field direction (Celnikier et al., 1987; Tu & Marsch, 1995). At higher frequencies, however, there is evidence for steepening at wave numbers $\approx 2\pi/l_i$. In high speed solar wind streams, there seems to be some evidence for spectral flattening at high frequencies prior to the inner scale.

### 3.2.4 Anisotropic scattering

An important result from angular broadening observations is that the solar wind turbulence is highly anisotropic. Scatter-broadened images of distant celestial sources viewed on foreground of the solar wind at small elongations from the Sun tend to be strongly anisotropic with axial ratio as high as $\sim 16$ (Anantharamaiah et al., 1994; Coles & Harmon, 1989; Coles et al., 1987).

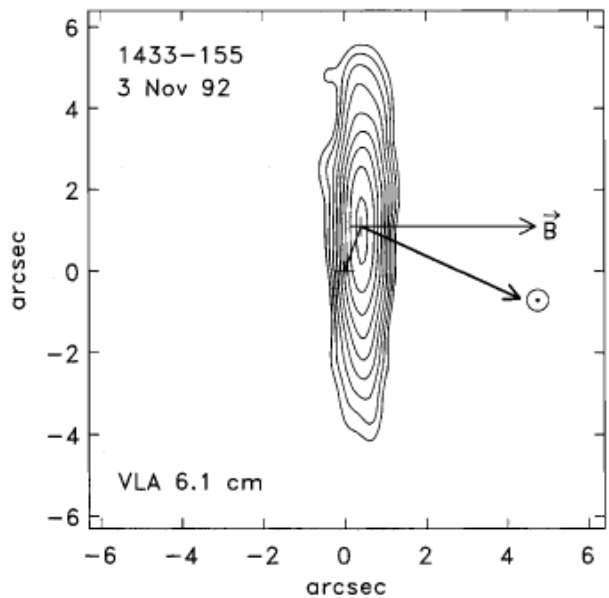

Figure 3.5: (Adopted from Bastian, 2000) An example of a source with an anisotropic brodening due to turbulent irregularities in the solar wind. Anisotropy is evident is predominantly in the direction of the magnetic field. The direction of the Sun is indicated by $\odot$

The scatter-broadened images are observed to be elongated along the direction of the (predominantly radial) large-scale magnetic field. One must therefore consider the effect of anisotropy while calculating the phase structure function for the scattering in these cases. The thin screen geometry is appropriate for this situation (Coles et al., 1987; Coles & Harmon, 1989).

Define a cartesian coordinate system $x, y, z$, with $z$ along the line of sight. For plane wave propagation, which is relevant for radiation from distant background celestial sources viewed against the turbulent solar wind scattering, density inhomogeneities are concentrated in a thin screen of thickness $\Delta L$, located at $z = 0$ between the source and the observer. In this case, the transverse coordinates $x$ and $y$ are in the plane of the scattering screen, which is perpendicular to the line of sight. The $x$ coordinate is taken to be along the projection of the local magnetic field vector into the $x - y$ plane and at small elongations, it is observed that scatter broadened images are typically stretched along the $x$ direction. This can be treated using a formalism where the underlying turbulent eddies are also elongated in the $x$ direction.

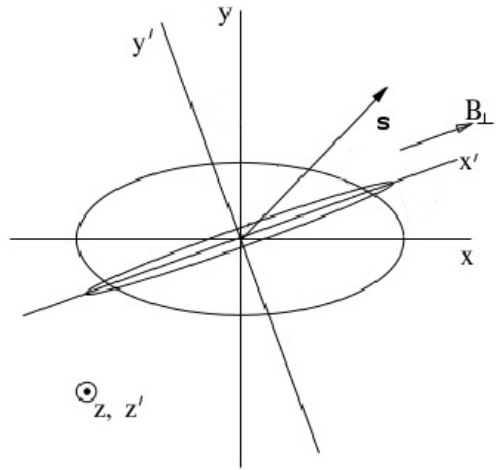

Figure 3.6: Geometry for the anisotropic scattering. z is along the line of sight and x' is such that the direction of $B_\perp$ is always along x'. The projection of the density structure appears as eccentric ellipse elongated along x'

Narayan & Hubbard (1988) developed a theory of refractive scintillation that includes anisotropy. They model the turbulent spectrum as a power law with the power law index $\alpha \leq 4$, but do not include a cutoff due to dissipation. We extend the formulation of Narayan & Hubbard (1988) to include the exponential cut-off together with the power law spectrum. The anisotropic generalization of the power law spectrum with exponential cutoff

can be written as :

$$S_n(z, \boldsymbol{\kappa}) = C_N^2(R) \, (\rho^2 \kappa_x^2 + \kappa_y^2)^{-\alpha/2} \exp[-(\rho^2 \kappa_x^2 + \kappa_y^2)(\mathbf{l_i}/2\pi)^2]. \qquad (3.15)$$

The axial ratio $\rho$ ($> 1$) measures the degree of anisotropy, which can be interpreted in the following way : if the density blob in the screen has length $l$ in the y direction, it is elongated to $\rho l$ in the x direction. In writing (4.1) we have assumed, following Narayan & Hubbard (1988), that :

- The axial ratio, $\rho$ is independent of the spatial scale .

- The focusing action of the density gradient in the scattering screen is negligible.

We assume that the inner scale ($\mathbf{l_i} = 2\pi/\mathbf{k_i}$) is anisotropic such that $l_{ix} = \rho l_{iy}$.

We replace the usual Cartesian coordinate system with the coordinates $\kappa_r$ and $\xi$ defined as :

$$k_r = (\rho^2 k_x^2 + k_y^2)^{1/2} \; ; \; \xi = tan^{-1}\left(\frac{k_y}{\rho k_x}\right) \qquad (3.16)$$

Accordingly, the area element will be scaled as :

$$d^2\kappa = \frac{\kappa_r}{\rho} \, d\kappa_r \, d\xi \qquad (3.17)$$

Using the coordinate transformation defined in 3.17, we obtain the following expression for the correlation of density fluctuations :

$$\langle \delta N_e(\mathbf{r})\delta N_e(\mathbf{r}+\mathbf{s})\rangle = \frac{1}{\rho}\int_0^\infty \kappa_r\, d\kappa_r\, S_n(z,\kappa_r) \int_0^{2\pi} d\xi\, \exp[i\kappa_r|\mathbf{s}|\cos\xi] \quad (3.18)$$

$$= \frac{1}{\rho}\int_0^\infty \kappa_r\, d\kappa_r\, J_0(\kappa_r s)\, S_n(z,\kappa) \;, \quad (3.19)$$

where $J_0$ is the Bessel function of the first kind. Thus for the scattering screen of thickness $\Delta L$ the phase structure function (3.12), using (4.1) and (3.19), can be written as :

$$D_\phi(z,\mathbf{s}) = \frac{1}{\rho}8\pi^2 r_e^2 \lambda^2 \Delta L C_N^2(z) \int_0^\infty [1-J_0(\kappa_r s)]\kappa_r^{1-\alpha}\exp[-(\kappa_r l_i/2)^2]\,d\kappa_r. \quad (3.20)$$

Integrating over $\kappa_r$ we obtain :

$$D_\phi(z,\mathbf{s}) = \frac{1}{\rho}\frac{8\pi^2 r_e^2 \lambda^2 \Delta L}{2^{\alpha-2}(\alpha-2)}\Gamma\left(1-\frac{\alpha-2}{2}\right)C_N^2(z)l_i^{\alpha-2}(z)$$
$$\times \left\{{}_1F_1\left[-\frac{\alpha-2}{2},1,-\left(\frac{s}{l_i(z)}\right)^2\right] - 1\right\} \;. \quad (3.21)$$

Cairns (1998) has showen that the PWE formalism for the angular broadening (Chapter 2) can be extended to include non-zero, spatially varying ratio $f_p/f < 1$, where $f_p$ is the plasma frequency and $f = 2\pi c/\lambda$ is the radiation frequency. Consequently the expression for the propagation integral (3.21) and thus scattering angle $\theta_s$ can be modified by including a factor $[1 - f_p(z)^2/f^2]$ inside the path integral. Thus Eq. (3.21) is modified to :

$$D_\phi(z,\mathbf{s}) = \frac{1}{\rho}\frac{8\pi^2 r_e^2 \lambda^2 \Delta L}{2^{\alpha-2}(\alpha-2)}\Gamma\left(1-\frac{\alpha-2}{2}\right)\frac{C_N^2(z)l_i^{\alpha-2}(z)}{(1-f_p^2(z)/f^2)}$$
$$\times \left\{{}_1F_1\left[-\frac{\alpha-2}{2},1,-\left(\frac{s}{l_i(z)}\right)^2\right] - 1\right\} \;, \quad (3.22)$$

where $_1F_1$ denotes the confluent hypergeometric function also known as the Kummer function. Defining $G = s/l_i$ the approximation $s \ll l_i$ corrsponds to $G \to 0$ and the approximation $s \gg l_i$ to $G \to \infty$ :

$$\lim_{\zeta \to \infty} 1F_1[a, b, \zeta] = \frac{\Gamma(b)(-\zeta^{-a})}{\Gamma(b-a)}$$
$$\lim_{\zeta \to 0} 1F_1[a, b, \zeta] = 1 + \frac{a}{b} \quad (3.23)$$

Therefore we find the following limiting forms of the phase structure function (3.22) for anisotropic scattering :

$$D_\phi(z, \mathbf{s}) = \frac{1}{\rho} \frac{4\pi^2 r_e^2 \lambda^2 \Delta L}{2^{\alpha-2}} \, \Gamma\left(1 - \frac{\alpha-2}{2}\right) \frac{C_N^2(z) l_i^{\alpha-4}(z)}{(1 - f_p^2(z)/f^2)} s^2 \quad (3.24)$$

for $s \ll l_i$, and :

$$D_\phi(z, \mathbf{s}) = \frac{1}{\rho} \frac{8\pi^2 r_e^2 \lambda^2 \Delta L}{2^{\alpha-2}(\alpha-2)} \frac{\Gamma\left(1 - (\alpha-2)/2\right)}{\Gamma\left(1 + (\alpha-2)/2\right)} \frac{C_N^2(z) s^{\alpha-2}}{(1 - f_p^2(z)/f^2)} \quad (3.25)$$

for $s \gg l_i$.

In the equations (3.24) and ()3.25, $|\mathbf{s}| = (\frac{s_x^2}{\rho^2} + s_y^2)^{\frac{1}{2}}$, and $l_i = (\frac{l_{ix}^2}{\rho^2} + l_{iy}^2)^{\frac{1}{2}}$.

The anisotropic coherence length is defined as

$$D_\phi(z, \mathbf{s_0}) = D_\phi(z, s_{0x}, s_{0y}) = 1 \,. \quad (3.26)$$

We note that the root of (3.22) is $|\mathbf{s_0}| = (s_{0x}^2 + s_{0y}^2)^{\frac{1}{2}}$. Following Narayan & Hubbard (1988), we assume that the coherence length in x direction $s_{0x}$ is elongated by the factor of $\rho$ relative to the coherence length $s_{0y}$ in the y direction; in other words,

$$s_{0x} = \rho s_{0y} \tag{3.27}$$

Psuedo-codes for implementing the GSF and numerically evaluating the coherence length are given in the appendix. Similarly we assume that : $l_{ix} = \rho l_{iy}$. Equations (3.26) and (3.27) give the following expressions for the semi-major axes of the scatter-broadened image projected on the scattering screen in terms of the coherence length $s_{0x}$ and $s_{0y}$ :

$$\theta_{cx} = (2\pi s_{0x}/\lambda)^{-1} , \tag{3.28}$$

$$\theta_{cy} = (2\pi s_{0y}/\lambda)^{-1} = \rho \theta_{cx} . \tag{3.29}$$

Note that expressions (3.20) − (3.25) depend on $\rho$ only through the factor $1/\rho$, while the scattering angles (3.28) and (3.29) in the $x$ and $y$ direction respectively differ by a factor of $\rho$. We recover the isotropic results for $\rho = 1$.

### 3.2.5 Isotropic scattering

If the underlying turbulent eddies are isotropic, isotropic scattering is a reasonable assumption. The extent of anisotropy observed in scatter-broadened images at small solar elongations is determined more by the variation in the direction of the large scale magnetic field with respect to the line of sight than by the degree of anisotropy of the density fluctuations (Chandran & Backer, 2002). Using an anisotropic Goldreich-Sridhar spectrum, Chandran & Backer (2002) have shown that the scatter broadened images will be isotropic if the direction of the large-scale magnetic field is substantially aligned with the line of sight. This is intuitively obvious, since the plasma response would be gyrotropic about the large-scale magnetic field. Specifically, if $\gamma$ is the angle between the magnetic field and the line of sight, they show that if :

$$\gamma \ll (s/l_{out})^{1/3} \;, \tag{3.30}$$

where s is the baseline and $l_{out}$ is the outer scale of the turbulence, the dominant contribution to the turbulent spectrum comes from the values of $k_x$ and $k_y$ satisfying $k_x^2 + k_y^2 \simeq s^{-2}$.

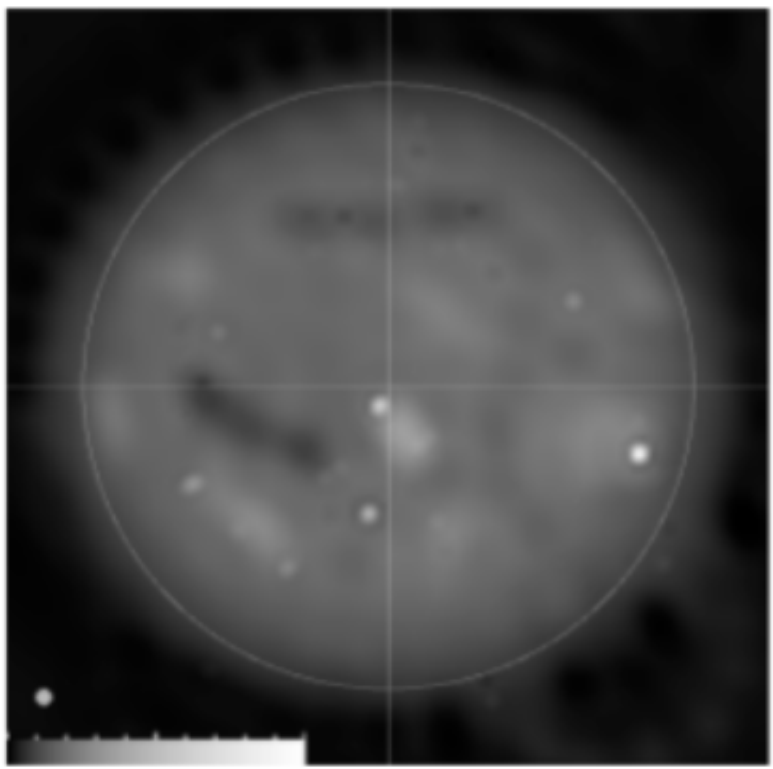

Figure 3.7: (From Mercier et al. (2006)) Image of radio sources in the solar corona obtained by combining visibilities from GMRT and NRH. The sources near the disc center are nearly isotropic

In case of spherical wave propagation, where the observer is looking through the corona down at a source on the Sun, the line of sight from the Earth to the Sun is radial, and $\gamma$ satisfies (3.30) amply. In this situation the effects of anisotropic scattering are likely to be minimal. Images of scatter-broadened sources near disk center in the solar corona are indeed not very anisotropic (Zlobec et al., 1992; Mercier et al., 2006), validating this idea (Figure 3.7). Stronger support for this argument is provided by the fact that type I radio bursts are strongly circularly polarised near disk center (Wentzel, 1997), and become less so near the limb. Since quasi-transverse magnetic field regions (in this case, regions of horizontal magnetic field) serve as depolarization sites, this implies that sources that are substantially near disk center do not encounter such regions, at least above the level where the emission originates; in other words, the magnetic field along the line of sight is largely radial (Wentzel, 1997). As discussed above, this implies that the scattering process will be isotropic. There is also some evidence for the fact that the inner scale itself is isotropic, in which case the assumption of isotropy is well justified for scales comparable to or less than the inner scale (Armstrong et al., 1990; Bastian, 1994). For an isotropic turbulent density spectrum with an inner scale $l_i$, (4.1) reduces to :

$$S_n(z, \kappa) = C_N^2(z)\, \kappa^{-\alpha} \exp[-(\kappa l_i/2\pi)^2] \;, \qquad (3.31)$$

where the spatial wavenumber $\kappa = (\kappa_x^2 + \kappa_y^2)^{\frac{1}{2}}$.

Therefore the phase structure function, (3.22) can be rewritten as (Coles & Harmon, 1989) :

$$D_\phi(z, s) = \frac{8\pi^2 r_e^2 \lambda^2 \Delta L}{2^{\alpha-2}(\alpha-2)} \Gamma\left(1 - \frac{\alpha-2}{2}\right) \frac{C_N^2(z) l_i(z)^{\alpha-2}}{(1 - f_p^2(z)/f^2)}$$
$$\times \left\{ {}_1F_1\left[-\frac{\alpha-2}{2}, 1, -\left(\frac{s}{l_i(z)}\right)^2\right] - 1 \right\} \;, \qquad (3.32)$$

and the corresponding asymptotic approximations (3.24) and (3.25) take the form (Coles et al., 1987; Subramanian & Cairns, 2011):

$$D_\phi(z, s) = \frac{4\pi^2 r_e^2 \lambda^2 \Delta L}{2^{\alpha-2}} \Gamma\left(1 - \frac{\alpha-2}{2}\right) s^2 \frac{C_N^2(z) l_i(z)^{\alpha-4}}{(1 - f_p^2(z)/f^2)} \;, \qquad (3.33)$$

for $s \ll l_i$ and

$$D_\phi(z,s) = \frac{8\pi^2 r_e^2 \lambda^2 \Delta L}{2^{\alpha-2}(\alpha-2)} \frac{\Gamma\left(1-(\alpha-2)/2\right)}{\Gamma\left(1+(\alpha-2)/2\right)} \frac{C_N^2(z) s^{\alpha-2}}{(1-f_p^2(z)/f^2)} \ . \quad (3.34)$$

for $s \gg l_i$.

The isotropic coherence length $s_0$ is defined by

$$D_\phi(s_0) = 1 \ . \quad (3.35)$$

For a given wavelength $\lambda$, the extent to which an ideal point source is broadened is given in terms of the isotropic coherence length $s_0$ as

$$\theta_c = (2\pi s_0/\lambda)^{-1} \ . \quad (3.36)$$

Just as the anisotropic coherence length $\mathbf{s}_0 = (s_{0x}, s_{0y})$ can be readily calculated from the asymptotically correct expressions (3.24) and (3.25) for the structure function (3.22), the isotropic coherence length $s_0$ can be calculated easily for the asymptotic approximations (3.33) and (3.34) of the phase structure function (3.32).

Several authors e.g., (Bastian, 1994; Subramanian & Cairns, 2011) have used these asymptotic expressions to obtain estimates of angular broadening of sources in the solar corona. However, there are limitations associated with using these asymptotic expressions. Specifically, equations (3.24) and (3.25) or (3.33) and (3.34) do not meet seamlessly at $s = l_i$, which suggests that in situations where the baseline is comparable to the inner scale, the asymptotic approximations cannot give reliable results (Coles et al., 1987; Subramanian & Cairns, 2011). In what follows, we compute the scatter broadening angles $\theta_{cx}$ and $\theta_{cy}$ for the anisotropic case using the full expressions (3.22) and compare them with those obtained with the limiting expressions (3.24) and (3.25).Similarly, for the isotropic case we compare scatter broadening angles $\theta_c$ obtained using full expression (3.32) with those obtained by using the limiting expressions (3.33) and (3.34). For this purpose we need to specify a model for the background electron density and for the amplitude of density turbulence $C_N^2$.

### 3.2.6 Density Models

To estimate angular broadening we need to integrate the random phase fluctuations along the line of sight. The lower limit of this integral is the fundamental plasma level ($f = f_p$) which depends on the background electron density, $n_e$. A model for $f_p$ is needed to fully describe refractive index and inner scale effects. Since $f_p^2 \propto n_e$ a model for the ambient electron density $n_e(z)$ is required.

One density model we use is due to Cairns et al. (2009), which we will call the "wind-like" density model from now on. The electron density as a function of heliocentric distance $R$ (which is measured in units of $R_\odot$) is given by:
$$n_e(z) = 1.58 \times 10^{27} \times (z-1)^{-2} \text{ cm}^{-3} . \qquad (3.37)$$

It is important to note that the wind-like density model includes an offset, $z_0 (= 1)$ in $(R_\odot)$, due to plasma sources being near to the photosphere. Cairns et al. (2009) shows that for $z_0 \sim 1.05 - 2 R_\odot$ the local density profile is often wind like $n_e \propto (z-1)^{-2}$. This model is known to be accurate close to the solar limb and also transitions smoothly to $z^{-2}$ profile at large distances.

In fact, the wind-like density model only specifies that the density should be proportional to $(z-1)^{-2}$; we obtain the proportionality constant of $1.58 \times 10^{27}$ cm$^{-3}$ by demanding that the density predicted by this model equal that predicted by the Leblanc et al. (1998) density model at 1 AU. Another useful model is the commonly used 4-fold Newkirk density model, which has a different proportionality constant (equal to $4.2 \times 10^4$ cm$^{-3}$).

## 3.2.7 Amplitude of Density Turbulence : $C_N^2(z)$

In order to characterize the amplitude $C_N^2$ of the turbulent density spectrum $S_n$ (eq 3.31), we use a model originally proposed by Armstrong & Woo (1980) and later refined by Spangler & Sakurai (1995). This model is based on VLBI observations in the outer corona and solar wind. Spangler & Sakurai (1995) obtained the following expression for $C_N^2$ as a linear fit to scattering data between $10R_\odot - 50R_\odot$ (Figure 3.8):

$$C_N^2(z) = 1.8 \times 10^{10} \left(\frac{z}{10R_\odot}\right)^{-3.66} \text{m}^{-20/3} . \tag{3.38}$$

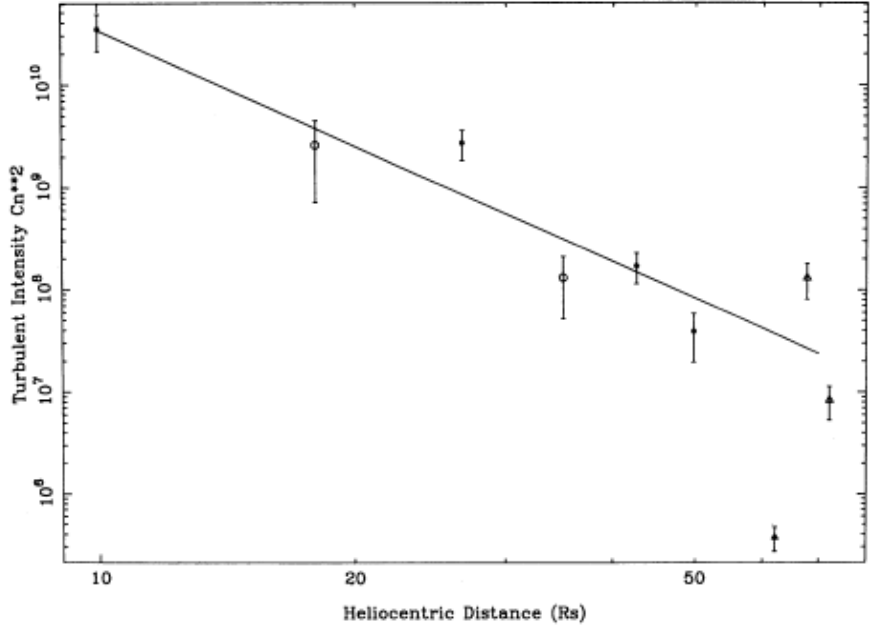

Figure 3.8: (Adapted from Spangler & Sakurai (1995)) Amplitude of density turbulence $C_N^2$ as a function of heliocentric distance $z$ in $R_\odot$

The dimensions of $C_N^2(z)$ in general are $\text{m}^{-\alpha-3}$, where $\alpha$ is the power law index characterizing the inertial range of the turbulent density spectrum $S_n$ in (3.31). It may be noted that (3.38) is valid only for a Kolmogorov spectrum ($\alpha = 11/3$). There is considerable evidence supporting the idea that inertial range density fluctuations in the solar wind follow the Kolmogorov scaling (Montgomery et al., 1987; Goldstein et al., 1995; Lithwick & Goldreich, 2001;

Dastgeer & Zank, 2007; Marsch & Tu, 1990; Tu & Marsch, 1995; Coles & Harmon, 1989).

Although we mostly use the Kolmogorov scaling ($\alpha = 11/3$) in this work, there is some evidence for flattening of the spectrum (at heliocentric distances of a few $R_\odot$) from scales around 100 km down to the inner scale (Coles & Harmon, 1989; Bastian, 1994), which might be evidence for the dispersion range that occurs prior to the dissipation range. The extent of this flattening strongly depends upon the phase of the solar cycle and the speed of the solar wind (Manoharan et al., 1994). The evolution of the flattening feature with heliocentric distance is currently not known.

We have addressed the issue of spectral flattening by using $\alpha = 3$ in (3.31), as in Bastian (1994). Since (3.38) as it stands is valid only for $\alpha = 11/3$, we need to re-calculate $C_N^2$ using equation (9) of Spangler & Sakurai (1995) with $\alpha = 3$. We obtained a least square fit to the plot of the newly calculated $C_N^2$ against the impact parameter $R_0$ (distance of closest approach), which yields the following expression for $C_N^2$ for $\alpha = 3$:

$$C_N^2(z) = 8.1 \times 10^{12} \left(\frac{z}{10 R_\odot}\right)^{-3.66} \text{m}^{-6} \ . \tag{3.39}$$

Although we mostly use (3.38), (which holds for $\alpha = 11/3$), we also discuss modifications to our results arising from the use of (3.39).

## 3.3   The relevance of the GSF

The principle aim of this chapter is to investigate the constraints on the appropriate form of the structure function in terms of the estimates of coherence scale ($s_0$) obtained by using the GSF (3.22) and (3.32) and the corrosponding asymptotic branches (given by (3.24) and (3.25) for anisotropic scattering and (3.33) and (3.34) for isotropic scattering). To explore the circumstances under which the GSF is significantly more accurate than the asymptotic branches, we start by comparing Eqs (3.24) and (3.25) with Eq. (3.22) for anisotropic scattering and Eqs (3.33) and (3.34) with Eq. (3.32) for isotropic scattering. To address this question we will use the coherence length $\mathbf{s_0}$. In case of anisotropic scattering recall that the coherence length $\mathbf{s_0} = (s_{0x}, s_{0y})$ is related to the phase structure function via (3.26), and the scattering angles $\theta_{cx}$ and $\theta_{cy}$ (which is the semi-major axes of an image corresponding to an ideal point source subject to scatter broadening) are related to $s_{0x}$ and $s_{0y}$ via (3.28) and (3.29), respectively. We have calculated the results for $s_{0x}$ and $s_{0y}$ but to avoid confusion we discuss only the y-component of the coherence length, namely $s_{0y}$. The results for the x-component of the coherence length, $s_{0x}(=\rho s_{0y})$ are identical.

In order to compare the angular broadening predictions of the GSF with those predicted by the asymptotic branches, we plot the relative error introduced in the coherence length $s_{0y}$ when either of the asymptotic approximations (3.24) or (3.25) of the GSF (3.22) is used, as a function of the inner scale $l_{iy}$. Denoting coherence scale obtained by using the GSF : $s_0(\text{GSF})$ and coherence scale obtained by using asymptotic branches : $s_0(\text{asymptotic})$, we define the relative error in $s_0$ as follows :

$$\% \text{ relative error in } s_0 = \frac{s_0(\text{GSF}) - s_0(\text{asymptotic})}{s_0(\text{GSF})} \times 100 \% \qquad (3.40)$$

An advantage in working with the relative error in $s_{0y}$ is that our conclusions are independent of the observing frequency for plane wave propagation. We carry out a similar exercise for isotropic scattering, which is relevant for the spherically diverging wavefront. In this situation, (3.32) is modified to include an integral along the line of sight (§3.3.2). The lower limit of the line-of-sight integral, which is the plasma level, is a function of an observing frequency. The spherical wave calculations are therefore expected to be sensitive to the observing frequency. When the relative error is significant the predictions of the GSF disagree with those of the asymptotic branches, and the converse is true when the relative error is negligible. The inner scale $l_i$ is maintained as a free parameter in §3.2 – 3.4 . When using the GSF, we note that the coherence length $\mathbf{s_0}$ needs to be calculated using a numer-ical root finding procedure. On the other hand, one can obtain an explicit analytical expression for $\mathbf{s_0}$ when using the asymptotic branches. For non-zero, spatially varying ratios $f_p(z)/f$, the expressions for the GSF and the asymptotic branches for an anisotropic case are modified by the inclusion of the factor $[1 - f_p^2(z)/f^2]^{-1}$ in (3.22), (3.25) and (3.24). It can be easily shown that (3.21) is recovered in the limit $f_p(z)/f \to 0$ (i.e. $f_p(z) \ll f$) and constant $f_p/f$). Similarly, for the isotropic case, the GSF and the corresponding asymptotic branches are modified to ((3.32), (3.33) and (3.34)). Consequently, in the limit $f_p(z)/f \to 0$ the angular broadening results are independent of the ratio $f_p/f$ (Cairns, 1998).

### 3.3.1 Plane wave propagation

Plane wave propagation is relevant to the situation where one is observing a distant cosmic source against the background of the solar wind. In this situation, anisotropic scattering is important, especially for sources at small solar elongations. In what follows, we compute the coherence length by using the anisotropic phase structure function discussed in § 2.1.1, and compare it with those obtained using the asymptotic approximations.

The coherence length $s_{0y}$ is calculated by using (3.26) and (3.27). Figures 3.9 and 3.10 show the relative error in the predictions of the GSF and either asymptotic branch of $s_{0y}$ as a function of the inner scale $l_{iy}$.

We obtained $s_{0y}$ as a root of the GSF, (3.22) and compared with $s_{0y}$ from the asymptotic branches, (3.24) and (3.25) for three different values for axial ratio $\rho$. The solid line is for $\rho = 1$, dotted line for $\rho = 5$ and the dashed line for $\rho = 10$. The results shown in Figures 3.9 and 3.10 are for a representative solar elongation of $10R_\odot$. It is evident that anisotropy effects are not very significant; varying $\rho$ by a factor of 10 results in a difference of $< 10\%$ in the relative error.

Figure 3.9 shows that the region where the relative error is significant decreases with increasing anisotropy. For $\rho = 1$, (solid line) the relative error increases sharply for $l_{iy} \leq 200$km; in this region, the $s_y \ll l_{iy}$ branch is inadequate and the GSF should be used. For $\rho = 5$, (dotted line) the relative error becomes significant for $l_{iy} \leq 100$km and for $\rho = 10$, (dashed line), the relative error is significant for $l_{iy} \leq 80$km.

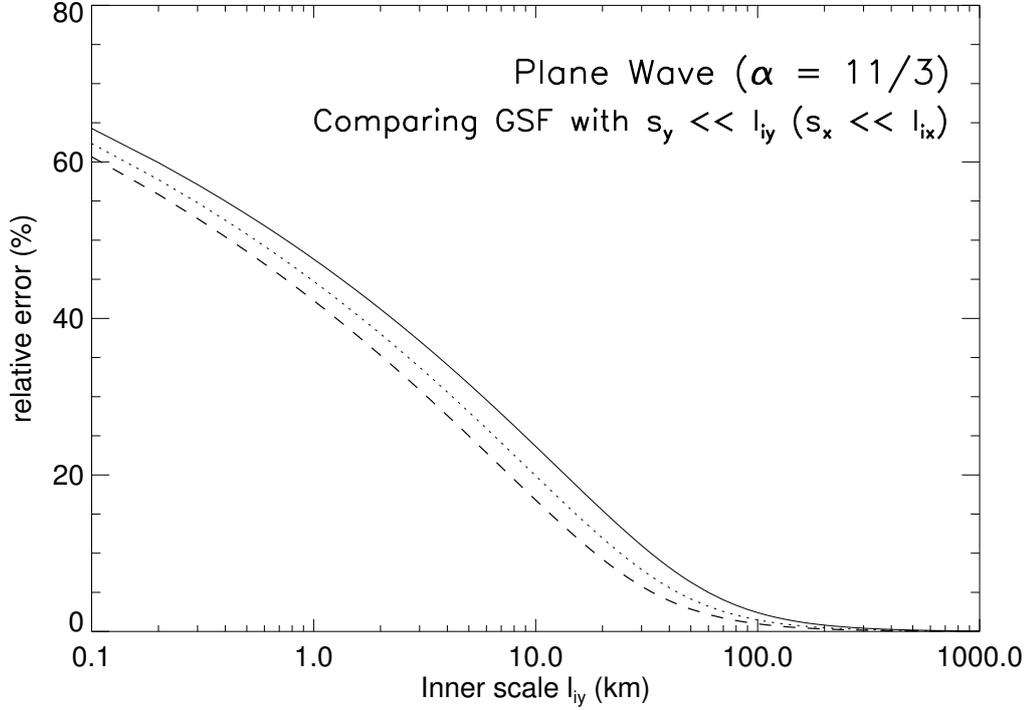

Figure 3.9: Relative error in the coherence length $s_{0y}$ ($s_{0x}$) as a function of $l_{iy}$ ($l_{ix}$) when the asymptotic branch $s_y \ll l_{iy}$ ($s_x \ll l_{ix}$), (3.24) is used. The calculations are for plane wave propagation through the corona and solar wind and for a representative solar elongation of $10R_\odot$. The solid line uses the degree of anisotropy, $\rho = 1$, the dotted line is for $\rho = 5$ and the dashed line uses, $\rho = 10$.

Figure 3.10 displays the corresponding relative error for the asymptotic branch $s_y \gg l_{iy}$. It is evident that the extent of the region for which the relative error is significant increases with the degree of anisotropy. For $\rho = 1$, (solid line) the relative error is significant for $l_{iy} \geq 10$km, implying the asymptotic branch $s_y \gg l_{iy}$ is inadequate and the GSF should be used. For $\rho = 5$, the relative error increases for $l_{iy} \geq 8$km and for $\rho = 10$, the relative error is significant for $l_{iy} \geq 4$km.

From Figures 3.9 and 3.10 it is clear that, for plane wave propagation through the solar wind and corona, the coherence length $s_{0y}$ (and therefore the broadening angle computed from it) computed via the GSF agrees with the asymptotically correct expressions (i.e. the relative error is negligible) only for $l_{iy} < 4$km or $l_{iy} > 200$km. In other words, for the degree of anisotropy ranging from 1−10, the statement $s_y \ll l_{iy}$ is valid for $l_{iy} \geq 200$ km, and the predictions of (3.24) hold well. Similarly, $s_y \gg l_{iy}$ is valid for $l_{iy} \leq 4$km, and the predictions of (3.25) will be accurate for values of the inner scale satisfying this condition. At a solar elongation of 10 $R_\odot$, we find that the GSF predictions disagree with those of the asymptotic branches for 4km $\leq l_{iy} \leq 200$ km. Although the results shown in Figures 3.9 and 3.10 hold only for a solar elongation of 10 $R_\odot$, we have also investigated this aspect for other elongations. We find that the $s_y \ll l_{iy}$ asymptotic branch is valid for elongations $< 5R_\odot$, while the $s_y \gg l_{iy}$ asymptotic branch is valid for elongations $> 20R_\odot$. In summary for solar elongations between 5 and 20 $R_\odot$ the GSF needs to be used.

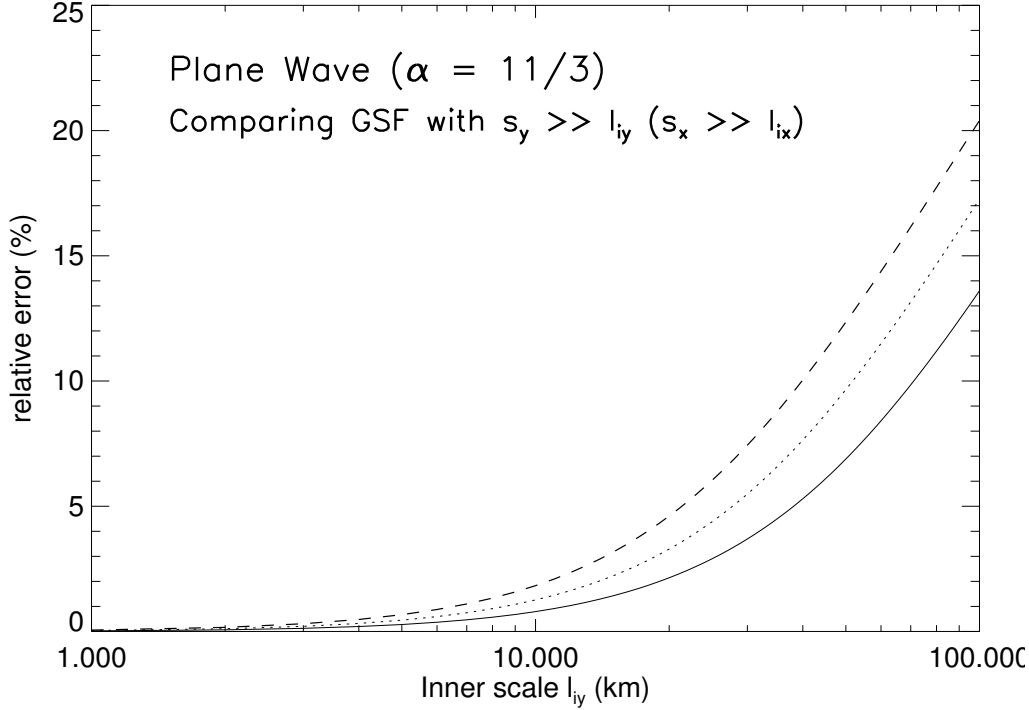

Figure 3.10: Relative error in the coherence length $s_{0y}$ ($s_{0x}$) as a function of $l_{iy}$ ($l_{ix}$) when the asymptotic branch $s_y \gg l_{iy}$ ($s_x \gg l_{ix}$), (3.25) is used. The calculations are for plane wave propagation through the corona and solar wind and for a representative solar elongation of $10 R_\odot$. The solid line uses the degree of anisotropy, $\rho = 1$, the dotted line is for $\rho = 5$ and the dashed line uses, $\rho = 10$.

### 3.3.2 Spherical wave propagation

For sources embedded in the solar corona, spherical propagation effects are important (figure 3.4). Since the assumption of isotropic scattering is justified in this situation (§ 2.2), the coherence lengths are computed using the formulation outlined in § 2.1.2.

For spherical wave propagation we need to use the effective baseline $s_{\text{eff}} = sR/(R_1 - R_0)$, where $R_1$ is the distance of the observer from the source and $R_0$ is the distance from which scattering is assumed to be effective. We consider $R_0$ to be equal to the fundamental plasma emission level; for 327MHz, with the wind-like density model $R_0$, is located at 0.0156 $R_\odot$ above the photosphere.

In this case, the line of sight from the source embedded in the solar corona to the observer (at the Earth) spans heliocentric distances ranging from the height of fundamental plasma emission (where $f_p(R) = f$) to 1 AU. One therefore needs to explicitly integrate equation (3.32) along the line of sight with $R$ being the integration variable. This aspect is different from the plane wave case (§ 3.1.1). Thus for the spherical wave propagation (3.32) should be modified to :

$$D_\phi(z,s) = \frac{8\pi^2 r_e^2 \lambda^2}{2^{\alpha-2}(\alpha-2)} \Gamma\left(1 - \frac{\alpha-2}{2}\right) \int_{z_0}^{L} \frac{C_N^2(z) l_i(z)^{\alpha-2}}{(1 - f_p^2(z)/f^2)}$$
$$\times \left\{ {}_1F_1\left[-\frac{\alpha-2}{2}, 1, -\left(\frac{sR/(R_1-R_0)}{l_i(z)}\right)^2\right] - 1 \right\} \mathrm{d}z \qquad (3.41)$$

and the corresponding asymptotic branches e.g. (Coles et al., 1987; Bastian, 1994; Subramanian & Cairns, 2011) by :

$$D_\phi(z,s) = \frac{4\pi^2 r_e^2 \lambda^2}{2^{\alpha-2}} \Gamma\left(1 - \frac{\alpha-2}{2}\right) \left(\frac{s}{L-z_0}\right)^2$$
$$\times \int_{z_0}^{L} \frac{z^2 C_N^2(z) l_i(z)^{\alpha-4}}{(1 - f_p^2(z)/f^2)} \, dz \, , \qquad (3.42)$$

for $s_{\text{eff}} \ll l_i$ and

$$D_\phi(z,s) = \frac{8\pi^2 r_e^2 \lambda^2}{2^{\alpha-2}(\alpha-2)} \frac{\Gamma(1-(\alpha-2)/2)}{\Gamma(1+(\alpha-2)/2)} \left(\frac{s}{L-z_0}\right)^{\alpha-2}$$
$$\times \int_{z_0}^{L} \frac{z^{\alpha-2} C_N^2(z)}{(1 - f_p^2(z)/f^2)} \, dz \, . \qquad (3.43)$$

for $s_{\text{eff}} \gg l_i$.

Furthermore, the lower limit of integration (which is the fundamental plasma level) depends on the observing frequency; it therefore follows that the relative error in the coherence length also depends on the observing frequency. We use the structure function given by (3.41) and (3.35) to find the coherence length predicted by the GSF. Similarly, we use (3.42), (3.43) and (3.35) to find the coherence lengths corresponding to the $s_{\text{eff}} \ll l_i$ and $s_{\text{eff}} \gg l_i$ branches respectively.

Figure 3.11 and 3.12 shows the relative error in the coherence length $s_0$ corresponding to the asymptotic branch $s_{\text{eff}} \ll l_i$ for three different frequencies. Figure 3.11 shows that for 150MHz (dashed line), the $s_{\text{eff}} \ll l_i$ branch is inadequate for $l_i \leq 10$ km, whereas for 327MHz (dotted line) and 600MHz (solid line) the $s_{\text{eff}} \ll l_i$ branch is inadequate for $l_i \leq 20$ km and for $l_i \leq 60$ km respectively.

Figure 3.12 shows that, for an observing frequency of 150MHz (dashed line), the GSF predictions disagree with those of the $s_{\text{eff}} \gg l_i$ branch for $l_i \geq 0.1$ km. On the other hand, for 327MHz (dotted line) and 600MHz (solid line) the GSF predictions disagree with those of the $s_{\text{eff}} \gg l_i$ branch

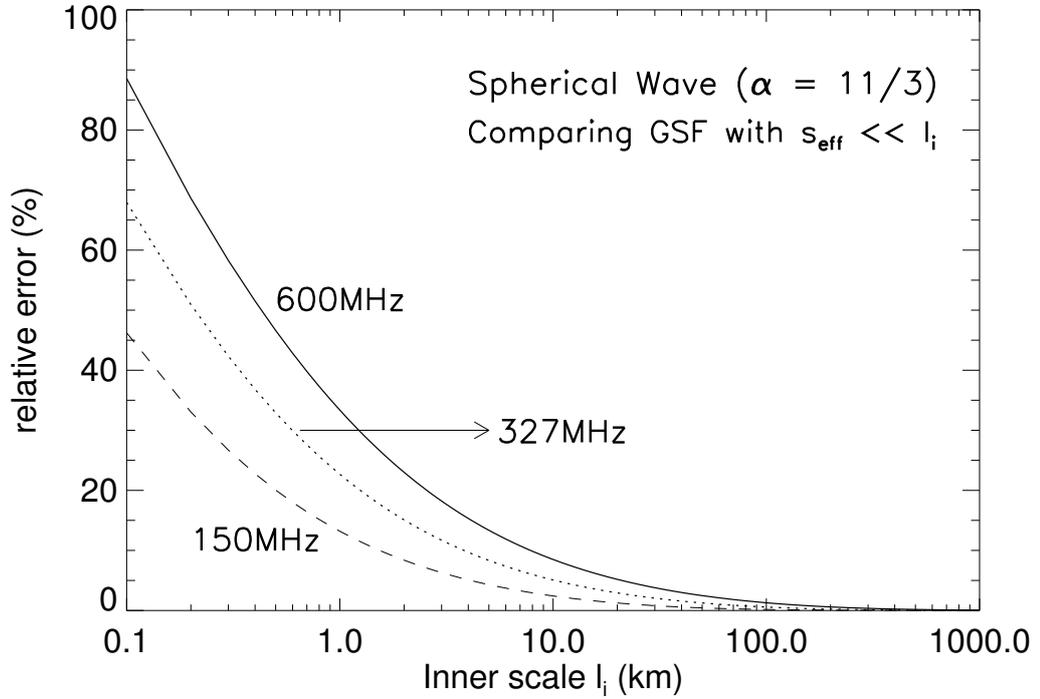

Figure 3.11: Relative error in the coherence length $s_0$ as a function of $l_i$ when the asymptotic branch $s_{\text{eff}} \ll l_i$, (3.33) is used. The calculations are for spherical wave propagation appropriate for sources embedded in the corona. The solid line uses an observing frequency, $f = 600$MHz, the dotted line uses $f = 327$MHz and the dashed line uses $f = 150$MHz.

for $l_i \geq 0.4$ km and for $l_i \geq 1$km respectively.

To summarize, we find that the range of the inner scales for which the relative error is significant (i.e., the predictions of the GSF disagree with those of the asymptotic branches) is a weak function of the observing frequency. For observing frequencies ranging from 150 MHz to 600 MHz the GSF predictions disagree with (and are more accurate than) those of the asymptotic branches for $0.1$ km $\leq l_i \leq 60$ km.

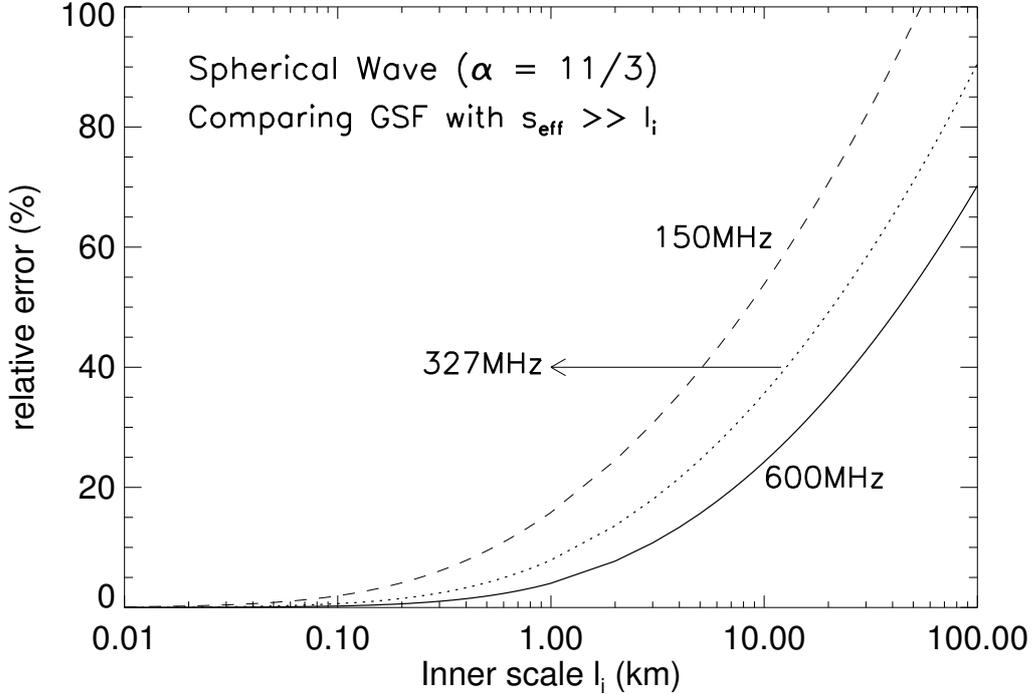

Figure 3.12: Relative error in the coherence length $s_0$ as a function of $l_i$ when the asymptotic branch $s_{\text{eff}} \gg l_i$, (3.34) is used. The calculations are for spherical wave propagation, appropriate for sources embedded in the the corona. The solid line uses an observing frequency, $f = 600$MHz, the dotted line uses $f = 327$MHz and the dashed line uses $f = 150$MHz.

### 3.3.3 Effect of local flattening of the turbulent spectrum

As mentioned earlier, there is some evidence for the flattening of the power spectrum of density turbulence between scales $\approx 100$ km and the inner scale (Coles & Harmon, 1989). This may be a manifestation of the so called "dispersion range" (Bruno & Carbone, 2013). It is not clear how this feature evolves with heliocentric distance. Although our current formalism cannot accommodate two power laws and an exponential turnover, we follow Bastian (1994) in using $\alpha = 3$ (instead of the Kolmogorov $\alpha = 11/3$) for the entire spectrum. As discussed in § 2.2, the appropriate expression to use for $C_N^2$ is then (3.39).

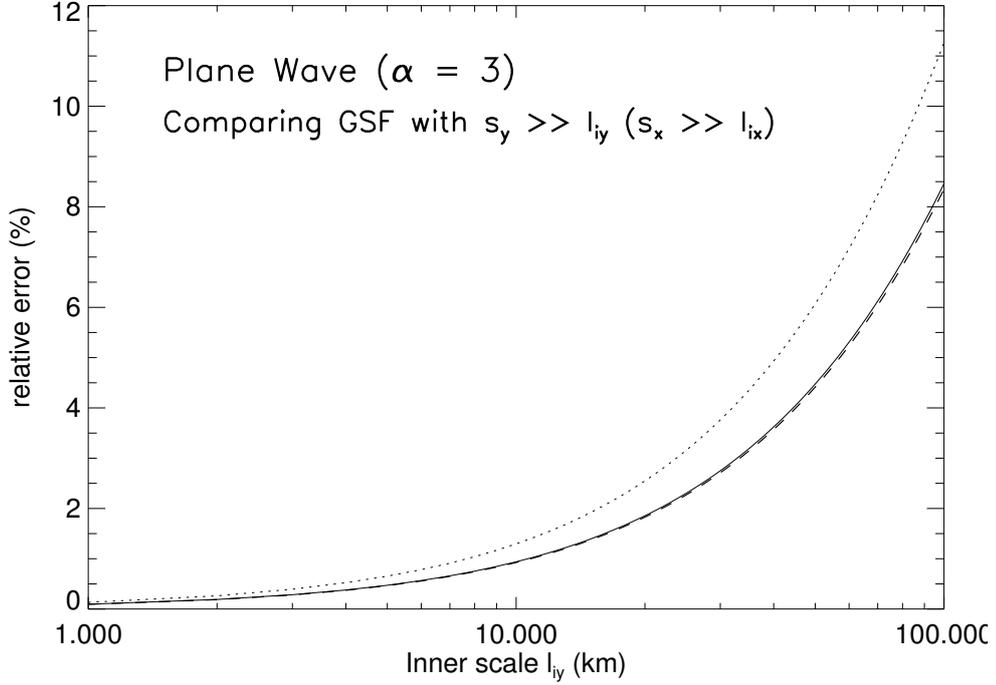

Figure 3.13: Relative error in the coherence length $s_{0y}$ ($s_{0x}$) as a function of $l_{iy}$ ($l_{ix}$) when the asymptotic branch $s_y \ll l_{iy}$ ($s_x \ll l_{ix}$), (3.24) is used with the power law index ($\alpha = 3$). The calculations are for plane wave propagation through the corona and solar wind and for a representative solar elongation of $10 R_\odot$. The dotted line uses the degree of anisotropy, $\rho = 1$, the solid line is for $\rho = 5$ and the dashed line uses, $\rho = 10$.

With these modifications, Figures 3.13 and 3.14 show that for plane wave propagation (at an elongation of 10 $R_\odot$ and for $1 < \rho < 10$) the relative error in the coherence length $s_0$ is significant for 1 km $\leq l_i \leq$ 1000km. Furthermore, the region of disagreement is insensitive to the value of $\rho$ when $\rho \geq 5$.

For spherical wave propagation, Figures 3.15 and 3.16 show that, the range of $l_i$ for which the disagreement is significant depends on the observing frequency. For 327 MHz we find that the GSF predictions for angular broadening observed at the Earth disagree with those of the asymptotic branches for 0.1 km $\leq l_i \leq$ 100 km. Thus, the range of $l_i$ over which the GSF and the asymptotic branch predictions disagree is larger for $\alpha = 3$ as compared to $\alpha = 11/3$.

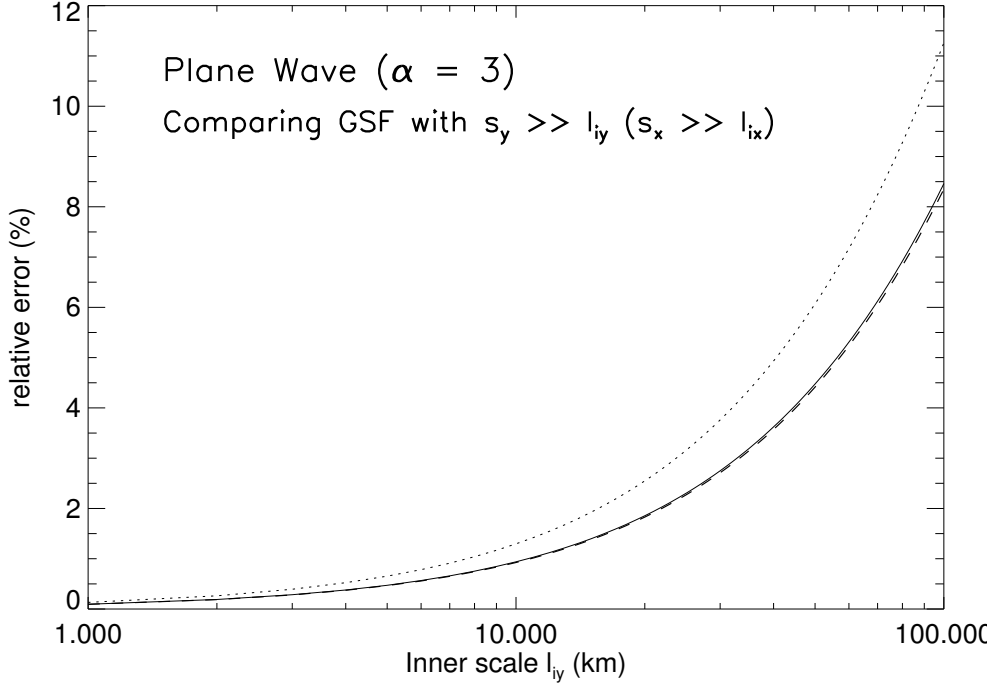

Figure 3.14: Relative error in the coherence length $s_{0y}$ ($s_{0x}$) as a function of $l_{iy}$ ($l_{ix}$) when the asymptotic branch $s_y \gg l_{iy}$ ($s_x \gg l_{ix}$), (3.25) is used with the power law index ($\alpha = 3$). The calculations are for plane wave propagation through the corona and solar wind and for a representative solar elongation of $10 R_\odot$. The dotted line uses the degree of anisotropy, $\rho = 1$, the solid line is for $\rho = 5$ and the dashed line uses, $\rho = 10$.

### 3.3.4 When are inner scale effects important?

We have established in Figures $3.9 - 3.12$ that it is essential to use the GSF for $4 \text{km} \leq l_i \leq 200 \text{km}$ for plane wave propagation and for $0.1 \text{km} \leq l_i \leq 60 \text{km}$ for spherical wave propagation. We next investigate the sensitivity of the predicted source size to $l_i$. For spherical wave propagation we calculate the scattering angle $\theta_c$ using (3.36) with the GSF (3.41) and for plane wave propagation the scattering angle $\theta_{cy}$ is calculated using the GSF (3.22) and (3.29). We take the inner scale $l_i$ (for spherical wave propagation) and $l_{iy}$ (for plane wave propagation) to be a free parameter.

Figure 3.17 shows the extent of angular broadening for an ideal point source as a function of $l_i$, for plane wave and spherical wave propagation at an observing frequency of 327 MHz. For plane wave propagation, these calculations are carried out at an elongation of 10 $R_\odot$ with a screen thickness $\Delta L = 0.5 R_\odot$ and $\rho = 1, 5$ and 10. It is clear from Figure 3.17 that for plane wave propagation, the extent of scatter broadening depends upon the value of the inner scale only for $l_{iy} \geq 10$ km. This gives the upper limit on the values of $l_i$ below which the results are independent of the inner scale. We find that this upper limit is a function of the degree of anisotropy, and it declines for larger values of $\rho$. For plane wave propagation, we find that inner scale effects are important only for heliocentric distances $\leq 20 R_\odot$. This result is consistent with our finding that the GSF can be approximated by the $s_y \gg l_{iy}$ and $s_x \gg l_{ix}$ asymptotic branch for solar elongations $> 20 R_\odot$; this branch (3.25) does not involve the inner scale. On the other hand, for spherical wave propagation Figure 3.17 shows that the scatter broadening angle is sensitive to the inner scale for $l_i \geq 1$ km.

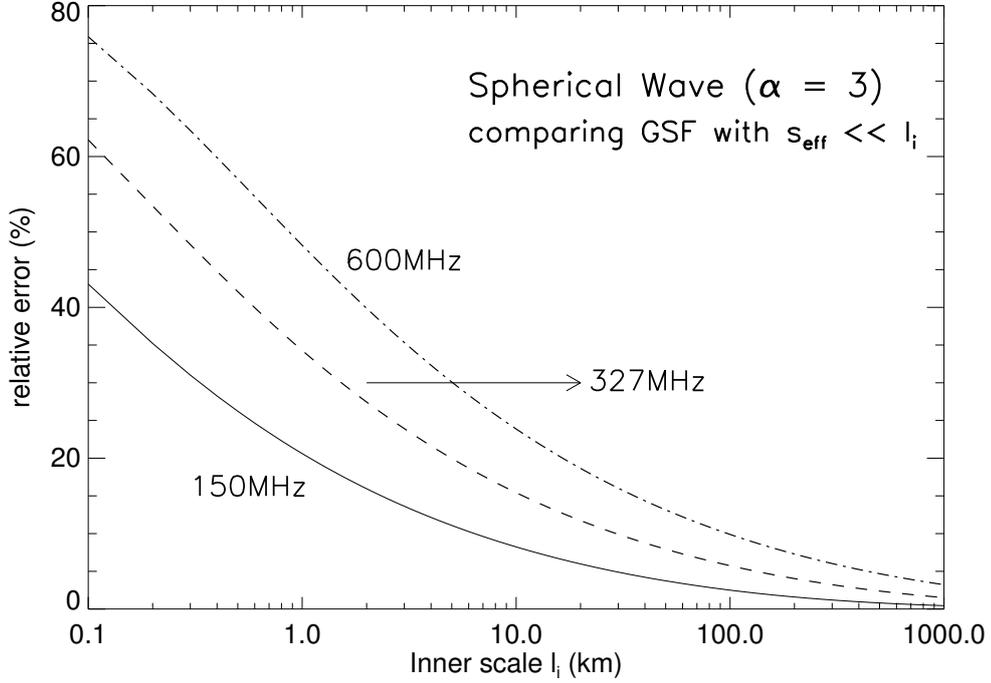

Figure 3.15: Relative error in the coherence length $s_0$ as a function of $l_i$ when the asymptotic branch $s_{\text{eff}} \ll l_i$, (3.33) is used with the power law index ($\alpha = 3$). The calculations are for spherical wave propagation, appropriate for sources embedded in the the corona. The solid line uses an observing frequency, $f = 150$MHz, the dashed line uses $f = 327$MHz and the dot-dashed line uses $f = 600$MHz.

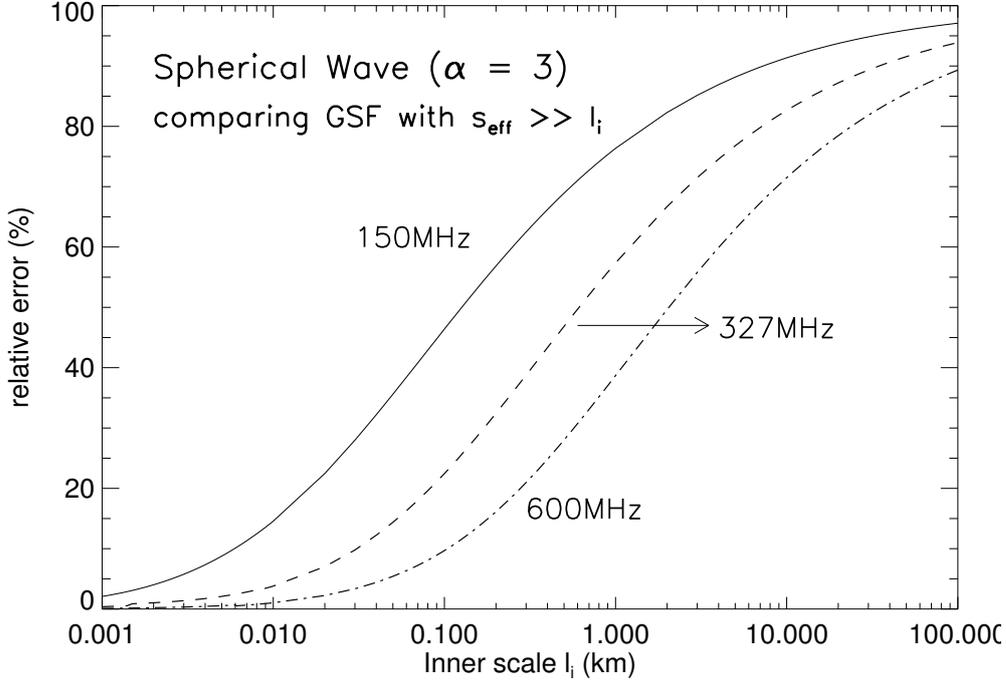

Figure 3.16: Relative error in the coherence length $s_0$ as a function of $l_i$ when the asymptotic branch $s_{\text{eff}} \gg l_i$, (3.34) is used with the power law index ($\alpha = 3$). The calculations are for spherical wave propagation, appropriate for sources embedded in the the corona. The solid line uses an observing frequency, $f = 150$MHz, the dashed line uses $f = 327$MHz and the dot-dashed line uses $f = 600$MHz.

To summarize, for $f = 327$ MHz, inner scale effects can generally be considered to be important (in the sense that the source size using the GSF is sensitive to the actual value of the inner scale) if $l_i \geq$ few hundred meters to a few km. We have carried out similar calculations for $f = 1500$MHz; for this frequency, we find that the source size is sensitive to the inner scale if $l_i \geq$ a few km to 100 km.

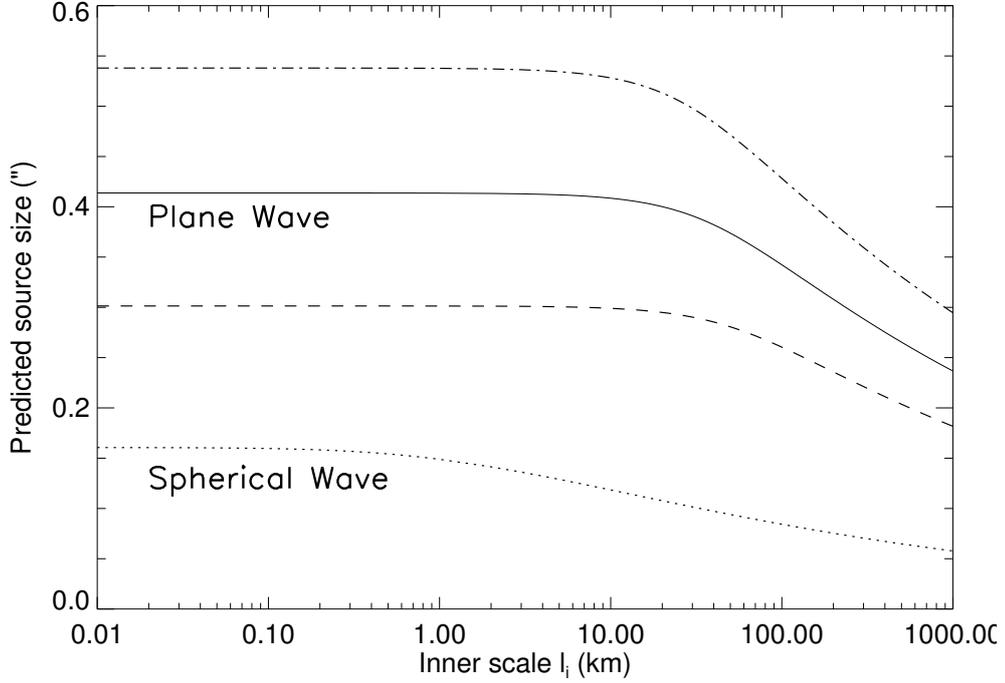

Figure 3.17: The predicted broadening $\theta_c$ as a function of the inner scale $l_i$ at an observing frequency of 327 MHz and Kolmogorov power law index $\alpha = 11/3$. The dashed, solid and dot-dashed lines are for $\theta_{cy}$ computed using GSF with plane wave propagation at a solar elongation of 10 $R_\odot$ and the degree of anisotropy $\rho = 1, 5$ and 10, respectively, and screen thickness $\Delta L = 0.5 R_\odot$, given by (3.22) while the dotted line is for $\theta_c$ computed using GSF with spherical wave propagation using (3.41).

### 3.3.5 Inner scale models

We have thus established the range of inner scale values for which the GSF needs to be used (§ 3.1.1 and § 3.1.2), using the inner scale as a free parameter. We now evaluate the inner scale in the corona and the solar wind using three different physical prescriptions. The first prescription is one where the inner scale arises from proton cyclotron damping by MHD waves (Coles & Harmon, 1989; Harmon, 1989; Verma, 1996; Yamauchi et al., 1998; Leamon et al., 1999, 2000; Smith et al., 2001; Bruno & Trenchi, 2014) :

$$l_i(R) = 684 \times n_e(R)^{-1/2} \quad \text{km} , \qquad (3.44)$$

where $n_e$ is the electron number density in cm$^{-3}$. For the second prescription, we identify the inner scale with the proton gyroradius (Bale et al., 2005; Alexandrova et al., 2012):

$$l_i(R) = 1.02 \times 10^2 \mu^{1/2} T_i^{1/2} B(R)^{-1} \quad \text{cm} , \qquad (3.45)$$

where $\mu$ ($\equiv m_i/m_p$) is the ion mass in terms of the proton mass, $T_i$ is the proton temperature in eV and B is the Parker spiral magnetic field in the ecliptic plane (Williams, 1995). However, recent work seems to suggest that the dissipation could occur at scales as small as the electron gyroradius (Alexandrova et al., 2012; Sahraoui et al., 2013). The third prescription we consider assumes that the inner scale is the electron gyroradius $\rho_e$, given by:

$$l_i(R) = 2.38 \times T_e^{1/2} B(R)^{-1} \quad \text{cm} , \qquad (3.46)$$

where $T_e$ is the electron temperature in eV.

Figure 3.18 shows the inner scale obtained using these three prescriptions as a function of heliocentric distance. It is useful to compare the predictions of the inner scale models with the range of inner scales for which we claim that the GSF needs to be used. For plane wave propagation, a distant cosmic source is located at a given solar elongation (which we take to be the same as the heliocentric distance for the purposes of this discussion) behind the solar wind scattering screen. At this heliocentric distance, the angular broadening prediction using the GSF is more accurate than that of either of the asymptotic branches if $0.3 \text{km} \leq l_i \leq 300$ km (§ 3.1.1). The light grey region in Figure 3.18 denotes this region; it indicates the range of inner scales for which the GSF predictions are more accurate than those of the asymptotic branches for plane wave propagation for distant cosmic sources located at solar elongations between $5R_\odot$ and $20R_\odot$ and having axial ratios $1 < \rho < 10$. The $s \ll l_i$ asymptotic branch is adequate for elongations $< 5R_\odot$ (dark grey region in Figure 3.18) , while the $s \gg l_i$ asymptotic branch is adequate for elongations $> 20R_\odot$. To summarize, Figure 3.18 reveals that, for distant cosmic sources (for which plane wave propagation is appropriate) located

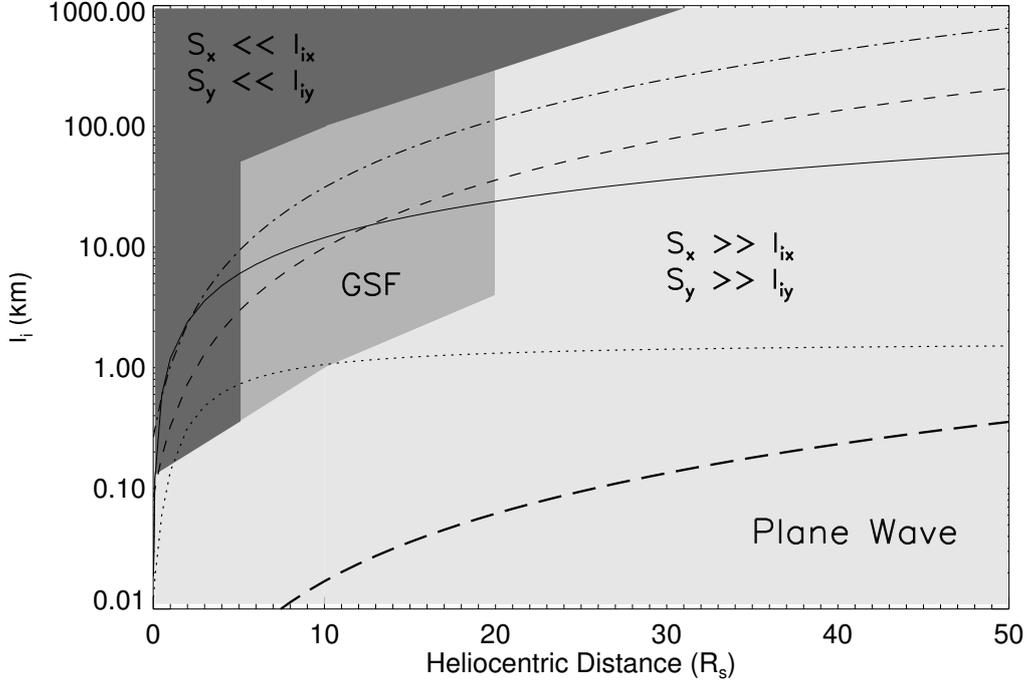

Figure 3.18: The inner scale $l_{iy}$ ($l_{ix}$) in km as a function of heliocentric distance in radius of Sun ($R_s$), for plane wave propagation. The dashed and dot-dashed lines show the proton gyroradius (3.45) using proton temperatures of $10^5$ and $10^6$ K respectively. The solid and dotted lines show the inner scale governed by proton cyclotron damping (3.44) using the wind-like density model (3.37) and the fourfold Newkirk density model respectively. The thick dashed line shows the electron gyroradius (3.46) using an electron temperature of $10^5$K. The light grey region denotes the range of (distant) source elongations and inner scale values for which the GSF yields predictions that are substantially more accurate than those of the asymptotic branches.

at solar elongations between $5R_\odot$ and $20R_\odot$ and axial ratios $1 \leq \rho \leq 10$, the GSF would need to be used if the inner scale is the proton gyroradius or is due to proton cyclotron resonance. These results are summarized in Table 3.1

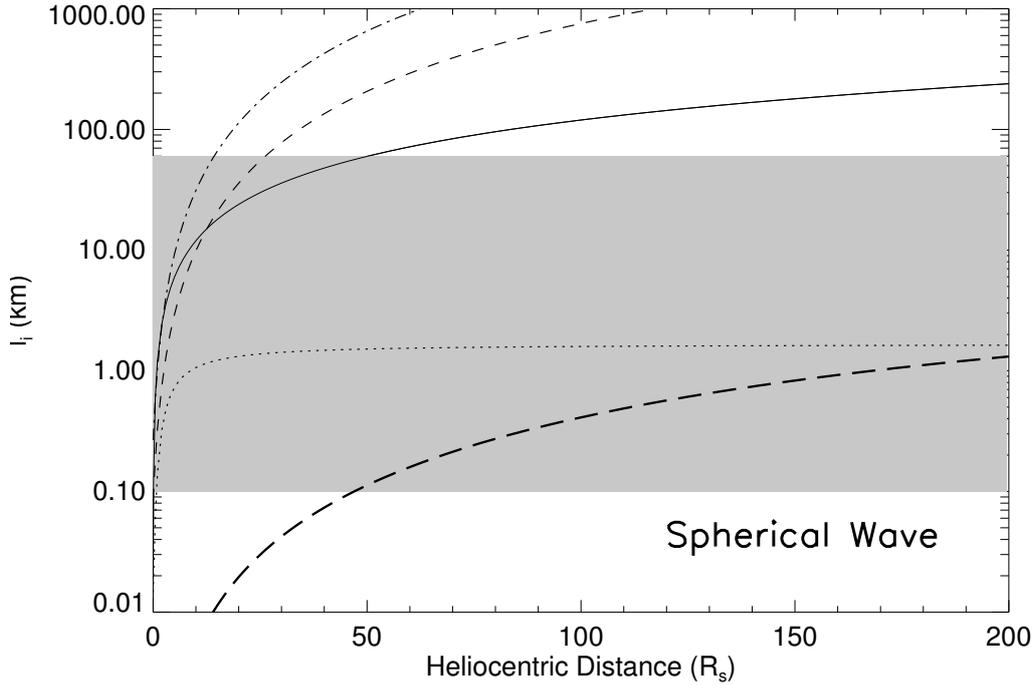

Figure 3.19: The inner scale $l_i$ in km as a function of heliocentric distance in radius of Sun ($R_s$), for spherical wave propagation. The dashed and dot-dashed lines show the proton gyroradius (3.45) using proton temperatures of $10^5$ and $10^6$ K respectively. The solid and dotted lines show the inner scale governed by proton cyclotron damping (3.44) using the wind-like density model (3.37) and the fourfold Newkirk density model respectively. The thick dashed line shows the electron gyroradius (3.46) using an electron temperature of $10^5$K. The light grey region denotes the range of (distant) source elongations and inner scale values for which the GSF yields predictions that are substantially more accurate than those of the asymptotic branches.

Table 3.1: Plane wave propagation with degree of anisotropy $\rho$ between 1–10 : when does the GSF need to be used? (GSF required for 0.3km $\leq l_i \leq$ 300km).

| Solar Elongation $R_\odot$ | Proton Gyroradius $T_p = 10^5$ K | $T_p = 10^6$ K | Inner scale model (km) proton cyclotron damping wind like density model | 4* Newkirk density model | electron Gyroradius $T_e = 10^5$ K | $T_e = 10^6$ K |
|---|---|---|---|---|---|---|
| $< 5$ | $s \ll l_i$ is valid | $s \ll l_i$ is valid | $s \ll l_i$ is valid | $s \ll l_i$ is valid | | |
| $5 < R < 20$ | GSF required | GSF required | GSF required | GSF required | $s \gg l_i$ is valid throughout | $s \gg l_i$ is valid throughout |
| $> 20$ | $s \gg l_i$ is valid | $s \gg l_i$ is valid | $s \gg l_i$ is valid | $s \gg l_i$ is valid | | |

Table 3.2: Spherical wave propagation : when does the GSF need to be used?.

| | Proton Gyroradius | | Inner scale model (km) proton cyclotron damping | | electron Gyroradius | |
|---|---|---|---|---|---|---|
| | $T_p = 10^5$ K | $T_p = 10^6$ K | wind like density model | 4* Newkirk density model | $T_e = 10^5$ K | $T_e = 10^6$ K |
| Is GSF required ? (GSF required if $0.1\text{km} \leq l_i \leq 60\text{km}$) | Yes | Yes | Yes | Yes | No, GSF required only from 70 $R_\odot$ to the Earth | No, GSF required only from 70 $R_\odot$ to the Earth |

We carry out a similar exercise for spherical wave propagation. In this situation, the source is embedded in the solar corona and the observer is at the Earth, looking at the source through the turbulent medium (figure 3.4). Figure 3.19 shows the inner scale obtained using the three prescriptions (Eqs. 3.44-3.46) as a function of heliocentric distance. The linestyles are the same as those used in Figure 3.18. As explained in § 3.1.2, for spherical wave propagation at observing frequencies ranging from 150 MHz to 600 MHz, the predictions of the GSF are more accurate than those of the asymptotic branches for $0.1 \text{km} \leq l_i \leq 60 \text{km}$. This region is represented by a grey band in Figure 3.19. It is well known that most of the scattering takes place well within $30 R_\odot$ (Subramanian & Cairns, 2011). We can claim that the angular broadening estimates using the GSF will be more accurate than those of the asymptotic branches if the grey band in Figure 3.19 encloses the inner scale predicted by a specific model for heliocentric distances $\leq 30\ R_\odot$. Figure 3.19 show that this is the case (i.e., the GSF needs to be used for accurate broadening estimates) if the inner scale is governed by proton-cyclotron damping or is given by proton gyroradius. If, on the other hand, the inner scale is the electron gyroradius with $T_e = 10^5$K, the inner scale values predicted by this model overlap the grey band in Figure 3.19 only for heliocentric distances $\geq 30\ R_\odot$. For $T_e = 10^6$K, this is true for heliocentric distances $\geq 50\ R_\odot$. Thus, if the inner scale is given by the electron gyroradius, we cannot claim that the GSF predictions will be substantially more accurate than the predictions of the asymptotic branches. These results are summarized in Table 3.2.

## 3.4 Summary and conclusions

The amplitude of MHD turbulence in the extended solar corona and solar wind, especially near the inner (dissipation) scale, is a subject that is of considerable interest in a variety of applications. We investigate it using predictions for the angular broadening of radio sources. Typical estimates of angular broadening due to refraction and scattering by density turbulence use approximations to the structure function that are valid for situations where the interferometer spacing is $\gg$ or $\ll$ than the inner scale $l_i$. We use a general structure function (GSF) that does not use these approximations. We consider both plane wave propagation, which is appropriate for distant cosmic sources observed against the background of the solar wind, and spherical wave propagation, which is appropriate for sources embedded in the solar corona. For plane wave propagation we consider an anisotropic density turbulence spectrum comprising a Kolmogorov power law ($\alpha = 11/3$) spectrum multiplied by an exponential turnover at the inner scale. For spherical wave propagation, isotropic scattering is a well justified assumption. We demonstrate that angular broadening predictions using the general structure function agree with those obtained using the appropriate asymptotic expressions in the limits $s \ll l_i$ and $s \gg l_i$. For plane wave propagation, for sources observed at elongations between 5 and 20 $R_\odot$ and with the degree of anisotropy $1 \leq \rho \leq 10$, we find that the GSF is substantially more accurate than the asymptotic branches for 4 km $\leq l_{ix}, l_{iy} \leq$ 200 km. These results are independent of observing frequency as well as the amplitude of the density turbulence ($C_N^2$), and only weakly dependent on the degree of anisotropy ($\rho$). For spherical wave propagation, however, the results are found to be weakly dependent on the observing frequency. For observing frequencies ranging from 150 MHz to 600 MHz, the predictions of the GSF are more accurate than those of the asymptotic branches if 0.1km $\leq l_i \leq$ 60 km. If the spectrum is taken to be flatter ($\alpha = 3$), the range of $l_i$ for which the GSF predictions disagree with those of the asymptotic branches is larger. Importantly, the range over which the GSF predictions are substantially more accurate than those of the asymptotic approximations for plane wave propagation (light grey band in figure 3.18) is well within the expected values of the inner scale for the proton cyclotron damping and the proton gyroradius models for the inner scale. For plane wave propagation, we find that angular broadening predictions using the GSF are sensitive to the value of the inner scale for distant cosmic source located at elongations $\leq 20 R_\odot$. For spherical wave propagation, which is applicable when a source embedded in the solar corona is viewed at the Earth, the GSF is more accurate if the inner scale is due to proton cyclotron damping or is given by the proton gyroradius.

Using the GSF with spherical wave propagation to calculate the predicted extent of broadening of an ideal point source, we find that angular broadening is sensitive to the value of $l_i$ (in other words, inner scale effects are significant) if $l_i \geq$ a few to a few tens of km for $f = 327$ MHz. For an observing frequency of 1500 MHz, inner scale effects are important if $l_i \geq$ a few to 100 km.

The rate at which energy in solar wind turbulence damps on ions is an important question cutting across sub-disciplines. While some progress has been made in this regard, its still not clear if there is enough energy in the cascade near the dissipation scale for direct perpendicular heating (Cranmer & Van Ballegooijen, 2003). This question can be addressed via accurate estimates of the amplitude of density turbulence ($C_N^2$). Observations of angular broadening of radio sources are typically reliable means of constraining $C_N^2$. Recent conclusions regarding the magnitude of density fluctuations (relative to the background density) in the heliosphere (Bisoi et al., 2014) are also expected to help in constraining $C_N^2$. However, such estimates have traditionally been made using expressions for the structure function that are only valid in limits where the interferometric baseline used for observing are either $\gg$ or $\ll$ the dissipation scale. We have used the general structure function and quantified the errors arising from the use of these approximations. Our results underline the necessity of using the GSF for quantitative estimates of angular broadening.

# Chapter 4

# Density fluctuations and heating of the solar wind

*We consider previously published radio wave scattering and IPS data of measurements of density turbulence in the solar wind. Taken together, these measurements yield density turbulence spectra spanning a wide range of spatial scales, including the important high frequency region where dissipation is expected to take place. The density fluctuations are inferred using a combination of recently developed theoretical tools (Chapter III) to analyze radio wave scattering data and existing analysis methods to treat interplanetary scintillation data. Hypothesizing that the density fluctuations are due to kinetic Alfvén waves, we constrain the rate at which the extended solar wind is heated due to turbulent dissipation. Our results provide the first estimates of the turbulent heating rate all the way from the Sun to the Earth.*

## 4.1 Introduction

The problem of how a corona is heated to $\sim$ a million degrees has been a subject of intense investigation for a few decades. A related, somewhat lesser known, and equally important problem is that of heating / energy deposition in the extended solar wind. Parker's theory of the solar wind predicts an adiabatic temperature profile, which requires a radial dependence of temperature as $T(r) \propto r^{-4/3}$. However the temperature profile deduced from the in-situ measurements (Helios, Pioneer and Voyager) over heliocentric distances from $0.3 - 100$ AU does not agree with the adiabatic temperature profile. Fits to measurements of the radial temperature profile in the ecliptic plane shows that the temperature follows a radial profile given by e.g., (Bruno & Carbone, 2013) :

$$T(\mathbf{r}) \sim T_0 (r_0/r)^\xi$$

where the exponent $\xi \to [0.7, 1]$, with $\xi \leq 1$ for $r < 1$AU and flattens to $\sim 0.7$ for $r \geq 30$AU.

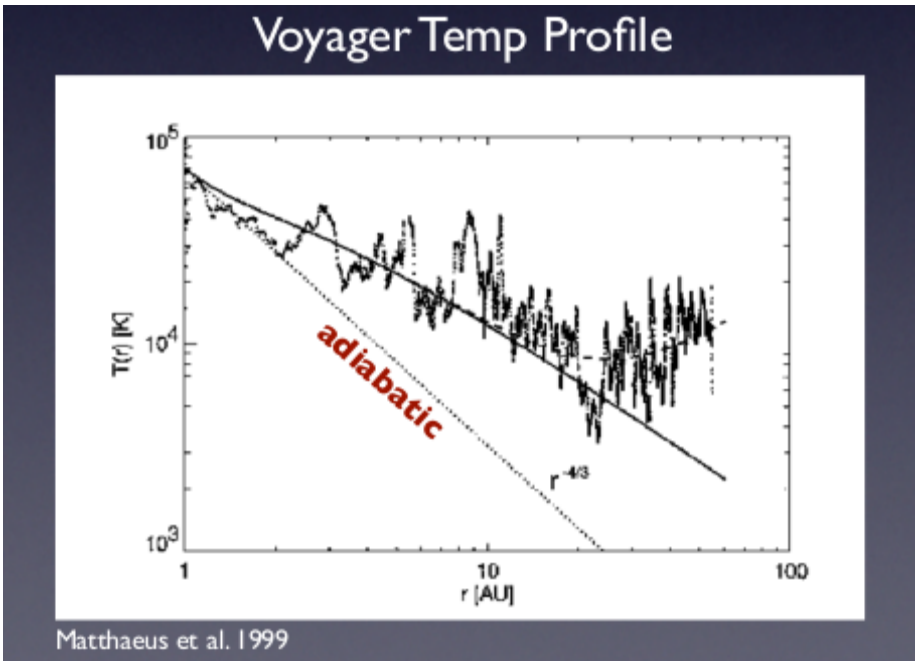

Figure 4.1: Observed vs predicted temperature profile : Radial temperature profile as a function of heliocentric distance (in AU). The dotted line shows the expected temperature profile due to Parker theory of adiabatic expansion. The dashed line shows the average trend in the actual temperature profile indicating slow down for r > 1 AU. (Adapted from Matthaeus et al. (1999b))

This clearly indicates that the temperature decays more slowly than that predicted by adiabatic expansion (Figure 4.1). A number of such observations in the inner heliosphere indicate that the solar wind undergoes distributed heating. This suggests that there must be some mechanism at work within the solar wind plasma responsible for supplying the energy required to account for the observed heating.

Several heating mechanisms have been proposed, which can be broadly categorized as follows :

- Sweeping model of cyclotron heating (non-turbulent model) :
  In this model non-turbulent high frequency waves (KHz range) are invoked for the ion cyclotron heating. Waves with frequency $\omega < \omega_c$, ($\omega_c$ is the proton cyclotron frequency) 'rise up' in the solar corona and reach a height where $\omega \sim \omega_c$. At this point the wave experiences cyclotron damping and leads to direct ion heating.

- Turbulent heating :
  In this model low frequency waves undergo a turbulent cascade due to non linear interactions and transfer energy from large scales to smaller scales. At smaller scales viscous effects become dominant due to 'effective collisions' (e.g. proton cyclotron damping) and this leads to the dissipation of energy - which heats up the ions.

In this work we consider the later class of models, to explain the observed solar wind heating. Since the solar wind is a collisionless plasma (with mean free paths as large as 1 AU) the viscous coefficients that drive the dissipation are weak. This requires an efficient mechanism to transfer energy in the solar wind fluctuations to smaller scales. Solar wind turbulence, with energy cascading from large to small scales before ultimately dissipating at small scales, might provide an answer e.g. (Verma, 1996).

As discussed earlier, radio waves propagating through the solar wind are scattered due to density fluctuations (Chapter 2). Therefore the properties and evolution of density fluctuations are of considerable interest in both the inner and outer heliosphere (Cairns et al., 2000)

There is considerable evidence pointing to the fact that the solar wind behaves as a weakly compressible turbulent medium (e.g., (Zank & Matthaeus, 1992; Dastgeer & Zank, 2009, 2010; Zank et al., 2012), with a Kolmogorov-like ($\kappa^{-11/3}$) turbulent spectrum for density, magnetic field and velocity fluctuations in the inertial range. Dastgeer & Zank (2010) suggest a physical picture where the background pressure imbalance gives rise to density fluctuations, which are passively convected in a field of weakly compressible velocity fluctuations. Consequently it is difficult to compute the direct contribution of density fluctuations in the energy budget of the solar wind.

The density modulation index $\epsilon_N$ ($\equiv \delta N/N$) is crucial in understanding the transport properties of the solar wind. It is therefore important to evaluate the radial profile of $\epsilon_N$, especially in the near-Sun region.

Another important point regarding solar wind properties is to explore the relation between the density modulation index and the magnetic field modulation index ($\epsilon_{N_e} - \epsilon_B$), where $\epsilon_B \equiv \delta B/B$. It is generally accepted that the slow solar wind exhibits fully developed turbulence; it therefore provides an ideal platform to study this relation. In-situ measurements provide good estimates of the mean magnetic field $B_0$ for the slow solar wind and extensive radio observations can be used to deduce density fluctuations in near Sun solar wind. In combination they can provide reliable estimates of the rms fluctuations of the magnetic field.

The goals of this work are 1) to undertake a brief survey of interplanetary scintillation (IPS) and angular broadening observations in order to estimate the density modulation index as a function of heliocentric distance, and 2) to obtain an estimate of the extended heating rate in the solar wind using these observations.

In the next section we briefly discuss observations of IPS and the angular broadening of distant celestial sources viewed against the background of the turbulent solar wind. We make use of previously published data of radio scintillation observations and derive density fluctuations at the inner scale of the solar wind turbulence (Armstrong et al., 1990; Anantharamaiah et al., 1994; Spangler & Sakurai, 1995; Bisoi et al., 2014). We then present results for the density modulation index and the turbulent energy cascade rate calculated at dissipation scale and finally conclude with summary and interpretation of the results.

## 4.2 Radio Scintillations : brief background

Radio scintillation techniques are useful for studying the properties of the solar wind. These techniques include

- Angular broadening,
- Phase scintillations and
- Single- and multi-station interplanetary scintillations (IPS) of the radio sources.

### 4.2.1 Angular broadening

Scattering of electro-magnetic waves due to density turbulence in plasma leads to a wide variety of observed phenomena such as intensity scintillations, angular broadening, spectral broadening, pulse smearing, etc. The change in angular size of the source due to the scattering of electromagnetic waves in the solar wind is known as angular broadening. The scattering of electromagnetic waves is a consequence of a density turbulence in the solar wind. We have discussed how solar wind density turbulence can be characterized in terms of the structure function in chapter 3.

Using measurements of the structure function we can deduce various parameters that characterize spatial power spectrum. For instance, from the measured dependence of the structure function on the baseline $s$ we can get power law index $\alpha$ for the spatial power spectrum of the density fluctuations.

Using the magnitude of the structure function we get the path integrated value of the amplitude of the turbulence $C_N^2$.

When the interferometric baseline ($s$) is comparable to the inner/dissipation scale of the solar wind density turbulence, (i.e., $s \sim l_i$) the GSF should be used for accurate estimates of the angular broadening (Ingale et al., 2015). Consider a plane wave propagating from a distant radio source incident on a thin screen of density irregularities in the solar wind. Let $\Delta L$ be the thickness of the scattering screen. The scattering screen lies in the $x - y$ plane and the propagation direction is along $z$. The scattering medium can be characterized by the anisotropic power law spectrum (Ingale et al., 2015) :

$$S_n(z, \boldsymbol{\kappa}) = C_N^2(R)\, (\rho^2 \kappa_x^2 + \kappa_y^2)^{-\alpha/2} \exp[-(\rho^2 \kappa_x^2 + \kappa_y^2)(\mathbf{l_i}/2\pi)^2]. \qquad (4.1)$$

The GSF for this case is :

$$D_\phi(z, \mathbf{s}) = \frac{1}{\rho} \frac{8\pi^2 r_e^2 \lambda^2 \Delta L}{2^{\alpha-2}(\alpha-2)} \Gamma\left(1 - \frac{\alpha-2}{2}\right) \frac{C_N^2(z) l_i^{\alpha-2}(z)}{(1 - f_p^2(z)/f^2)}$$
$$\times \left\{ {}_1F_1\left[-\frac{\alpha-2}{2}, 1, -\left(\frac{s}{l_i(z)}\right)^2\right] - 1 \right\}. \qquad (4.2)$$

This enables us to compute amplitude of density turbulence $C_N^2$ as a function of heliocentric distance, which when used with (4.1) yields information on the variance of the density fluctuations.

### 4.2.2 Interplanetary scintillations (IPS)

Interplanetary scintillations (IPS) have played an instrumental role in deducing the properties of the solar wind over the wide rage of the heliocentric distances e.g., (Manoharan et al., 1994; Janardhan et al., 1996). We present here a brief introduction to the IPS and its properties. Variations in the in-tensity of an astronomical source over time scales of 0.1 second − 10 seconds are referred to as IPS. IPS observations therefore provide useful information about properties of the solar wind, such as density fluctuations and wind velocity.

Consider radiation of frequency $f$ coming from a distant "point" source incident as a plane wavefront on the thin screen of of solar wind having density irregularities. The role of the thin screen is to introduce phase variations between the emergent wavefronts.

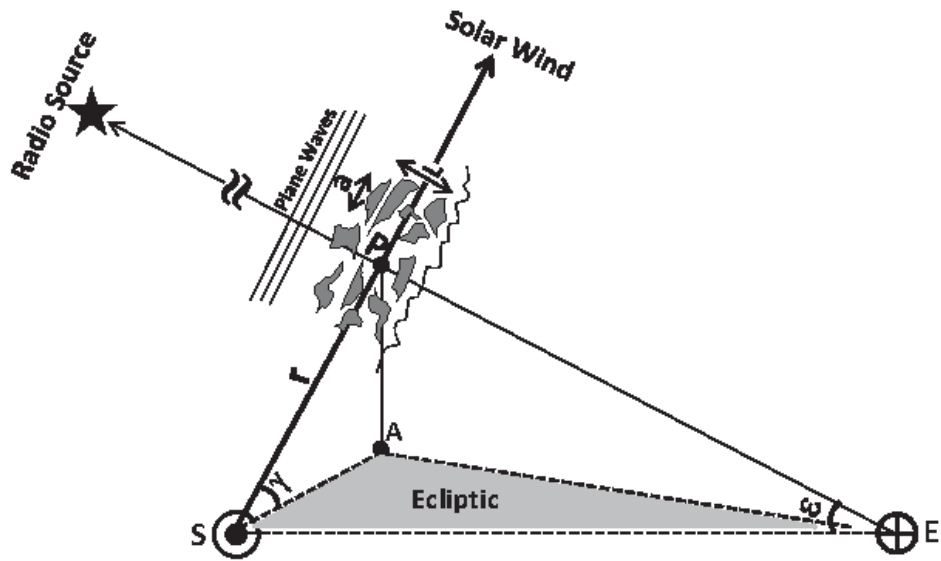

Figure 4.2: Adapted from Bisoi et al. (2014), figure illustrates the geometry of IPS observations.

The phase difference between scattered and unscattered waves varies with distance from the screen. Phase changes are gradually converted to amplitude variations by constructive and destructive interference. As the wavefront continues in the direction of propagation, interference leads to the formation of a spatial intensity pattern. Beyond the first Fresnel zone the intensity modulations are fully developed and follow the spectrum of the refractive index (equivalently, density) fluctuations. This allows us to express the amplitude of intensity variations in terms of density fluctuations $\delta N_e$. It is useful to note at this point that due to the action of the Fresnel filter, one can probe, using IPS observations, the density structure at the scales $\leq 1000$km (Rao et al., 1974; Coles & Filice, 1985; Yamauchi et al., 1998; Fallows et al., 2008). Large scale density fluctuations (like those of coronal mass ejections (CMEs)) are automatically excluded in IPS observations.

The measurable quantity in IPS is the scintillation index $m$, which quantifies the degree to which a distant compact source experiences scintillations. It can be expressed as :

$$m^2 = \langle (I(t) - \langle I \rangle) \rangle / \langle I \rangle^2 \tag{4.3}$$

Where $\langle I \rangle$ is the mean intensity of the source. The value of $m$ depends principally upon

- Distance from the Sun
- Source structure
- Observing frequency and
- Solar wind density fluctuations.

IPS observations are particularly useful in the weak scattering regime, characterized by $\delta\phi \ll 1$, where $\delta\phi$ is the path integrated phase deviation along the line of sight of the source. The condition for strong scattering $\delta\phi \gg 1$ develops as the line of sight of the source approaches the Sun and the electron density ($N_e$) increases.

It is particularly simple in the weak scattering approximation to demonstrate the relation between the scintillation index $m$ and the density fluctuations $\delta N_e$. The four point correlation function of the electric field gives intensity scintillations which defines the scintillation index (Lee & Jokipii, 1975III). Starting from the parabolic wave equation for the statistical moments of the medium (chapter 2) we can derive the expression for the fourth order moment ($\Gamma_{2,2}$). It is possible to write $\Gamma_{2,2}$ in terms of the linear combination of the second order moments ($\Gamma_{1,1}$) which are related to the phase structure function (2.21) and thus to the phase fluctuations $\delta\phi$. Therefore the scintillation index $m$ can be written as (Lee & Jokipii, 1975III) :

$$m^2 = \langle(I(t) - \langle I \rangle)\rangle/\langle I \rangle^2 = 1 - \exp[-2(\delta\phi)^2] \quad (4.4)$$

In the weak scattering approximation, ($\delta\phi \ll 1$) we can neglect the second and higher order terms and write (4.4) as :

$$m = \sqrt{2}\delta\phi \quad (4.5)$$

Phase fluctuations can be expressed in terms of density fluctuations by (Salpeter, 1967; Bisoi et al., 2014) :

$$\delta\phi = (2\pi)^{\frac{1}{4}}\lambda r_e (a\Delta L)^{\frac{1}{2}}[\langle \delta N_e^2 \rangle]^{\frac{1}{2}}, \quad (4.6)$$

where $\lambda$ is the observing wavelength, $r_e$ is the classical electron radius and $a$ is the typical scale size in the thin scattering screen of thickness $\Delta L$. Using (4.6) with (4.5) we can write density fluctuations $\delta N_e$ as :

$$\delta N_e = \frac{m}{(2)^{\frac{1}{2}}(2\pi)^{\frac{1}{4}}\lambda r_e (a\Delta L)^{\frac{1}{2}}} \quad (4.7)$$

Eq. (4.7) enables one to calculate the density fluctuations from $m$, given the parameter $a$. We use angular broadening observations and IPS observations to probe density structures at different solar elongations at different observing frequencies. For the near-Sun region (which is in the strong scattering regime characterized by rms phase fluctuations $\Delta \gg 1$) angular and spectral broadening observations are useful. In the weak scattering regime (i.e. for heliocentric distances $\geq 40 R_\odot$, where $\Delta\Phi \ll 1$ ) IPS observations provide an excellent guide to density fluctuations (Yamauchi et al., 1998; Bisoi et al., 2014). A knowledge of density fluctuations ($\delta N_e$) is required in order to obtain the velocity fluctuations which can then be used to derive the turbulent cascade rate or equivalently heating rate. We infer $\delta N_e$ using 1) the recently developed GSF (Ingale et al., 2015), chapter (3) to analyze radio wave scattering and 2) existing analysis methods for interplanetary scintillation (IPS) data.

## 4.3 Angular broadening and IPS observations : details

In what follows, we briefly describe and reanalyze previously published data regarding observations of angular broadening carried out by Armstrong et al. (1990); Anantharamaiah et al. (1994) phase scintillation observations due to Spangler & Sakurai (1995) and IPS observations from Bisoi et al. (2014).

### 4.3.1 Angular broadening observations

Armstrong et al. (1990) carried out their observations during solar occultation of 3C279 in October 1983 and 1985. which was during the solar minimum of cycle 21. Anantharamaiah et al. (1994) observed standard VLA calibrators during 2-6 November 1988, which was during the maximum of cycle 22.

The angular broadening observations can be directly interpreted as the spatial correlations in the electric field, **E**, also known as visibilities (chapter 2). The spatial scale of the electric field correlations depends upon the radio frequency $f$ and the distance of closest approach of the line of sight to the source (impact parameter $R_0$). To obtain useful measurements these scales must be comparable with the interferometric baselines available. In order to allow for the spatial scales of scattering field to be comparable to the available baseline lengths it is useful to employ different observing frequencies for different solar elongations. Armstrong et al. (1990) and Anantharamaiah et al. (1994) carry out multi-frequency observations between the solar elongations of 2 to 10 $R_\odot$ and 2 to 16 $R_\odot$ respectively.

The very large array (VLA) has baselines ranging from 7-35 km with observing frequencies between 327MHz and 24GHz. The advantage of using interferometric imaging is that they also provide a measure of the anisotropy of the irregularities. Scatter- broadened images of distant celestial sources viewed against the foreground of the solar wind at small solar elongations often exhibit strong anisotropy e.g., (Armstrong et al., 1990; Anantharamaiah et al., 1994). Furthermore, the observed correlations are found to be higher for baselines oriented in a direction close to the radial (defined as the source direction) and the correlations are lower for the baselines oriented perpendicular to the radial (Armstrong et al., 1990), indicating anisotropy, figure (4.3).

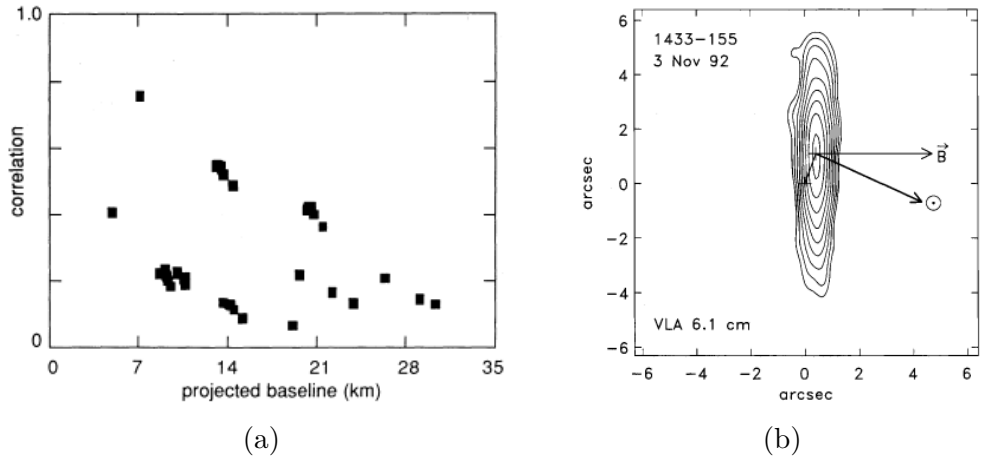

Figure 4.3: (Adapted from (Armstrong et al., 1990)). Figure (a) shows electric field correlation as a function of projected baseline. Figure (b) shows the scatter broadened image of a radio source showing anisotropy. The image is elongated in a direction perpendicular to the mean magnetic field.

We know that the medium can be characterized by the phase structure function $D_\phi(\mathbf{s})$, where $\mathbf{s}$ is the vector along an interferometric baseline. The relation between $D_\phi(\mathbf{s})$ and the spatial power spectrum $S_n(R, \mathbf{s})$ is given by (2.27). Therefore if the power spectrum is anisotropic, it will be reflected in the phase structure function, as is evident in the measurements of Armstrong et al. (1990) and Anantharamaiah et al. (1994).

**Measurements of $D_\phi(\mathbf{s})$**

The observed visibilities $V(\mathbf{s})$ can be directly interpreted in terms of the scattering parameters. The normalized visibilities give the mutual coherence function $\Gamma(\mathbf{s})$ which is related to the structure function by (Anantharamaiah et al., 1994) :

$$D_\phi(\mathbf{s}) = -2ln\Gamma(\mathbf{s}) \text{ where, } \Gamma(\mathbf{s}) = \frac{V(\mathbf{s})}{V(\mathbf{0})} \qquad (4.8)$$

The estimation of the $D_\phi(\mathbf{s})$ using observations of V($\mathbf{s}$) requires no assumptions about the strength of the scattering (Lee & Jokipii, 1975I; Armstrong et al., 1990).

The geometry considered is similar to that outlined in chapter 3. $z$ is the direction of the propagation, and scattering screen is in the $x-y$ plane, such that x-axis is along the direction perpendicular to the mean magnetic field (figure 3.6). The scattered broadened images are elongated along the x-axis. The structure function in this direction is denoted by the $D_{\text{major}}$ and that in perpendicular direction (y-axis) is denoted as the $D_{\text{minor}}$. The anisotropy is parametrized by the axial ratio $\rho \equiv x/y$. Armstrong et al. (1990) presented estimates of the anisotropic structure function and axial ratio ($\rho$) in their figures 4a and 4b which are reproduced here (figure (4.4), (4.5)).

Observations of angular broadening with various baselines shows that the anisotropy is scale dependent i.e. smaller scales are closer to being isotropic than larger scales (Armstrong et al., 1990).

It is also found that the anisotropy is larger for the smaller solar elongations and decreases significantly beyond $6R_\odot$ (Armstrong et al., 1990; Anantharamaiah et al., 1994). Figure 5 of Armstrong et al. (1990) reproduced here (figure 4.6), shows the axial ratio as function of solar elongation. It is evident from figure (4.6) that for solar elongations $> 15R_\odot$, the axial ratio $\rho \simeq 1$. For the data in Anantharamaiah et al. (1994) the values for $D_{\text{major}}$ and $D_{\text{minor}}$ can are obtained directly from their table (2). In both these cases

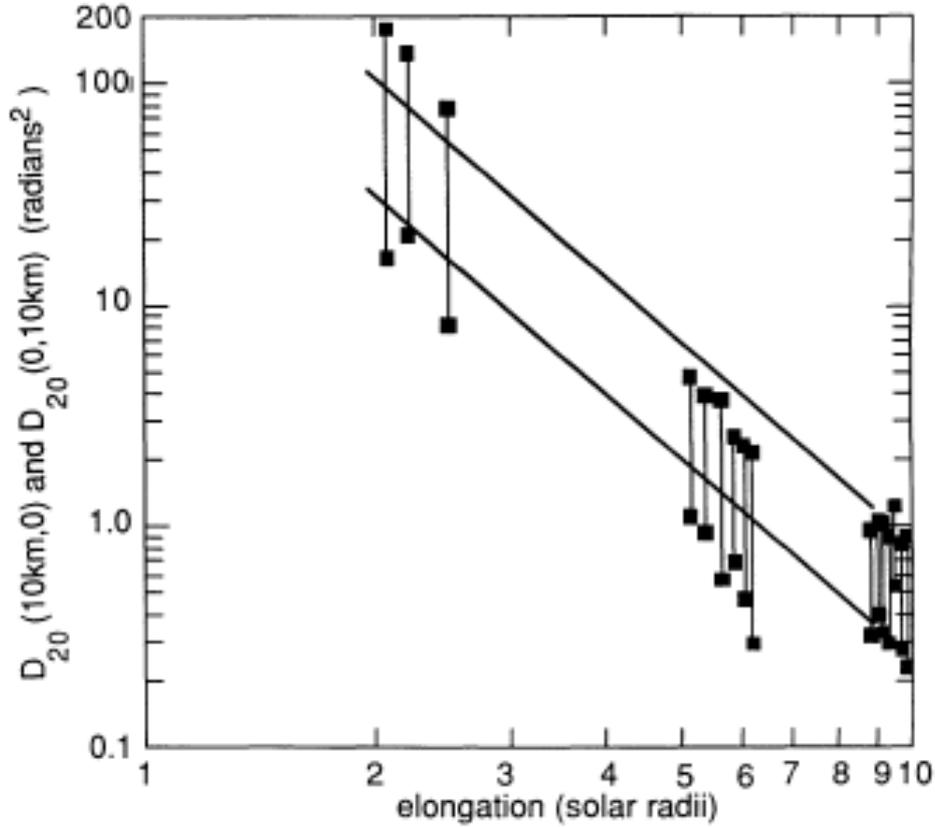

Figure 4.4: Adapted from Armstrong et al. (1990). The figure shows the anisotropic structure function scaled to the baseline of 10 km and wavelength of 20cm as a function of solar elongation. $D(10\text{km}, 0)$ corresponds to $D_{\text{major}}$ and $D(0, 10\text{km})$ corresponds to $D_{\text{minor}}$

all the measurements are scaled to the wavelength of $\lambda = 20$cm and a baseline of $s = 10$km.

Using observations of $D_\phi(\mathbf{s})$ along the major and the minor axes of the density structures we calculate resultant structure function in the $x - y$ plane as :

$$D_{\text{resultant}}(\mathbf{s}) = \sqrt{D_{\text{major}}^2 + D_{\text{minor}}^2} \qquad (4.9)$$

The corresponding error bars are listed in all cases except for the measurements of $D_\phi(\mathbf{s})$ by Armstrong et al. (1990), figure 4.4. Since the errors are not depicted in their paper, following Spangler & Sakurai (1995) we set the errors to be one-third of the values of $D_\phi(\mathbf{s})$. The derived values for the

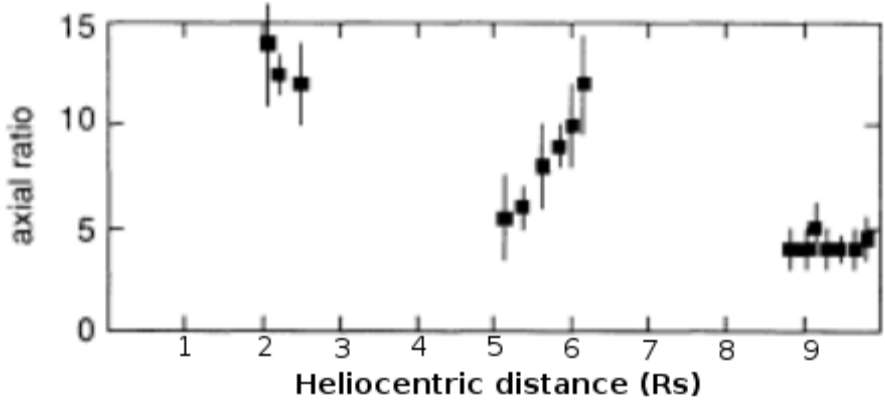

Figure 4.5: Adapted from Armstrong et al. (1990). The figure shows axial ratio as a function of heliocentric distance

measurements of the Armstrong et al. (1990) are given in table (4.1)

**Measurements of $D_\phi$ from Spangler & Sakurai (1995)**

Spangler & Sakurai (1995) use phase scintillation observations to probe the density turbulence in the solar wind at heliocentric distances between 10 to 49.8 $R_\odot$. The observations were carried out with the VLBA during July and August of 1991, which corresponds to the maximum of solar cycle 22. These are also multi-frequency observations. Observations of phase scintillations give measurements of the phase variance from which the phase structure function can be derived using $D_\phi(\mathbf{s}) = \langle[\delta\phi(\mathbf{r}) - \delta\phi(\mathbf{r}+\mathbf{s})]^2\rangle$, (Coles & Harmon, 1989). Here $\delta\phi(\mathbf{r})$ is the phase deviation along the line of sight. This gives information about the phase structure function as a function of the baseline $s$ for different impact parameters $R_0$. Figure 4 of Spangler & Sakurai (1995) showing their results for the measurements of the $D_\phi$ is reproduced here (Figure 4.7).

Figure (4.7) shows the measurements of $D_\phi$ at four different impact parameters ($R_0$) : 10, 26.7, 42.7 and 49.8 $R_\odot$ as a function of the baseline $s$. All the observations are scaled to a wavelength of 12.6 cm. This gives the distribution of $D_\phi(\mathbf{s})$ with $s$ for different solar elongations. Using these values for $D_\phi(\mathbf{s})$ which are functions of solar elongation alone we carry out the

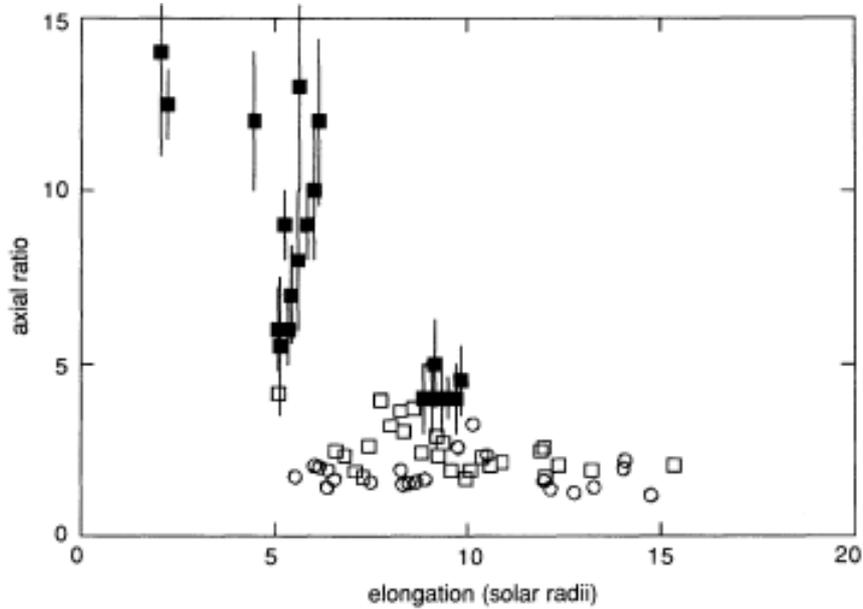

Figure 4.6: Adapted from Armstrong et al. (1990). The figure shows axial ratio $\rho$ as a function of solar elongation. Open boxes represents results of Blessing & Dennison (1981), during 1969-1971 (solar maximum), open circles represents data from Ward (1975), during 1972-1974 (solar minimum) and filled boxes represents data from Armstrong et al. (1990) during October 1983 and 1985 (Solar minimum)

following exercise :

- Step 1 : Baseline scaling - We scale all the observations for each solar elongation to a baseline of 10km. : This is carried out by using the GSF which includes the exponential cut-off (Ingale et al., 2015), instead of a phase structure function that is valid only for the inertial range and was used by Spangler & Sakurai (1995) in their calculations.

- Step 2 : Averaging over the baseline - From step one we obtained four sets of baselines corresponding to four different solar elongations. We consider the average of each set of baselines as a representative baseline at that perticular solar elongation.

This procedure yields values for $D_\phi(\mathbf{s})$ as a function of solar elongation. However there is no information about the errors associated with the measurements of $D_\phi$ given in Spangler & Sakurai (1995). Following Spangler &

Table 4.1: Resultant $D_\phi(\mathbf{s})$ derived using data points from Armstrong et al. (1990)

| distance | $D_{\text{resaultant}}$ | $\rho$ | error in $D_{\text{reaultant}}$ | error in $\rho$ |
|---|---|---|---|---|
| 2.08 | 177.15 | 14 | 19.15 | 3 |
| 2.21 | 138.33 | 12.5 | 15.13 | 1 |
| 2.45 | 78.43 | 12 | 8.83 | 2 |
| 5.17 | 4.88 | 5.5 | 0.63 | 2 |
| 5.4 | 3.99 | 6 | 0.52 | 1 |
| 5.66 | 3.7 | 8 | 0.48 | 2 |
| 5.88 | 2.62 | 9 | 0.34 | 1 |
| 6.07 | 2.33 | 10 | 0.31 | 2 |
| 6.22 | 2.17 | 12 | 0.3 | 2.2 |
| 8.86 | 1.00 | 4 | 0.14 | 1 |
| 9.13 | 1.09 | 6 | 0.15 | 0 |
| 9.34 | 0.93 | 4 | 0.13 | 1 |
| 9.54 | 1.27 | 4 | 0.17 | 0.6 |
| 9.74 | 0.87 | 4 | 0.12 | 0 |
| 9.88 | 0.92 | 4.5 | 0.13 | 1 |

Sakurai (1995) we set the errors to be one third of the mean values of $D_\phi$ obtained while averaging over the baseline (step 2). These measurements are scaled to a wavelength of $\lambda = 12.6$cm whereas previous measurements of $D_\phi(\mathbf{s})$ by Armstrong et al. (1990) and Anantharamaiah et al. (1994) have been scaled to $\lambda = 20$cm. To maintain consistency between the datasets we need to scale $D_\phi$ deduced from Spangler & Sakurai (1995) to the same wavelength i.e. to $\lambda = 20$cm. The wavelength dependence of the structure function is given by :

$$D_\phi(\mathbf{s}) \propto \lambda^2$$

Using this dependence we scale the $D_\phi(\mathbf{s})$ values at $\lambda = 12.6$cm to $D_\phi(\mathbf{s})$ at $\lambda = 20$cm.

As discussed earlier we can use $\rho = 1$ for all the observations of Spangler & Sakurai (1995) except the one at $10R_\odot$. We use a least square fit to the datapoints of figure (4.6) to obtain the value for $\rho$ at $10R_\odot$. The derived values are given in Table (4.2).

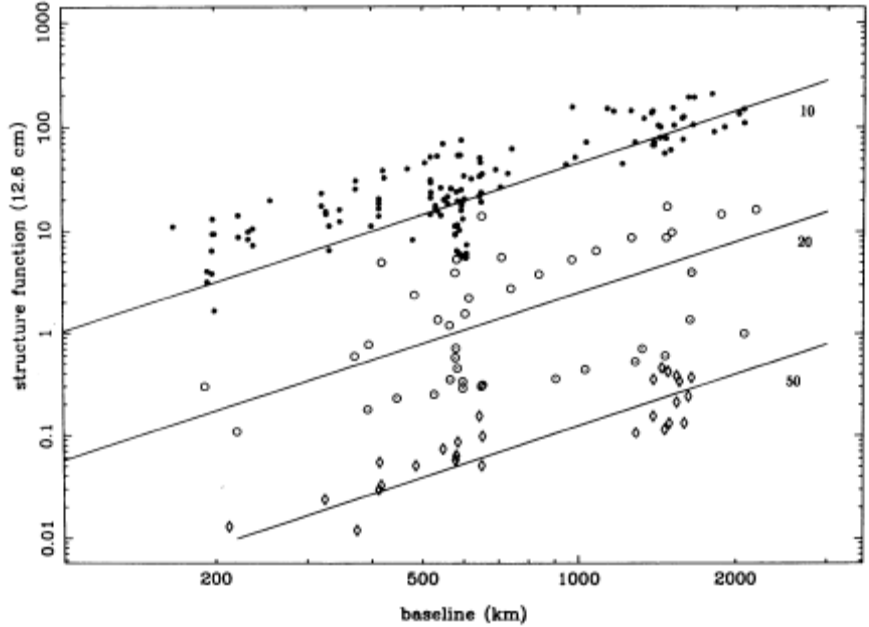

Figure 4.7: $D_\phi(\mathbf{s})$ as a function of baseline $s$ deduced from VLBI observations at different solar elongations and scaled to the wavelength of 12.6 cm. The filled circles represent observations at impact parameter $R_0 = 10 R_\odot$ while the open circles are for $R_0 = 26.7 R_\odot$, the dotted circles are for $R_0 = 42.7 R_\odot$ and the diamonds are for $R_0 = 49.8 R_\odot$. The solid lines are due the model of Coles & Harmon (1989) at $R_0 = 10, 20, 50 R_\odot$. Adapted from (Spangler & Sakurai, 1995))

In all cases discussed above, the measurements of $D_\phi(\mathbf{s})$ are scaled to a baseline of $s = 10$km which is well within the regime where GSF should be used (Ingale et al., 2015); also see figure 3.18).

**IPS Observations**

Observations of about 200 radio sources at 327MHz have been carried out at the multi-station IPS observatory of STEL, Japan (Kojima & Kakinuma, 1990; Asai et al., 1998). Every source is observed each day as it moves over the heliocentric distance of 0.2 AU ($\sim 40 R_\odot$) to 0.8 AU ($\sim 174 R_\odot$) over the period of one year. Bisoi et al. (2014) shortlisted 27 radio sources from these observations spanning the years 1998 to 2007 i.e., the entire solar cycle 23.

We use the data set for the scintillation index $m$ for the 27 "shortlisted"

Table 4.2: Resultant $D_\phi(\mathbf{s})$ derived using data points from Spangler & Sakurai (1995)

| distance | $D_{\text{resaultant}}$ | $\rho$ | error in $D_{\text{reaultant}}$ | error in $\rho$ |
|---|---|---|---|---|
| 10 | 0.96 | 4.44 | 0.13 | 1.37 |
| 26.7 | 0.09 | 1 | 0.01 | 0.3 |
| 42.7 | 0.01 | 1 | 0.002 | 0.3 |
| 49.8 | 0.003 | 1 | 0.0005 | 0.3 |

sources from Bisoi et al. (2014). This data set covers all of solar cycle 23 and spans heliocentric distances between 0.26 to 0.8 AU (i.e., $\sim 40 R_\odot - 174 R_\odot$). In order to calculate the density fluctuations (and therefore the density modulation index) as a function of heliocentric distance at the dissipation scale we processed the data as follows.

- Step 1 : Calculation of $\delta N_e$ - Equation (4.7) gives relation between the $m$ and $\delta N$. To compute $\delta N$ we need to specify $a$ − the spatial scale of interest in the scattering screen and $\Delta L$ − thickness of the scattering screen.

  *Choice of a* : To choose $a$, we note that, 327 MHz observations are known to be sensitive only to a particular (or narrow range of) spatial scales; between 10 km to 1000 km (Yamauchi et al., 1998; Bisoi et al., 2014). We therefore choose $a = 1000$km.

  *Thickness of scattering screen* : We assume that at a particular heliocentric distance ($z$) the thickness $\Delta L$ is of the order of that distance i.e., $\Delta L = z$.

- Step 2 : We calculate density modulation index ($\epsilon_{N_e} = \frac{\delta N_e}{N_e}$) using $\delta N_e$ from (step 1) and the background electron density ($N_e$) due to Leblanc et al. (1998). We adopt the procedure of Bisoi et al. (2014) who use ACE data for $N_e$ and normalize the Leblanc et al. (1998) density model with the value of $N_e$ at the Earth.

- Step 3 : Averaging $\epsilon_{N_e}$ - The Bisoi et al. (2014) data for $m$ (and therefore derived data of $\epsilon_{N_e}$) contains spatial as well as temporal dependences. To obtain $\epsilon_{N_e}$ as function of heliocentric distance alone we average $\epsilon_{N_e}$ over the entire time period of $1998-2007$ at each heliocentric distance. This finally gives $\epsilon_{N_e}(z)$.

- Step 4 : Error calculation - Following Bisoi et al. (2014) we calculate the standered deviation from the time averaged value of $\epsilon_{N_e}$ at each heliocentric distance.

## 4.4   Data selection

We further divide the measurements according to the phase of the solar cycle; i.e., we separate the measurements during solar minima from those during solar maxima. The observations of Armstrong et al. (1990) were carried out during solar minima. whereas those of Anantharamaiah et al. (1994) and Spangler & Sakurai (1995) were carried out during solar maxima. The IPS observations of Bisoi et al. (2014) spans an entire solar cycle; we have therefore carefully separated observations corresponding to solar maximum from those corresponding to the minimum period.

We select only ecliptic sources for the following reason : The slow solar wind turbulence is usually thought to be fully developed, so that the radiation emitted from a distant ecliptic source is dominantly effected by the slow solar wind It is well known that the properties of the solar wind depends on its flow speed. Spangler & Sakurai (1995) shows that the bulk of radio wave scattering occurs in the slow solar wind. Therefore we only consider ecliptic sources so as to primarily study turbulence in the slow solar wind.

## 4.5   Results

In this section we present the calculations and our results obtained on the basis of the formulation outlined in (§5.5.1) and (§5.5.2) using observational inputs from (§3).

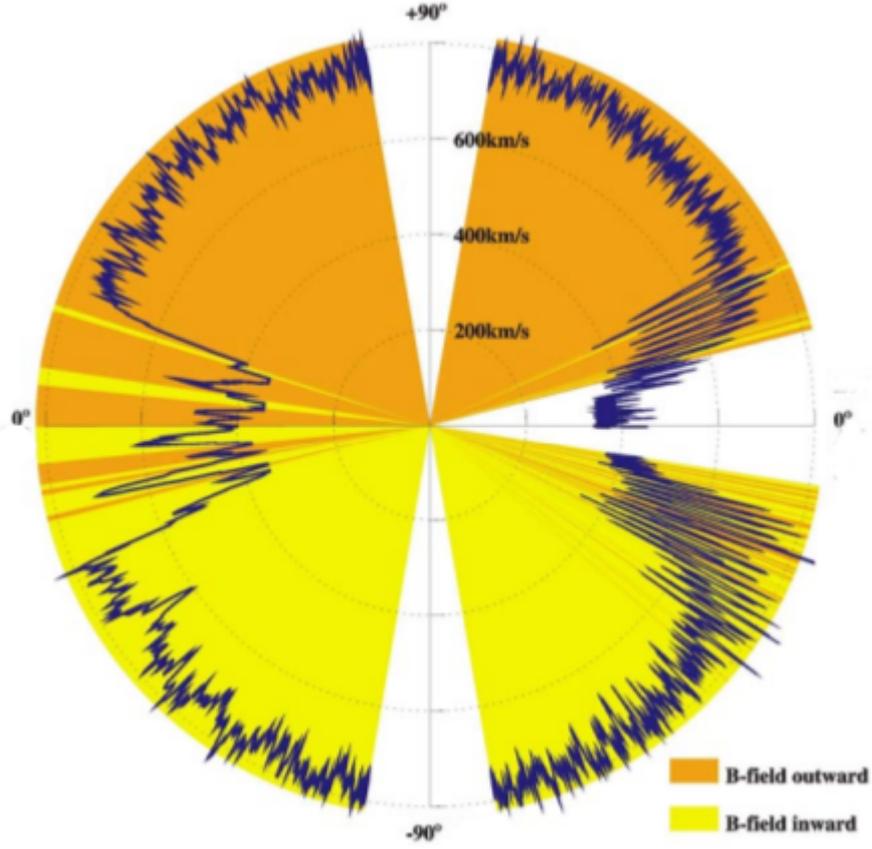

Figure 4.8: solar wind speed distribution adapted from Schwenn (2001), (Courtesy J Woch, Max-Planck-Institute für Aeronomie). Slow solar wind originates in the ecliptic region.

### 4.5.1 Density modulation index

The GSF given by (4.2) includes the effect of anisotropy via the axial ratio ($\rho$) and the refractive index effect $(1 - f_p(z)^2/f^2)$ as well. The symbols have their usual meaning defined in chapter 3. Equation (4.2) is used to calculate amplitude of density fluctuations $C_N^2$ as a function of heliocentric distance.

Defining $f(z, \alpha)$ as :

$$f(z,\alpha) = \frac{l_i^{\alpha-2}(z)}{(1 - f_p^2(z)/f^2)} \left\{ {}_1F_1\left[-\frac{\alpha-2}{2}, 1, -\left(\frac{s}{l_i(z)}\right)^2\right] - 1 \right\} .$$

We can write $C_N^2(z)$ as :

$$C_N^2(z) = D_\phi(z,\mathbf{s})\rho \left[\frac{8\pi^2 r_e^2 \lambda^2 \Delta L}{2^{\alpha-2}(\alpha-2)}\Gamma\left(1-\frac{\alpha-2}{2}\right)f(z,\alpha)\right]^{-1} \quad (4.10)$$

In calculating $C_N^2(z)$ we note that estimates of $D_\phi(\mathbf{s})$ using observations of the angular and spectral broadening, are scaled to a wavelength of $\lambda = 20$ cm and the interferometric baseline $s = 10$km. We take $\Delta L = z$ as mentioned earlier. There is some evidence that the power spectrum for the density fluctuations exhibit flattening at spatial scales between 1000km and the inner scale ($l_i$) (Coles & Harmon, 1989). The power law index in this range of spatial scales is usually approximated by $\alpha = 3$. As we have scaled all the measurements to $s = 10$km which is in the range of the spatial scales where flattening is observed, we take $\alpha = 3$ in our calculations.

The observed spatial power spectrum of density irregularities shows an abrupt steepening at smaller scales, indicative of the existence of an inner scale ($l_i$). While there are typically multiple scales where steepening occures, proton cyclotron damping due to magnetohydrodynamic (MHD) waves is often invoked as the physical process to explain the appearance of the inner scale (Coles & Harmon, 1989; Harmon, 1989; Verma, 1996; Yamauchi et al., 1998; Leamon et al., 1999, 2000; Smith et al., 2001; Bruno & Trenchi, 2014). This mechanism yields the following expression for the inner scale :

$$l_i(z) = 684 N_e(z)^{-1/2} \text{ km}, \quad (4.11)$$

where $N_e$ is the background electron density. A background density model is required to calculate inner scale as well as the refractive index effect ($f_p(z) \propto N_e(z)$). We use the Leblanc et al. (1998) model for the background electron density which as a function of heliocentric distance (measured in the

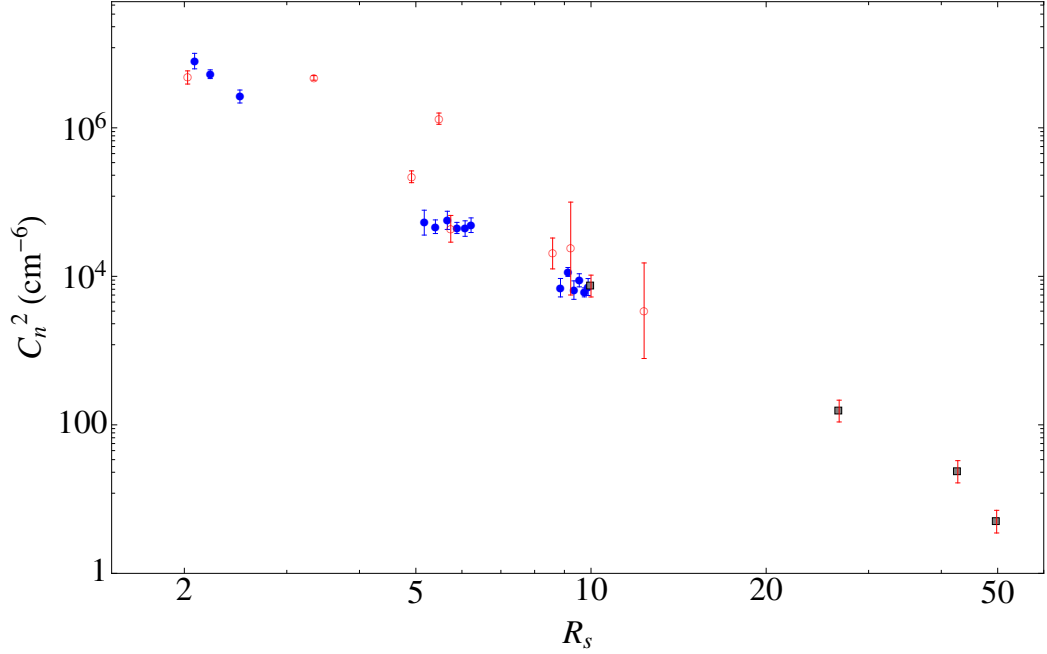

Figure 4.9: $C_N^2$ as a function of heliocentric distance $z$ in units of $R_\odot$. Filled circles denotes the $C_N^2$ derived using estimates of $D_\phi(\mathbf{s})$ by Armstrong et al. (1990), the open circle indicate $C_N^2$ used for estimates of $D_\phi(\mathbf{s})$ by Ananthara- maiah et al. (1994) and the filled boxes are for Spangler & Sakurai (1995). Data points in blue corresponds to solar minimum and data points in red corresponds to solar maximum.

units of solar radius) is given by :

$$N_e(z) = 3.3 \times 10^5 z^{-2} + 4.1 \times 10^6 z^{-4} + 8^7 z^{-6} \text{ cm}^{-3} \qquad (4.12)$$

Figure 4.9 shows a scatter plot of the amplitude of density turbulence $C_N^2$ as a function of heliocentric distance obtained by combining estimates of $D_\phi(\mathbf{s})$ from various radio observations during solar maximum and minimum. Having specified the parameters required to calculate $C_N^2(z)$ using equation (4.10) we now turn to the calculation of the spectral density given by eq. (4.1). We focus on the inner scale since most of the power in the perpendicular cascade is present at $\kappa_\perp \simeq \kappa_i$ $(2\pi/l_i)$ and most of the dissipation occurs for $\kappa_\perp > \kappa_i$ (Chandran et al., 2009). We can further simplify expression (4.1) by using the observations of anisotropy by Armstrong et al. (1990). Figure

(6) of Armstrong et al. (1990) illustrates two situations regarding the crucial question of whether the inner scale is isotropic or anisotropic. They conclude that the inner scale is likely to be isotropic. Dastgeer & Zank (2004) also found in their simulations of nearly incompressible turbulence that small scale structures are isotropic. We can therefore consider the inner scale to be isotropic. In this situation $\kappa_\perp \to \kappa_i$, the axial ratio $\rho \to 1$, which implies that :

$$\kappa_\perp^2 = \rho^2 \kappa_x^2 + \kappa_y^2 \to \kappa_x^2 + \kappa_y^2$$

Thus spectral density ($S_n$) at the inner scale (i.e. for $\kappa_\perp = k_i$) can be written as :

$$S_n(z, \kappa_i) = C_N^2(z) \kappa_i^{-3} e^{-1} \qquad (4.13)$$

We know that the Fourier transform of the variance of the density fluctuations $\langle \delta N_e^2 \rangle$ is the spatial power spectrum $S_n$. The density fluctuations at the inner scale ($\delta N_{k_i}$) can be written as (Chandran et al., 2009) :

$$\delta N_{\kappa_i}^2(z) \sim 4\pi \kappa_i^3 S_n(z, \kappa_i) \qquad (4.14)$$

Combining equations (4.10) and (4.13) thus enables us to compute density fluctuations using measurements of $D_\phi(\mathbf{s})$ derived from angular and spectral broadening observations in the near Sun region ($2 - 50 R_\odot$). Thus given density fluctuations at the inner scale (4.14) and the background electron density $N_e$, the density modulation index at the inner scale ($\epsilon_{k_i}$) can be defined as :

$$\epsilon_{N_e}(z, \kappa_i) \equiv \frac{\delta N_{\kappa_i}(z)}{N_e(z)} \qquad (4.15)$$

Figure (4.10)shows the modulation index $\epsilon_{N_e}(z, \kappa_i)$ at the inner scale as a function of heliocentric distance ($z$) in $R_\odot$ for the observations during solar

maximum of the solar cycle 23.

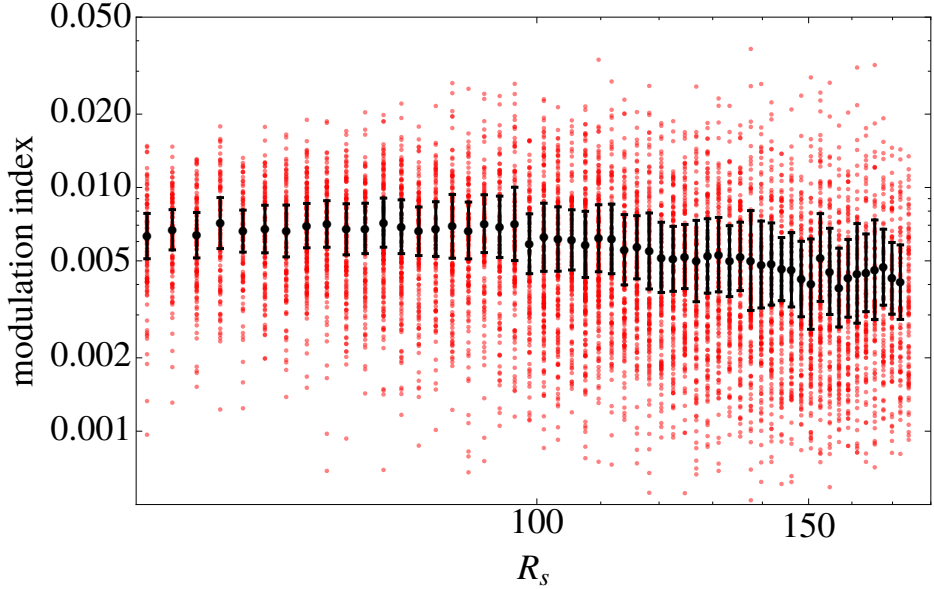

Figure 4.10: Density modulation index at the inner scale as a function of heliocentric distance for the observations during solar maximum. Red data points corresponds to the $\epsilon_{N_e}(z, \kappa_i)$ over the period of 1998 to 2007 for a particular solar elongation, whereas black data points are time averaged values of $\epsilon_{N_e}(z, \kappa_i)$ associated with the error bars calculated in (§5.4)

It is clear from figure (4.10) that the mean values of $\epsilon_{N_e}(z, \kappa_i)$ (filled black circles) at the inner scale during solar maximum don't vary much over $40 - 174 R_\odot$. Similarly, figure (4.11) shows the modulation index $\epsilon_{N_e}(z, \kappa_i)$ at the inner scale as a function of the heliocentric distance ($z$) in $R_\odot$ for observations during the minimum of cycle 23.

Figure (4.11) shows that the mean value of $\epsilon_{N_e}(z, \kappa_i)$ at the inner scale (filled black circles) varies only by a small amount over the heliocentric distance $40 - 174 R_\odot$ during solar minimum. A comparison of the scatter plots, (Figures 4.10 and 4.11) shows that the density modulation index at the inner scale depends only weakly on the phase of the solar cycle as also noted by Bisoi et al. (2014).

Figure (4.12) represents the density modulation index at the inner scale as a function of heliocentric distance, calculated using radio scattering and IPS observations. The density profile (4.12) is assumed in deriving $\epsilon_{N_e}(z, \kappa_i)$.

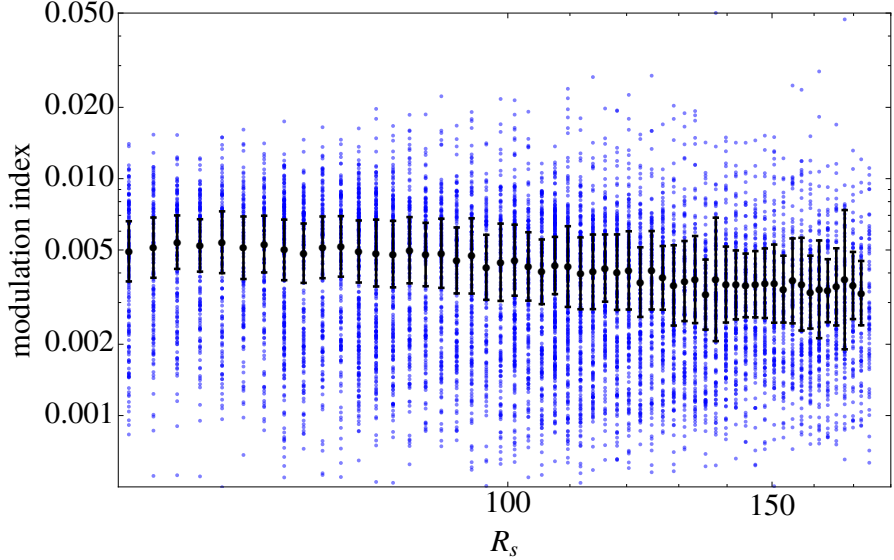

Figure 4.11: Density modulation index at inner scale as a function of heliocentric distance for the observations during solar minimum. Blue data points corresponds to the $\epsilon_{N_e}(z, \kappa_i)$ over the period of 1998 to 2007 for a particular solar elongation, whereas black data points are time averaged values of $\epsilon_{N_e}(z, \kappa_i)$ with error bars calculated as specified in (§5.4)

Blue data points are for solar minimum and red data points are for solar maximum. Filled circles are due to observations from Armstrong et al. (1990), open squares indicate data points due to Anantharamaiah et al. (1994), dotted circle are for data from Spangler & Sakurai (1995) and open circles represents data points due to Bisoi et al. (2014). The density modulation index ($\epsilon_N(z)$) increases for R $< 10 R_\odot$, reaching a maximum at $10 R_\odot$, ($\epsilon_N(10 R_\odot) \simeq 10\%$), and decreases at larger distances. This behavior roughly agrees with the recent observations of Miyamoto et al. (2014). Miyamoto et al. (2014) calculated the fractional density amplitude, equivalent to density modulation index $\epsilon_N$, using a spacecraft radio occultation technique. They found that $\epsilon_N$ increases with heliocentric distance $< 5 R_\odot$ and reaches the maximum value of $\sim 0.3$. For larger heliocentric distances ($> 5 R_\odot$) definite trend could not be inferred. The background electron density (4.12) is a monotonically decreasing function of heliocentric distance. Therefore the peak observed in $\epsilon_N$ at $\sim 10 R_\odot$ in figure (4.12) suggests that the density fluctuations override the behavior of the background electron density in the near-Sun region to produces the observed behavior of $\epsilon_N$.

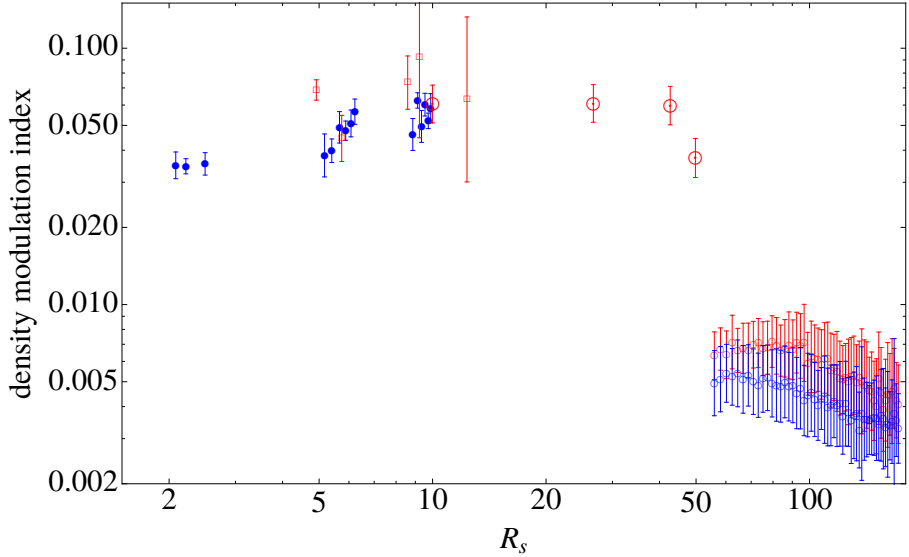

Figure 4.12: Density modulation index at the inner scale as a function of heliocentric distance. Blue data points are for solar minimum and red data points correspond to solar maximum. Filled circles are the observations of Armstrong et al. (1990), open squares indicate data points from Ananthara- maiah et al. (1994) dotted circle represent data from Spangler & Sakurai (1995) and open circles represent data points from Bisoi et al. (2014)

## 4.6 Extended solar wind heating

We obtained density fluctuations and density modulation index at the inner scale. This can be used to compute the heating rate in the extended solar wind. We consider the compressive kinetic alfvén wave turbulence and obtain the turbulent energy cascade rate at the inner scale. Details of the KAW turbulence are discussed in the A.

### 4.6.1 Density fluctuations in KAW

At wavenumbers corresponding to $\kappa_\perp \rho_i \gg 1$, MHD Alfvén waves, via nonlinear interactions, transfer their energy to KAWs. As the wave-vector becomes increasingly oblique, $(\kappa_\perp \gg \kappa_\parallel)$, KAWs develop compressibility and induce parallel magnetic field fluctuations. At this point it is possible for KAW to exchange energy with compressive slow modes, which contributes passively,

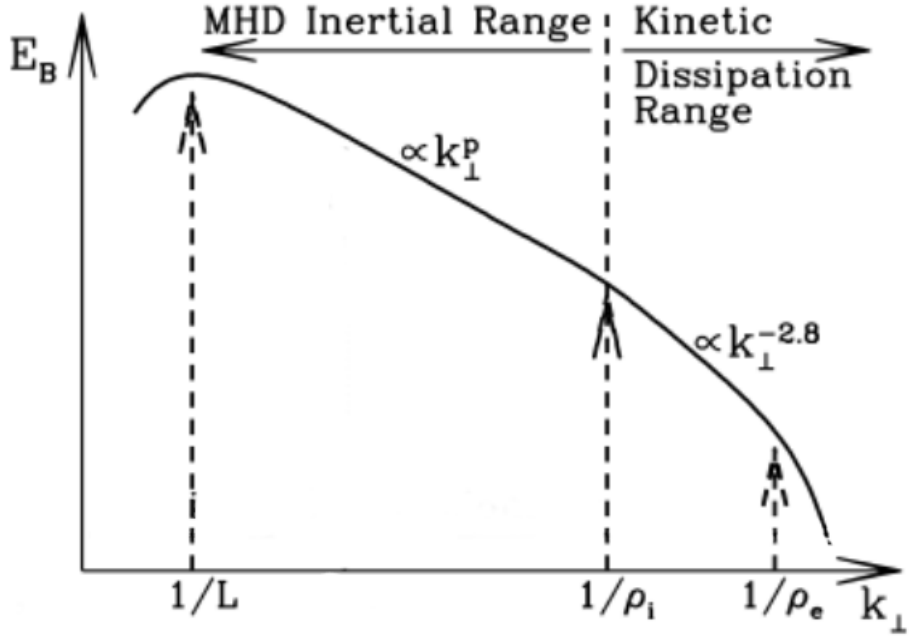

Figure 4.13: Adapted from Howes, 2015; this figure shows the MHD turbulence spectrum of magnetic energy ($E_B$) as a function of $\kappa_\perp$. The KAW regime is marked with wavenumbers $> 1/\rho_i$. $\rho_e$ indicates the electron gyroradius

and becomes highly compressive. As the Alfvén wave turbulence cascades to higher $\kappa_\perp$ the "active" KAW contribution to the density fluctuations increases rapidly, so that the density fluctuation spectrum "rises" at KAW scales (Chandran et al., 2009).

(Harmon, 1989) showed that this dominance of "active" KAW contribution to the density fluctuations can be seen in the spatial power spectrum of density irregularities in the solar wind derived from the radio scintillation observations. He showed that for $\kappa_\perp \approx 10^{-2}$ km$^{-1}$ KAW compressibility dominates the density fluctuation spectrum and produces the observed flattening (with power law index $\alpha = 3$; Figure 4.15) in an otherwise steep (Kolmogorov power law) power spectrum (Coles & Harmon, 1989; Harmon, 1989).

Hollweg (1999) derived the dispersion relation for the KAW using the two fluid model and assuming $\kappa_\perp \gg \kappa_\parallel$. This equation relates density fluctuations ($\delta N_e/N_e$) to the velocity fluctuations ($\delta v/v_A$) in KAW, where $v_A$ is the Alfvén speed.

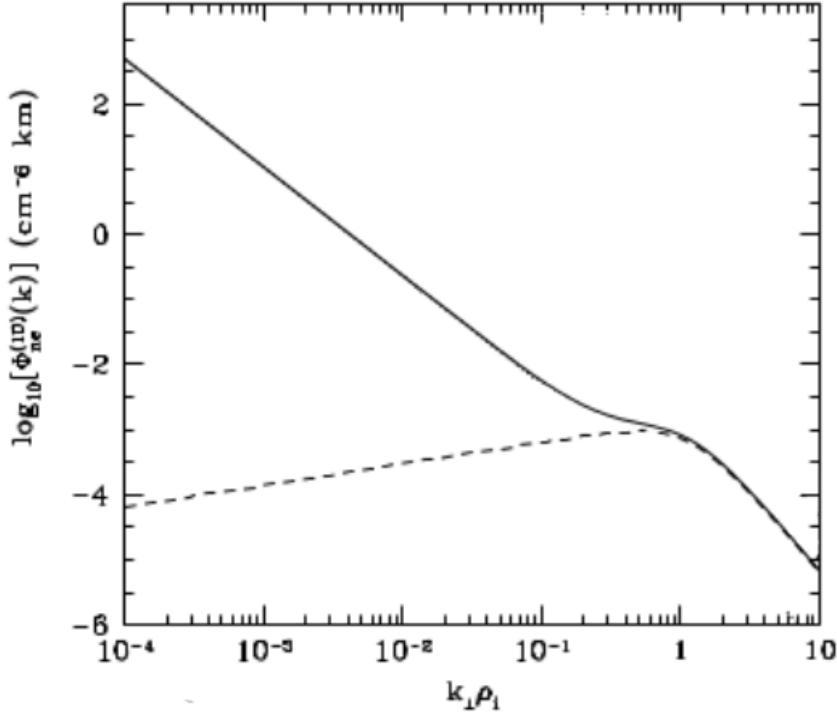

Figure 4.14: Adapted from (Chandran et al., 2009), this figure shows 'predicted' one dimensional spatial power spectrum of density turbulence as a function of perpendicular wavenumber ($\kappa_\perp$). "active" KAW component is shown as the dotted line. Notice the rise of the power spectrum for $\kappa_\perp \rho_1 > 1$, which is the KAW regime.

$$\left|\frac{\delta N_e}{N_e}\right| = \frac{\kappa_\perp d_i}{1 + \gamma_i \kappa_\perp^2 \rho_i^2} \left|\frac{\delta v}{v_A}\right| \quad (4.16)$$

The quantity $\delta v/v_A$ is a measure of the first order amplitude of the Alfvén wave (Hollweg, 1999; Chandran et al., 2009), and $\gamma_i$ is the adiabatic index of the ions.

This dispersion relation can help us use the measurements of density fluctuation to constrain the KAW turbulence in the dissipation range and can be useful in computing turbulent heating.

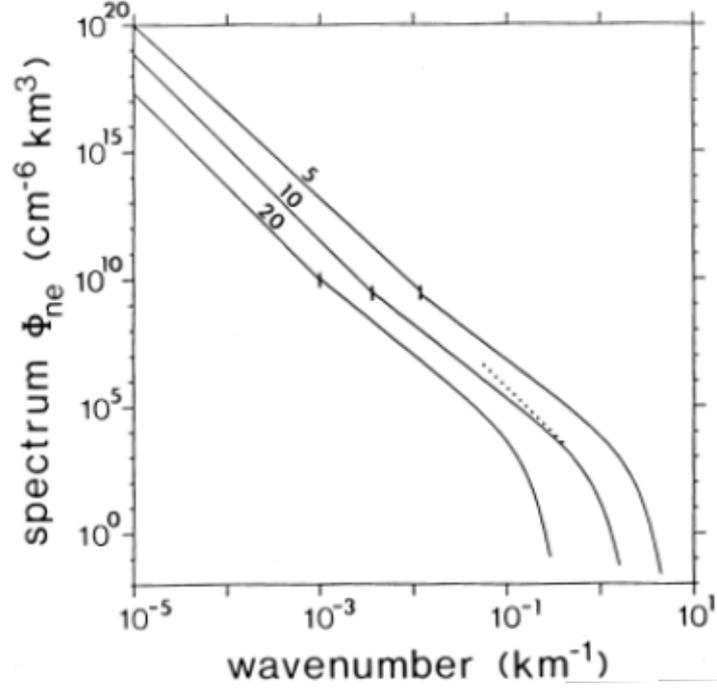

Figure 4.15: Adapted from Coles & Harmon (1989). This figure shows the model density fluctuation power spectra derived from radio observations. The flattening is observed between $10^{-3}$ and $10^{-1}$ km$^{-1}$.

Using data from radio wave scattering and IPS observations we have derived the density fluctuations at inner scale ($\delta N_{k_i}$). Assuming that the density fluctuations at the inner scale ($\kappa_i^{-1}$) arise entirely from KAWs we can use equation (A.17) for the KAW to estimate velocity fluctuations at the inner scale. Following (Chandran et al., 2009) we can write (A.17) as :

$$\delta v_{k_i}(z) = \left( \frac{1 + \gamma_i \kappa_i(z)^2 \rho_i(z)^2}{\kappa_i(z) d_i(z)} \right) \epsilon_{N_e}(z, \kappa_i) V_A(z) \qquad (4.17)$$

The symbols have their usual meaning specified in (§5.2). We assume that the adiabatic index, $\gamma_i = 1$ (Chandran et al., 2009). The ion gyroradius is given by :

$$\rho_i(z) = 1.02 \times 10^2 \mu^{1/2} T_i^{1/2} B(z)^{-1} \text{ cm} \qquad (4.18)$$

Here $\mu$ denotes the ion mass expressed in terms of the proton mass; for our purposes, $\mu \sim 1$. $T_i$ is the ion temperature in eV; we assume it to be 86.22 eV (corresponding to $10^6$ K) for $R < 50 R_\odot$ and 8.622 eV (corresponding to $10^5$ K) for R$> 50 R_\odot$. $B(z)$ is the Parker spiral magnetic field in the ecliptic plane (Williams, 1995). The inner scale model due to Coles & Harmon (1989) can be used to compute $\kappa_i(z) = 2\pi/l_i(z)$. The ion-inertial length is given by $d_i = V_A/\omega_c$, where $\omega_c$ is the proton cyclotron frequency and $V_A$ is the Alfvén velocity given by :

$$V_A(z) = 2.18 \times 10^{11} \mu^{-1/2} N_e(z)^{-1/2} B(z) \text{ cm/sec} \quad (4.19)$$

The ion-inertial length ($d_i$) in terms of the background electron density is given by :

$$d_i(z) = 1/\kappa_i(z) = 228 N_e(z)^{-1/2} \text{ km} \quad (4.20)$$

Thus we obtain velocity fluctuations $\delta v_{k_i}(z)$ using (4.17). This enables us to compute the turbulent energy cascade rate $\epsilon_{k_i}(z)$ at the inner scale (Chandran et al., 2009) :

$$\epsilon_{k_i}(z) = c_0 \rho_p \kappa_i(z) \delta v_{k_i}(z)^3 \text{ erg cm}^{-3} \text{s}^{-1} , \quad (4.21)$$

where $c_0$ is a dimensionless constant. The value of $c_0$ is not precisely known but we follow Howes et al. (2008) and set $c_0 = 0.25$. The proton density $\rho_p = m_p N_e(z)$ gm cm$^{-3}$, where $m_p$ is the proton mass in gm.

### 4.6.2 Turbulent heating rate in the extended solar wind

The turbulent energy cascade rate at the inner scale $\epsilon_{k_i}(z)$ is obtained as a function of heliocentric distance using (4.21) and for different phases of the solar cycle. We interpret this as an upper limit to the solar wind heating rate. The result is shown in figure (4.16). The red points are for solar maximum and the blue points are for solar minimum. Data points derived using observations of Armstrong et al. (1990) are indicated by the filled circles. Data points derived using observations of Anantharamaiah et al. (1994) are denoted by the diamonds. Data points derived using observations of Spangler & Sakurai (1995) are denoted by the open boxes. Data points derived using observations of Bisoi et al. (2014) are denoted by the open circles. We carried out a non-linear least square fitting to the data points in Figure (4.16), the result is shown in figure (4.17).

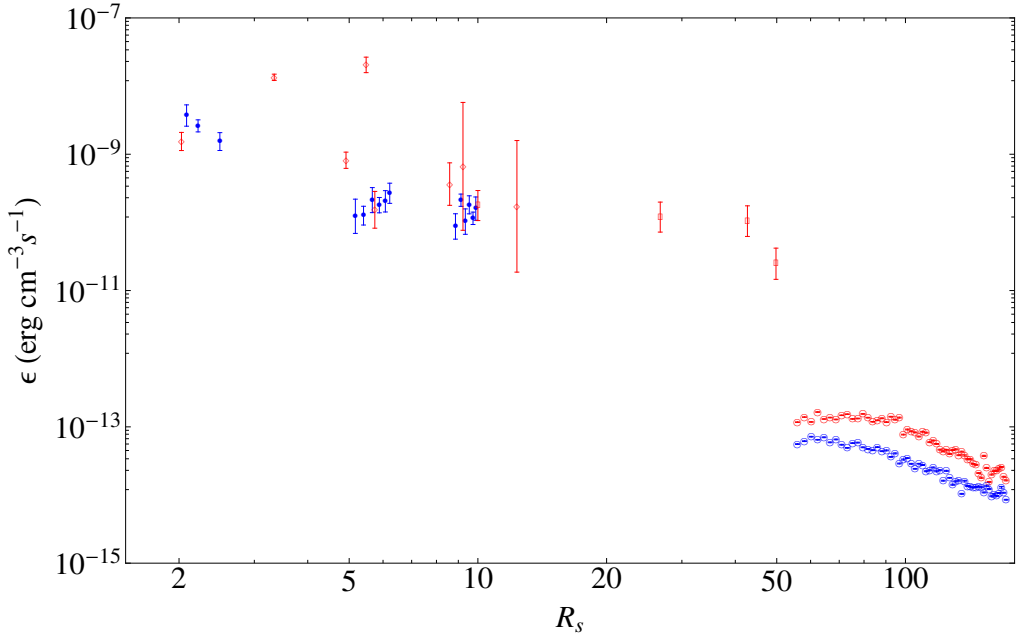

Figure 4.16: Turbulent cascade rate $\epsilon_{k_i}(z)$ at inner scale as a function of heliocentric distance ($R_s$) in the units of solar radius. Data points in blue indicate $\epsilon_{k_i}(z)$ during solar minimum and red data points indicates $\epsilon_{k_i}(z)$ during solar maximum. Filled circles use data from Armstrong et al. (1990), diamonds use data from Anantharamaiah et al. (1994) and open boxes use data from Spangler & Sakurai (1995) measurements of $D_\phi(\mathbf{s})$. Open circles use data from Bisoi et al. (2014).

Our results indicate that the cascade rate is sensitive to the phase of the solar cycle. At a given heliocentric distance ($\epsilon_{k_i}(z)$), for solar maximum is higher than that of for solar minimum. For heliocentric distances ranging from 2 to 174 $R_\odot$, we find that the cascade rate during solar maximum ranges from $3 \times 10^{-8}$ ergcm$^{-3}$s$^{-1}$ at $2R_\odot$ to $10^{-14}$ ergcm$^{-3}$s$^{-1}$ at $174R_\odot$, whereas during solar minimum the cascade rate varies between $3 \times 10^{-9}$ and $10^{-15}$ ergcm$^{-3}$s$^{-1}$. A non-linear least square fit to data points in Figure (4.16) shows that $\epsilon_{k_i}(z)$ has a power law dependence on heliocentric distance $z$ ($\epsilon_{k_i}(z) \to z^{-\beta}$) with the power law index $\beta \sim 1.85 \pm 0.09$. We compare the estimate of $\epsilon_{k_i}(z)$ with the previous findings at $5R_\odot$ and 1AU. Our estimate of $\epsilon_{k_i}(z)$ at 1AU ($\sim 10^{-16}$) during solar minimum is consistent with the findings of Leamon et al. (1999). Chandran et al. (2009) compute the upper limit to $\epsilon_{k_i}(z)$ at 5 $R_\odot$ and 1AU. They scale the value of $S_n$ due to Coles & Harmon (1989), at the inner scale $\kappa_i = 10^{-1}$km$^{-1}$ and at $5R_\odot$, to a value appropriate for coronal holes and find $\epsilon_{k_i}(z) \lesssim 1.5 \times 10^{-8}$ erg cm$^{-3}$s$^{-1}$. This result is for solar minimum (since the observations of Coles & Harmon (1989) coincides with the period of solar minimum of cycle 21). For the slow solar wind the value of $S_n$ is a factor $\approx 15$ greater than that for the coronal holes Coles et al. (1995). We therefore directly use the estimate of $S_n$ obtained from figure (4) of Coles & Harmon 1989 at $\kappa \sim \kappa_i$ (which is 15 times the value of $S_n$ used by Chandran et al. (2009). Using the ecliptic magnetic field (Williams, 1995) we obtained $\rho_i$, $d_i$ and $v_A$ at $5R_\odot$, equation (4.21) then predicts $\epsilon_{k_i}(5R_\odot) \gtrsim 1.1 \times 10^{-7}$ erg cm$^{-3}$ s$^{-1}$. Our estimate of $\epsilon_{k_i}(z) \sim 10^{-9}$ erg cm$^{-3}$ s$^{-1}$ at $5R_\odot$ can be compared with this upper limit. It is clear that the upper limit is above the value we obtained by using radio scintillation observations.

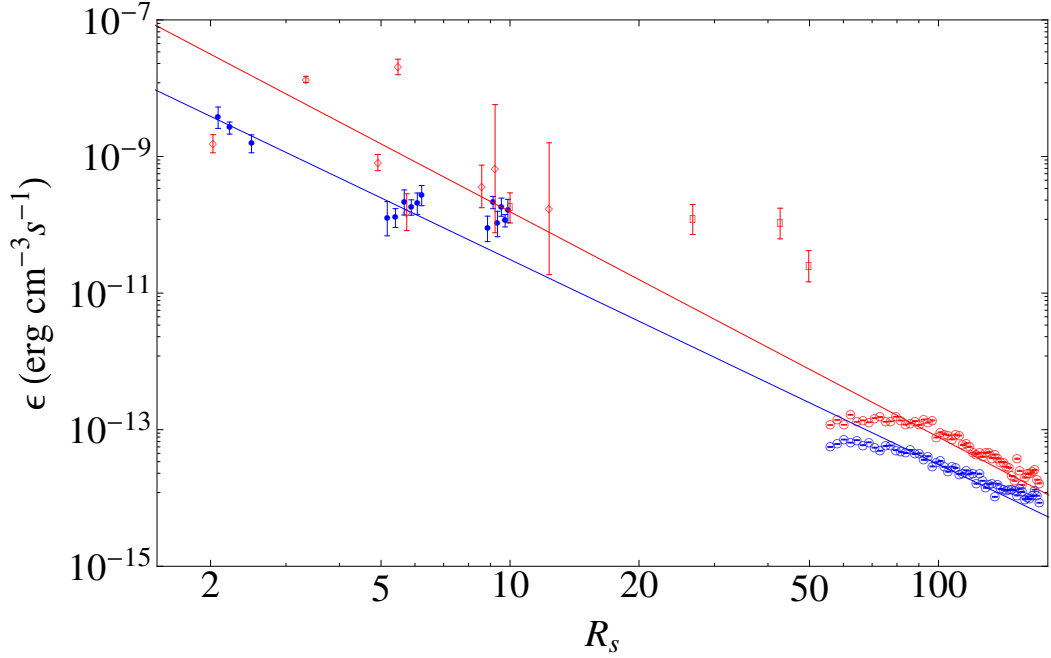

Figure 4.17: Turbulent cascade rate $\epsilon_{k_i}(z)$ at the inner scale as a function of heliocentric distance in the units of $R_\odot$. The red solid line indicates the fitting for data points obtained during solar maximum and the blue solid line denotes the fitting for data points obtained during solar minimum.

Verma et al. (1995) derived an expression for the evolution of temperature gradient in the spherically symmetric solar wind. Using this relation Vasquez et al (2007) obtained the turbulent energy dissipation rate as a function of heliocentric distance. Setting the appropriate values for the ion temperature and solar wind speed at 1 AU with this model, yields $\epsilon_k \sim 1.7 \times 10^{-16}$ erg cm$^{-3}$ s$^{-1}$ which agrees well with our estimate of $\epsilon_{k_i} \sim 10^{-16}$ erg cm$^{-3}$ s$^{-1}$ at 1 AU.

Allen et al. (1998) and Esser et al. (1997) have constructed a model for the parametrized heating rate and calculated the value of $\epsilon_{k_i}(z)$. At $5R_\odot$ the range of values of the turbulent energy cascade rate, due to these models, is between $2 \times 10^{-10}$ erg cm$^{-3}$ s$^{-1}$ and $1.4 \times 10^{-8}$ erg cm$^{-3}$ s$^{-1}$. We find that our value of the $\epsilon_{k_i}(z)$ at $5R_\odot$, ($\sim 10^{-9}$ erg cm$^{-3}$ s$^{-1}$), is within this range. The discussion above suggests that the radio observations can constrain the turbulent heating rate over the wide range of heliocentric distances spanning $2 - 174 R_\odot$.

## 4.7  Summary and Conclusion

The heliocentric distance profile of the density modulation index $\epsilon_N$ in the near Sun region and the heating of the extended solar corona and solar wind have been a subject of considerable interest. Numerous observations indicate strong perpendicular ion heating (Cranmer & Van Ballegooijen, 2003) within a few $R_\odot$ as well as non adiabatic temperature profile at heliocentric distances between $\sim 0.3$ to 50 AU (Richardson & Smith, 2003; Lamarche et al., 2014). This observed heating likely plays a crucial role in accelerating the solar wind and considerably impacts plasma properties in the inner heliosphere.

Several heating mechanisms have been proposed, (including turbulent and non-turbulent heating), to explain these observations. We envisage a scenario, where obliquely propagating Alfvén waves undergoes an anisotropic ($\kappa_\perp \gg k_\parallel$) turbulent cascade developing high wavenumber as well as high compressibility at $\kappa_\perp \rho_i \gg 1$ (Hollweg, 1999; Chandran et al., 2009). This enables us to use observations of density fluctuations in computing the heating rate.

We use existing observations of radio wave scattering and interplanetary scintillations to calculate the amplitude of density fluctuations. Taken together, these measurements yield density turbulence spectra spanning a wide range of spatial scales, including the important high frequency region where dissipation is expected to take place. The density fluctuations are inferred using a combination of recently developed theoretical tools (Ingale et al., 2015) to analyze radio wave scattering data and existing analysis methods to treat interplanetary scintillation data. We use radio scattering data from (Armstrong et al., 1990; Anantharamaiah et al., 1994; Spangler & Sakurai, 1995) and IPS data from Bisoi et al. (2014). We selected only ecliptic sources and divided data according to the phase of the solar cycle.

We estimate the density modulation index $\epsilon_N$ using (4.12) and (4.14) at the inner scale. We found that the $\epsilon_N$ increases for heliocentric distance $< 10 R_\odot$ reaching maximum $\sim 10\%$ and decreases for larger distances. This behavior roughly agrees with the recent observations by Miyamoto et al. (2014). Our values for $\epsilon_N$ are an order of magnitude lower than the values obtained by Miyamoto et al. (2014). This difference in the magnitude of $\epsilon_N$ can be attributed mainly to the different spatial scales to which the two studies are primarily sensitive. The spatial scales we employ are an order of magnitude larger than those used by Miyamoto et al. (2014). If we consider the scales similar to those used by Miyamoto et al. (2014) we find that our estimates of $\epsilon_N$ agree well with the values obtained by Miyamoto et al. (2014).

We evaluate the turbulent energy cascade rate using (4.21) from $2 R_\odot$ to $174\ R_\odot$ at the dissipation scale of the MHD turbulence, assuming that the density fluctuations are due to kinetic Alfvén waves. Our results provide useful upper limits to the rate at which the extended solar wind is heated

The cascade rate is found to depend on the phase of the solar cycle. It ranges from $3 \times 10^{-8}$ $\mathrm{erg\,cm^{-3}s^{-1}}$ during solar maximum to $\sim 10^{-15}$ $\mathrm{erg\,cm^{-3}s^{-1}}$ during solar minimum over $2-174R_\odot$. Our results are consistent with previous findings at $5R_\odot$ and 1 AU. A heating rate required to produce observed non-adiabatic temperature profile of protons and electrons can be estimated. Using Voyager 2 data e.g., (Matthaeus et al., 1999b) and parameters appropriate to 1 AU, Chandran et al. (2009) find that the required heating rate is $\epsilon_{k_i}(1AU) \sim 3 \times 10^{-16}$ $\mathrm{erg\,cm^{-3}s^{-1}}$. This matches closely with the heating rate we found, ($\sim 10^{-16}$ $\mathrm{erg\,cm^{-3}s^{-1}}$) at 1 AU during solar minimum. Our results are also in good agreement with Verma et al. (1995) and Vasquez et al (2007), who predicts the heating rate to be $\sim 10^{-16}$ $\mathrm{erg\,cm^{-3}s^{-1}}$ at 1 AU.

We therefore conclude that radio scintillation observations can provide much needed observational constraints to account for the observed heating and acceleration of the solar wind.

# Chapter 5

# Conclusions and future work

*In this chapter we summarize the main conclusions from this thesis. We also give a flavour of future work arising from the work done in this thesis.*

## 5.1 Conclusions

This thesis has dealt with the nature of density fluctuations near the inner scale and has shown that it impinges on several problems of current interest.

In particular, we have dealt with radio wave scattering due to density turbulence in plasma which leads to a wide variety of observed phenomena and is quantified by the structure function. Most treatments of radio wave scattering so far have employed asymptotic approximations to the structure function which valid in the limits where interferometric baseline, $s$ is either $\ll$ or $\gg$ than the inner scale, $l_i$ of the density turbulence spectrum. We have used an anisotropic density power spectrum characterized by a power law in the inertial range multiplied by an exponential cut-off at the inner scale. We have quantitatively demonstrated that the predictions of the general structure function (GSF) are more accurate than those of the asymptotic approximations for the region where $s \sim l_i$. This is of practical relevance, for $s$ is comparable to $l_i$ for several interferometers in current use, with commonly used models for $l_i$. Our results therefore emphasize the necessity of using the GSF for accurate quantitative estimates of radio wave scattering phenomena.

We combine published observations of radio wave scattering and interplanetary scintillations to infer density fluctuations at the inner scale of the turbulent spectrum. We analyze radio wave scattering using the GSF and existing analysis methods for interplanetary scintillations. We evaluate the density modulation index (defined as $\epsilon_N(z, \kappa_i)$, chapter 4) at the inner scale from the Sun to the Earth. Our results show an initial increase in the density modulation index for R$< 10 R_\odot$ followed by a decrease for larger distances. This behavior agrees qualitatively with the measurements of Miyamoto et al. (2014).

Hypothesizing that the density fluctuations at the inner/dissipation scale are due to kinetic Alfvén waves, we obtained an upper limit on the turbulent cascade rate at the inner scale. Our results are among the first instances where distributed solar wind heating has been computed from the near-Sun region to the Earth. Our results are consistent with previous findings for solar wind heating rates at 1 AU and support the claim that the dissipation of compressive KAW turbulence may be the principal mechanism for extended solar wind heating.

## 5.2 Future work

### 5.2.1 Density fluctuations in the inner solar wind

Several aspects of the behavior of density fluctuations in the inner solar wind remain unclear. We have seen that (chapter 4) radio scintillation techniques - angular broadening and interplanetary scintillations taken together can provide useful information density fluctuations from the Sun to the Earth. There is a fair amount of IPS observatons available for the outer solar corona, from which one can garner information regarding density fluctuations. Observations in the inner solar corona are rarer; an exception is recent work by Miyamoto et al. (2014) and Imamura et al. (2014). There are also a few angular broadening observations by Armstrong et al. (1990) and Anantharamaiah et al. (1994) closer to the Sun (2 to 16 $R_\odot$) which provide very useful information. Zank et al. (2012) using nearly incompressible magnetohydrodynamics (Dastgeer & Zank, 2009; Zank & Matthaeus, 1992) derived the transport equation for the variance of density fluctuations in the solar wind. However, there is a lack of concrete theory for the transport of density fluctuations and therefore the evolution of density turbulence spectrum with heliocentric distance. For $R > 50 R_\odot$ IPS observations are useful and serves as a very good proxy for the density fluctuations. However for $R < 50 R_\odot$ which is believed to be the region of strong scattering as well as the region where solar wind accelerates, there are very few observations. Thus for better understanding of the behavior of density fluctuations and thereby the properties of the solar wind in the near Sun region more observations are required.

### 5.2.2 Scintillation enhancement factor ($g$) and density fluctuations

A model for the background electron density $N_e$ is essential, e.g. to describe the refractive index effects for angular broadening (chapter 3). Also, inner scale effects depends on the model for the $N_e$. It is therefore crucial to have a reliable estimate of the background electron density as a function of heliocentric distance.

IPS observations are a good proxy for the electron density fluctuations $\Delta N_e$. Though the IPS technique does not measure background plasma density, Hewish et al. (1985) showed that the IPS measurements of the density fluctuations are sensitive to the variations in the background plasma density. They defined a quantity called scintillation enhancement factor $g$ as :

$$g = \frac{\Delta S}{\overline{\Delta S}}, \qquad (5.1)$$

where $\Delta S$ is the r.m.s. deviation of flux density at certain frequency and $\overline{\Delta S}$ is an average value for a given source at particular elongation. Hewish et al. (1985) found a strong correlation between the plasma density and $g$. The observations at a solar elongation of 90° satisfy the relation $g = $ (N cm$^{-3}$/9)$^{0.52\pm0.05}$.

This indicates that the quantity $g$ serves as a good proxy for the background plasma density. By exploring the relation between the density fluctuations and the scintillation enhancement factor $g$, (5.1), we can constrain the background density model in the solar corona and solar wind. This in turn will be useful to obtain an accurate estimate of the density modulation index $\epsilon_N$ for small solar elongations.

### 5.2.3 Relating the anisotropy of scatter-broadened images to that of the turbulent spectrum

Although there is intense research activity with regard to the anisotropy, there is relatively little work in relating the anisotropy of the observed scatter-broadened images to the anisotropy of the underlying turbulence. We have used some results pertaining to this aspect in this thesis. The work of Chandran & Backer (2002) investigates some other interesting issues such as the inclination of the line of sight to the large scale magnetic field, which can substentially influence the anisotropy (or lack thereof) of the observed scatter-broadened image. This aspect is especially important for sources in the solar corona. Chandran & Backer (2002) consider radio wave scattering due to density fluctuations elongated in the direction of the ambient magnetic field and characterized by the anisotropic Goldreisch Sridhar (GS) power spectrum. They derive an expression for the wave phase structure function $D_\phi$ and thereby the quantities characterizing radio wave scattering.

Chandran & Backer (2002) assume the GS power spectrum in the inertial range i.e. wavenumber satisfying $l_{\text{out}}^{-1} < \kappa_\perp l_i^{-1}$ with sharp cutoff at wavenumber $\kappa_\perp < l_{\text{out}}$ and $\kappa_\perp > l_i$, where $l_{\text{out}}$ and $l_i$ are the outer and inner scale of the density turbulence respectively. This spectrum therefore does not include a dissipation range of the density turbulence. We (Ingale et al., 2015) have demonstrated that the appropriate form of the phase structure function to be used in case of radio wave scattering due to density turbulence in the solar corona and solar wind, is the general structure function (GSF). The GSF requires the general form of the density power spectrum which includes the inertial range, (characterized by the power law), together with dissipation characterized by the exponential cutoff (Eq. 4.1).

We intend to include the exponential cut-off to the GS power spectrum used by the Chandran & Backer (2002) and generalize their treatment of radio wave propagation. This allows us to derive a general expression for the wave phase structure function and therefore help to generalize the results of Chandran & Backer (2002).

### 5.2.4 Validity of the WKB approximation for density fluctuations

The WKB approximation is frequently used to describe the transport of fluctuations in the MHD Alfvén waves. In this approximation small scale MHD fluctuations must satisfy the Alfvén wave dispersion relation (eq. A23) i.e. only the leading order terms contribute significantly. The order parameter is defined as $\delta_{\text{WKB}} = 1/\kappa L$, where $\kappa$ is the wavenumber that satisfies Alfvén wave dispersion relation and $L$ is the length-scale of the inhomogeneities in the flow (Barnes, 1979). Thus the WKB approximation is valid when the characteristic length scale ($\kappa^{-1}$) is smaller than the length scale of the inhomogeneities, $L$, i.e. $\delta_{\text{WKB}} \ll 1$.

However there exists several situations where the WKB approximation is not expected to be valid (Jokipii & Kota, 1989; Velli et al., 1989; Hollweg, 1990; Barnes, 1992). For the density fluctuations near the inner scale, addressed in (chapter 4) of this thesis, there are reasons to believe that the WKB paradigm may not be realistic. Bale et al. (2005) and Malaspina et al. (2010); use measurements of the electric field fluctuation spectrum. The spectrum is found to agree well with the MHD spectrum over the inertial range but shows an enhancement at the dissipation scale. This change is consistent with the dispersion relation of Kinetic Alfvén waves (KAWs), which deviates from that of the MHD Alfvén waves for $\kappa_\perp \rho_i \geq 1$. Malaspina et al. (2010) carried out observations over the frequency range of $7 - 152$ Hz and showed that for this range of frequencies the WKB approximation is not valid Kellog et al. (1999) Thus there is conclusive evidence for the fact that for scales where phase-mixing and active cascades (dissipation range) dominates and the WKB approximation is not valid.

It is therefore important to define precise limits for the applicability of WKB approximation for fluctuations in the solar wind.

### 5.2.5　Parallel electron heating

Strong anisotropy in MHD turbulence for large wavenumbers ($\kappa$) indicates dominant spectral cascade occurs in a direction perpendicular to the background magnetic field. However Cranmer & Van Ballegooijen (2003) found that as much as 30 % of the input cascade energy remains unaccounted for in the total Alfvén wave damping. They argue that this might be a consequence of abruptly turning down the Alfvén solution branch at high wavenumbers. We should consider the "leakage" of power to high values of $\kappa_\parallel$ and follow the complete turbulent cascade of energy on more than one dispersion branch.

It is therefore important to investigate the parallel heating rate. We know that an important characteristic of Alfvén wave is the existence of electric field fluctuations ($E_\parallel$) parallel to the background magnetic field. Any process that produces energy-conserving cascade, such as phase mixing or turbulence, leads to amplification of $E_\parallel$ associated with the Alfvén wave (Bian & Kontar, 2011). In a collisionless plasma this gives rise to enhanced electron Landau damping causing parallel electron heating. Thus parallel and perpendicular heating rates together gives a complete estimate of "mode-coupled' energy which is an important subject for future work.

# Appendix A

# Kinetic Alfvén Waves

*In the magnetohydrodynamic (MHD) approximation, Lorentz self-forces ($\mathbf{J} \times \mathbf{B}$ forces) in a plasma predict the existence of an incompressible wave mode propagating parallel to the mean magnetic field, causing transverse magnetic field oscillations, called the Alfvén wave. In this chapter we review the salient features of the physics of the Alfvén waves, paying particular attention to obliquely propagating waves as they cascade to large wavenumbers and develop a strong anisotropy. In this regime they are called Kinetic Alfvén Waves (KAW) and are known to exhibit compressibility. We briefly review the linear eigenfunction of KAW which relates density fluctuations with the fluctuations in the Alfvén velocity. This is then used to derive the heating rate at the dissipation scale of the KAW turbulence.*

## A.1 Plasma wave modes

In this section we present a sketch of the derivation of kinetic Alfvén wave (KAW) dispersion relation starting from the linear eigenmodes of the Magnetohydrodynamics (MHD) waves. In particular we concentrate on the KAW dispersion relation based on the two fluid model. We begin with the ideal

MHD equations which describe electrically conducting fluids subject to the presence of external and internal magnetic field and can be approximated as single-fluid of density $\rho$ and current density **j**. Though the scope of one-fluid MHD model restricted to low frequencies ($\omega \ll \omega_p$ or equivalently, the time scales involved should be $\gg$ the inverse of the plasma frequency ($\omega_p$)), its an elegant and clean dynamical theory which successfully predicts the existence of plasma wave modes. Ideal MHD is well described by the field lines moving exactly with the highly conducting fluid while magnetic stresses push on the fluid. The interplay between Maxwell and Reynold stresses effectively forces the dynamics to be formulated in terms of the large-scale bulk velocity **v** and magnetic field **B**, in addition to the usual hydrodynamic variables (e.g. pressure $p$ and density $\rho$). Since ideal MHD is a low frequency theory displacement current can be neglected. The governing equations of the ideal MHD also neglect transport coefficients (such as viscosity and resistivity). Continuity equation describe conservation of mass :

$$\frac{\partial \rho}{\partial t} = -\rho \nabla \cdot \mathbf{v} \tag{A.1}$$

The Faraday induction equation with a generalized Ohm's law, $\mathbf{E} + (\mathbf{v} \times \mathbf{B}) = 0$ can be written as,

$$\frac{\partial \mathbf{B}}{\partial t} = \nabla \times (\mathbf{v} \times \mathbf{B}) \tag{A.2}$$

The momentum equation for ideal MHD is just the familiar hydrodynamic equation with the addition of "Lorentz self-force" given by $\mathbf{j} \times \mathbf{B}$. The current density **j** can be eliminated by using Ampere's law $\nabla \times \mathbf{B} = \mu_0 \mathbf{j}$, and we can write :

$$\rho \frac{\partial \mathbf{v}}{\partial t} + \rho (\mathbf{v} \cdot \nabla) \mathbf{v} = -\nabla p + (\nabla \times \mathbf{B}) \times \mathbf{B}/\mu_0 \tag{A.3}$$

The system of equations can be closed by considering the adiabatic pres-

sure law :

$$\frac{\mathrm{d}}{\mathrm{d}t}\left(\frac{p}{\rho^\gamma}\right) = 0 \tag{A.4}$$

Where $\gamma$ is the adiabatic index of the "plasma fluid". To define initial state of the single fluid plasma we assume stationary and homogeneous conditions which implies that magnetic stresses, average velocity, electric field and overall pressure vanish. We can decompose plasma density, velocity, magnetic and electric field in two parts - sums of their initial values and space and time dependent fluctuations e.g.,(Baumjohann & Treumann, 2004),

$$\begin{aligned}
\rho(\mathbf{r}, t) &= \rho_0 + \delta\rho(\mathbf{r}, t) \\
\mathbf{v}(\mathbf{r}, t) &= \delta\mathbf{v}(\mathbf{r}, t) \\
\mathbf{E}(\mathbf{r}, t) &= \delta E(\mathbf{r}, t) \\
\mathbf{B}(\mathbf{r}, t) &= \mathbf{B}_0 + \delta B(\mathbf{r}, t)
\end{aligned} \tag{A.5}$$

Our aim is to linearize the ideal MHD equations subjected to the perturbations given by (A.5), and thereby obtain the dispersion relation for the eigenmodes of the plasma. Since the MHD equations are non-linear the perturbations should be small enough for linearization to be valid. As we have assumed most of the quantities to have zero initial values, we can represent them by a single variable whose value remains free. This can be readily achieved if the ambient magnetic field is strong enough (which is often is the case in the astrophysical plasma), so that the fluctuations $\delta B(\mathbf{r}, t)$ are weaker than the stationary magnetic field, (Baumjohann & Treumann, 2004) :

$$|\delta\mathbf{B}| \ll B_0 \tag{A.6}$$

With this assumptions we can write the set of linearized equations for an ideal MHD system. The continuity equation (A.1) becomes :

$$\frac{\partial \delta \rho}{\partial t} + \rho_0 (\boldsymbol{\nabla} \cdot \delta \mathbf{v}) = 0 \tag{A.7}$$

Using a vector identity $\nabla \times (F \times G) = (G \cdot \nabla)F - G(\nabla \cdot F)$ in (A.2) and then applying perturbations in (A.5), Faraday's induction equation can be linearized (i.e. neglecting second and higher order terms) to yield,

$$\frac{\partial \delta \mathbf{B}}{\partial t} = (\mathbf{B} \cdot \boldsymbol{\nabla}) \delta \mathbf{v} - B_0 (\boldsymbol{\nabla} \cdot \delta \mathbf{v}) \tag{A.8}$$

The plasma is typically unable to damp out the rapid temperature variations caused by the fluctuations; one can therefore use the adiabatic pressure law (A.4). Linearizing the pressure fluctuation $\delta p$ gives :

$$\frac{\partial \delta p}{\partial t} = v_s^2 \frac{\partial \delta \rho}{\partial t} = -\rho_0 v_s^2 (\boldsymbol{\nabla} \cdot \delta \mathbf{v}) \tag{A.9}$$

where we make use of (A.7) and the quantity $v_s = (\gamma p_0 / \rho_0)^{1/2}$ is the sound speed.

Similarly, after linearizing, the momentum equation (A.3) can be written as,

$$\rho_0 \frac{\partial \delta \mathbf{v}}{\partial t} + \nabla \delta p + \frac{\mathbf{B_0}}{\mu_0} \times (\boldsymbol{\nabla} \times \delta \mathbf{B}) = 0 \tag{A.10}$$

Equations (A.7)−(A.9) represents a linear and homogeneous system of equations for $\delta \rho$, $\delta \mathbf{v}$ and $\delta \mathbf{B}$. Equations (A.8), (A.10) and (A.9) together form a closed system of first order differential equations in magnetic field, velocity and pressure. We derive a second order "wave equation" for one of the field variables. The physics is more transparent if we consider the second order wave equation of velocity fluctuations $\delta \mathbf{v}$ and eliminate the other two variables, pressure and magnetic field fluctuations, $\delta p$ and $\delta \mathbf{B}$ respectively.

Differentiating (A.10) with respect to time and using (A.8) and (A.9) we can write the second order wave equation for $\delta \mathbf{v}$ as follows :

$$\rho_0 \frac{\partial^2 \delta \mathbf{v}}{\partial t^2} - \rho_0 v_s^2 \boldsymbol{\nabla}(\boldsymbol{\nabla} \cdot \delta \mathbf{v}) + \frac{\mathbf{B_0}}{\mu_0}\{\boldsymbol{\nabla} \times [\boldsymbol{\nabla} \times (\delta \mathbf{v} \times \mathbf{B_0})]\} = \mathbf{0}. \qquad (A.11)$$

The large scale magnetic field $\mathbf{B}_0$ can be represented in terms of velocity units by using the Alfvén velocity (Alfvén, 1942) :

$$\mathbf{v}_A = \frac{\mathbf{B}_0}{\sqrt{\mu_0 \rho_0}} \qquad (A.12)$$

Thus we can write (A.11) using (A.12) :

$$\frac{\partial^2 \delta \mathbf{v}}{\partial t^2} - v_s^2 \boldsymbol{\nabla}(\boldsymbol{\nabla} \cdot \delta \mathbf{v}) + \mathbf{v_A} \times \{\boldsymbol{\nabla} \times [\boldsymbol{\nabla} \times (\delta \mathbf{v} \times \mathbf{v_A})]\} = \mathbf{0}. \qquad (A.13)$$

We now assume plane wave ansatz :

$$\delta \mathbf{v}(\mathbf{r}, t) = \delta \mathbf{v} \exp[i\boldsymbol{\kappa} \cdot \mathbf{r} - i\omega t] \qquad (A.14)$$

and replace the derivatives :

$$\boldsymbol{\nabla} \to i\boldsymbol{\kappa} \text{ and } \frac{\partial}{\partial t} \to i\omega t \qquad (A.15)$$

to obtain the dispersion relation :

$$-\omega^2 \delta \mathbf{v} + v_s^2 (\boldsymbol{\kappa} \cdot \delta \mathbf{v})\boldsymbol{\kappa} - \mathbf{v}_A \times \{\boldsymbol{\kappa} \times [\boldsymbol{\kappa} \times (\delta \mathbf{v} \times \mathbf{v}_A)]\} = 0 \qquad (A.16)$$

We can further expand the term in the curly brackets by repeated use of a vector triple product $\mathbf{E} \times (\mathbf{F} \times \mathbf{G}) = (\mathbf{E} \cdot \mathbf{G})\mathbf{F} - (\mathbf{E} \cdot \mathbf{F})\mathbf{G}$, which gives :

$$-\omega^2 \delta\mathbf{v} + (v_s^2 + v_A^2)(\boldsymbol{\kappa}\cdot\delta\mathbf{v})\boldsymbol{\kappa}$$
$$+ (\boldsymbol{\kappa}\cdot\mathbf{v}_A)[(\boldsymbol{\kappa}\cdot\mathbf{v}_A)\delta\mathbf{v} - (\mathbf{v}_A\cdot\delta\mathbf{v})\boldsymbol{\kappa} - (\boldsymbol{\kappa}\cdot\delta\mathbf{v})\mathbf{v}_A] = 0 \quad (A.17)$$

This is the desired dispersion relation for the MHD plasma wave modes. The presence of a large scale magnetic field causes a preferred direction in the MHD system. If we assume a uniform plasma with straight magnetic field lines, then the direction of the ambient magnetic field represents this preferred direction and it is the only direction of symmetry. We choose the direction of an ambient magnetic field to be along the z axis of our orthogonal coordinate system. Thus we can write :

$$\mathbf{B}_0 = B_0 \mathbf{e_z} \quad (A.18)$$

Consequently :

$$\mathbf{v}_A = v_A \mathbf{e_z} \quad (A.19)$$

The propagation vector can be decomposed into the parallel and perpendicular components defined with respect to the ambient magnetic field. From symmetry arguments, it is easy to see that the $x$ and $y$ coordinates are interchangeable and thus without loss of generality we can set one of the perpendicular components, say $\kappa_y = 0$. Then for a right handed coordinate system, a perpendicular component is parallel to the $x$ axis and we can write a wavevector as :

$$\boldsymbol{\kappa} = \kappa_\perp \mathbf{e}_x + \kappa_\parallel \mathbf{e}_z \quad (A.20)$$

where $\kappa_\perp = \kappa\sin\theta$ and $\kappa_\parallel = \kappa\cos\theta$.

Equation (A.17) is actually a set of three separate equations for $\delta\mathbf{v} = (\delta v_x, \delta v_y, \delta v_z)$, which are coupled equations and can be written in matrix

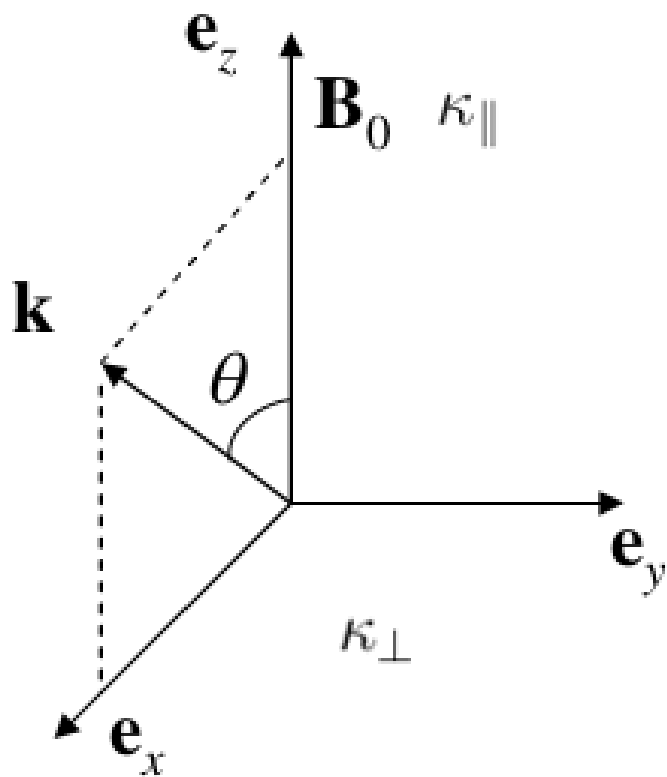

Figure A.1: The parallel and perpendicular wave-vectors are defined with respect to the direction of the ambient magnetic field **B** which is taken to be along the $z$ axis.

form :

$$\mathbb{A}\delta\mathbf{v} = -\omega^2\delta\mathbf{v} \tag{A.21}$$

where

$$\mathbb{A} = \begin{pmatrix} -\kappa_\parallel v_A^2 - (v_s^2 + v_A^2)\kappa_\perp^2 & 0 & -v_s^2\kappa_\parallel\kappa_\perp \\ 0 & -v_A^2\kappa_\parallel & 0 \\ -v_s^2\kappa_\parallel\kappa_\perp & 0 & -v_s^2\kappa_\parallel \end{pmatrix}$$

The problem of MHD modes is therefore the eigenvalue-eigenvector problem with eigenvalues provide dispersion relations. Since $\mathbb{A}$ is a $3 \times 3$ matrix, there are three fundamental MHD wave modes and any disturbance can be expressed as linear combination of these three basic modes. Meaningful solution can be obtained if :

$$|\mathbb{A} - \omega^2 \mathbb{I}| = 0 \tag{A.22}$$

it may be noted that the y-component of the velocity fluctuations decouples from the rest of the fields representing a wave with linear dispersion relation :

$$\omega_A = \pm \kappa_\parallel v_A \tag{A.23}$$

The wave propagates parallel to the ambient magnetic field and is a type of magnetohydrodynamic wave known as Alfvén wave.

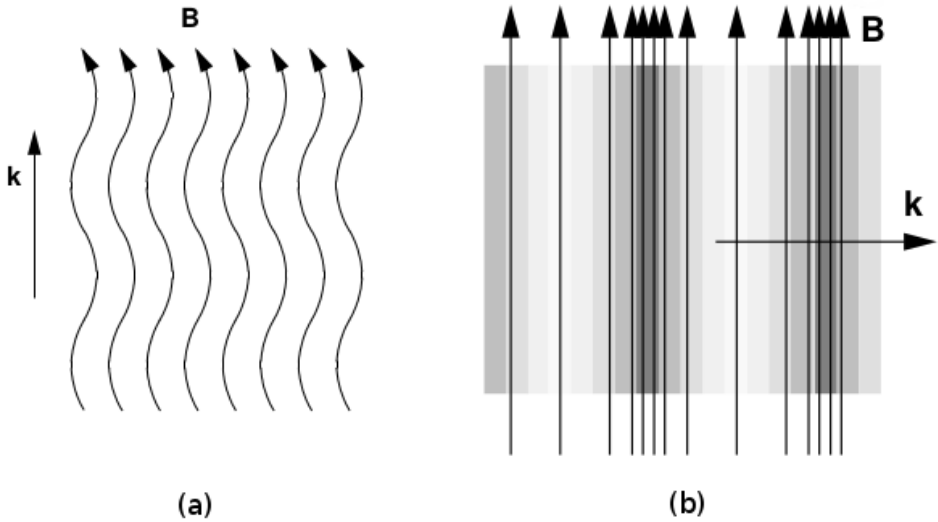

Figure A.2: (a) Alfvén waves propagating along the $B_0$. The fluid motion and the magnetic perturbations are normal to the field lines. (b) Magnetosonic wave propagates normal to $B_0$ compressing and releasing magnetic lines of force as well as conducting fluid tied to the field.

From (A.8) it is clear that the Alfvén waves are associated with the mag-netic field fluctuations parallel to the velocity component :

$$\frac{\delta \mathbf{B}_\perp}{B_0} = \frac{\delta \mathbf{v}_\perp}{v_A} \qquad (A.24)$$

and thus there are no electric field fluctuations along the ambient magnetic field (Baumjohann & Treumann 2012). Electric field fluctuations associated with the Alfvén waves are given by using Faraday law :

$$\delta \mathbf{E}_\perp = \frac{\delta \mathbf{B}_\perp}{v_A} \qquad (A.25)$$

The eigenvalues of the determinant formed by other four elements of the (A.22) gives the other two modes, known as magnetosonic modes. The solution to the eigenvalue equation for this determinant yields two roots :

$$\omega_{ms}^2 = \frac{\kappa^2}{2} \left\{ (v_s^2 + v_A^2) \pm \left[ (v_A^2 - v_s^2)^2 + 4 v_A^2 v_s^2 \frac{\kappa_\perp^2}{\kappa^2} \right]^{1/2} \right\} \qquad (A.26)$$

Where the root with positive sign is known as the fast magnetosonic wave and the root with negative sign is known as the slow magnetosonic wave. The behavior of the three modes can be conveniently expressed with the help of the phase velocity diagram.

Figure (A.3) shows the three magnetohydrodynamics wave modes represented by the vector arrows of their phase velocities.

## A.2    MHD in Elssäser variables

We now focus on the Alfvén wave modes as they are the fundamental modes of incompressible MHD. They are ubiquitous in the solar corona and solar wind (Tomczyk et al., 2007; Velli & Pruneti, 1997) and play an important role for the turbulent motions in the MHD plasma. The physics of the Alfvén waves is more transparent when expressed in terms of Elssäser variables (Elssäser, 1950). Elssäser variables are defined as the sum and difference of the bulk velocity **v** and the magnetic field **B**, expressed in the units of velocity :

$$\mathbf{z}^\pm = \mathbf{v} \pm \frac{\mathbf{B}}{\sqrt{\mu_0 \rho_0}} \tag{A.27}$$

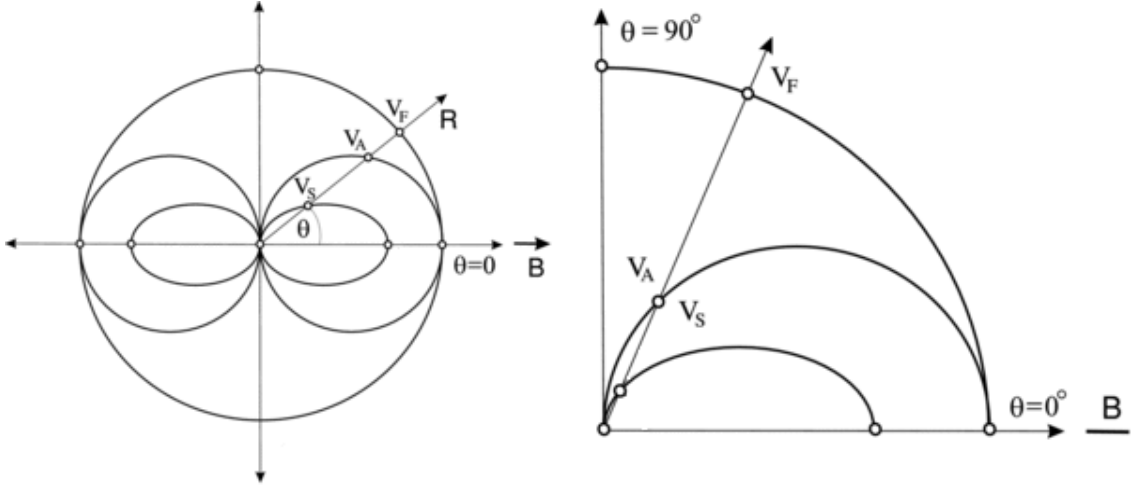

Figure A.3: Wave-normal diagram for the MHD wave modes. The length of the arrow from the origin to a point on the associated closed curve is proportional to the wave-phase velocity.

Equations (A.1)−(A.4) can then be expressed in the form of following set of equations :

$$\frac{\partial \mathbf{z}^\pm}{\partial t} \mp \mathbf{v}_A \cdot \boldsymbol{\nabla} \mathbf{z}^\pm = -\mathbf{z}^\mp \cdot \boldsymbol{\nabla} \mathbf{z}^\pm - \nabla P/\rho_0 \tag{A.28}$$

$$\boldsymbol{\nabla} \mathbf{z}^\pm = 0 \tag{A.29}$$

Where $P$ is the total pressure, $P_{\text{thermal}} + P_{\text{magnetic}}$. This formulation offers number of advantages and an important physical insights into the Alfvén turbulence. If one looks carefully at the above system of equations, the divergence free condition, (A.29) closes the system and we don't actually need to specify the equation of state. The divergence of (A.28) with (A.29)

yields for pressure :

$$\nabla^2 P = -\rho_0 \boldsymbol{\nabla} \cdot (\mathbf{z}^\mp \cdot \boldsymbol{\nabla} \mathbf{z}^\pm) \qquad (A.30)$$

An important thing to note here is that, both the terms on the right side of equation (A.28) are non-linear and pressure at any point responds instantaneously, as expected in case of incompressible flows. Linearization of (A.28) gives (Bruno & Carbone, 2013) :

$$\frac{\partial \mathbf{z}^\pm}{\partial t} = \pm(\mathbf{v}_A \cdot \boldsymbol{\nabla})\mathbf{z}^\pm \qquad (A.31)$$

Which predicts the existence of counterpropagating Alfvén wave modes, $z^\pm(\mathbf{x} \pm \mathbf{v}_A t)$ with respect to the ambient magnetic field. Thus we can understand (A.28) as follows, the second term (linear) on the left side gives the propagation of the Elssäser fields with the Alfvén speed. The first term on the right hand side specifies the non-linear interaction between the counterpropagating Alfvén waves and the second term on the right hand side ensures incompressibility.

### A.2.1  Non-Linear properties

The non-linear term $\mathbf{z}^\mp \cdot (\boldsymbol{\nabla} \mathbf{z}^\pm)$ in (A.28) is crucial in understanding the mechanism of turbulence in the MHD system. It reveals several important properties of incompressible MHD turbulence; for instance the vector form of the (A.28) demonstrates that the incompressible MHD turbulence is an inherently three dimensional phenomenon. Consider the non-linear term $\mathbf{z}^- \cdot (\boldsymbol{\nabla} \mathbf{z}^+)$ from equation (A.28); the contribution due to this term is non-zero only when both $z^- \neq 0$ and $z^+ \neq 0$, i.e. the two waves must propagate in opposite directions along the magnetic field. The presence of $z^+$ and $z^-$ at the same point of the space suggests that the $\mathbf{z}^-$ distorts the counterpropagating $\mathbf{z}^+$ wave and vice-versa. Consider the non-linear interaction between two plane waves with wavevectors, $\boldsymbol{\kappa}_1$ and $\boldsymbol{\kappa}_2$. For Alfvén waves, $\omega_1 = \kappa_{\|1} v_A$ and $\omega_2 = \kappa_{\|2} v_A$, with wave frequency $\omega_1, \omega_2 > 0$. The direction of the propa-gation along the magnetic field is therefore specified by the sign of $\kappa_\parallel$. It can be shown that the the non-linear term in (A.28) is maximized when perpendicular components of the counterpropagating Alfvén waves are orthogonal, i.e. $\boldsymbol{\kappa}_{\perp 1} \cdot \boldsymbol{\kappa}_{\perp 2} = 0$. This gives rise to the anisotropic turbulent cascade of energy, where turbulent energy is preferentially transferred to small scales perpendicular to the local magnetic field direction. Thus equation (A.28) demonstrates two crucial properties of the incompressible MHD turbulence,

- Non-linear interaction occurs only when counterpropagating waves "collide", this serves as a building block of MHD turbulence and,

- This non-linear interaction leads to an anisotropic cascade in the sense $\kappa_\perp \gg \kappa_\parallel$

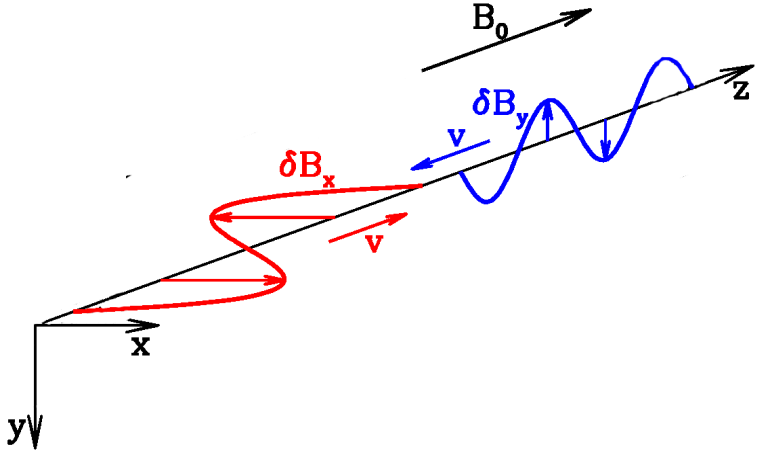

Figure A.4: The counter-propagating Alfvén waves parallel and anti parallel to the $B_0$. Alfvén waves overlap for non zero $(z^\mp \cdot \nabla) z^\pm$. (Adapted from **?**)

## A.2.2 Turbulent Cascade

Following significant developments in incompressible MHD in recent years through numerical simulations and laboratory experiments (Howes et al., 2012; Drake et al., 2013), we can present a qualitative picture of a process of "collision" of counterpropagating Alfvén waves (Howes et al., 2013). Howes et al., (2013) presents Elssäser variables in terms of Elssäser potentials defined by :

$$z^{\pm} = \hat{z} \times \nabla_{\perp} \zeta^{\pm} \tag{A.32}$$

Equation (A.28) with Elssäser potentials defined by (A.32) gives Elssäser potential equation. The asymptotic solution of which can be obtained in the weak turbulence limit given by :

$$\frac{z^{\pm}}{v_A} \sim \eta \ll 1 \tag{A.33}$$

We can express the solution by expanding $\zeta^{\pm}$ in the powers of $\eta$ :

$$\zeta^{\pm} = \zeta_0^{\pm} + \eta \zeta_1^{\pm} + \eta^2 \zeta_2^{\pm} + \eta^3 \zeta_3^{\pm} + \cdots \tag{A.34}$$

Where $\zeta_1^{\pm}$ is the primary solution which constitute the couterpropagating Alfvén wave modes, $\zeta_2^{\pm}$ is the secondary solution, $\zeta_3^{\pm}$ - tertiary solution and so on. Intuitive picture of the turbulence cascade is presented in following figure :

From figure we can understand the process of turbulent cascade as follows, the primary modes at $\mathcal{O}(\eta)$ corresponds to the counterpropagating Alfvén waves with wavevectors $\boldsymbol{\kappa}_1^+ = \kappa_{\perp}\hat{x} - \kappa_{\parallel}\hat{z}$ and $\boldsymbol{\kappa}_1^- = \kappa_{\perp}\hat{y} + \kappa_{\parallel}\hat{z}$ and frequency $\omega_0 = \kappa_{\parallel} v_A$. The primary modes are denoted by filed circles in Figure(). These two primary Alfvén modes interact with each other according to the non-linear terms on the right hand side of (A.28). This interaction generates a secondary mode at $\mathcal{O}(\eta^2)$ :

$$\boldsymbol{\kappa}_2^0 = \boldsymbol{\kappa}_1^+ + \boldsymbol{\kappa}_1^- = \kappa_{\perp}\hat{x} + \kappa_{\perp}\hat{y} \tag{A.35}$$

This is an inherently nonlinear and purely magnetic mode, denoted by filled triangle in figure (A.5). Note that the secondary mode has $\kappa_{\parallel} = 0$ and frequency $= 2\omega_0$. The primary modes then interact nonlinearly with the secondary mode and gives rise to tertiary mode with wavevector $\boldsymbol{\kappa}_3^+$ and $\boldsymbol{\kappa}_3^-$ at $\mathcal{O}\eta^3$. The process can be expressed as :

$$\begin{aligned}\boldsymbol{\kappa}_3^+ &= \boldsymbol{\kappa}_2^0 + \boldsymbol{\kappa}_1^+ = 2\kappa_\perp \hat{x} + \kappa_\perp \hat{y} - \kappa_\parallel \hat{z} \\ \boldsymbol{\kappa}_3^- &= \boldsymbol{\kappa}_2^0 + \boldsymbol{\kappa}_1^- = \kappa_\perp \hat{x} + 2\kappa_\perp \hat{y} + \kappa_\parallel \hat{z}\end{aligned} \quad (A.36)$$

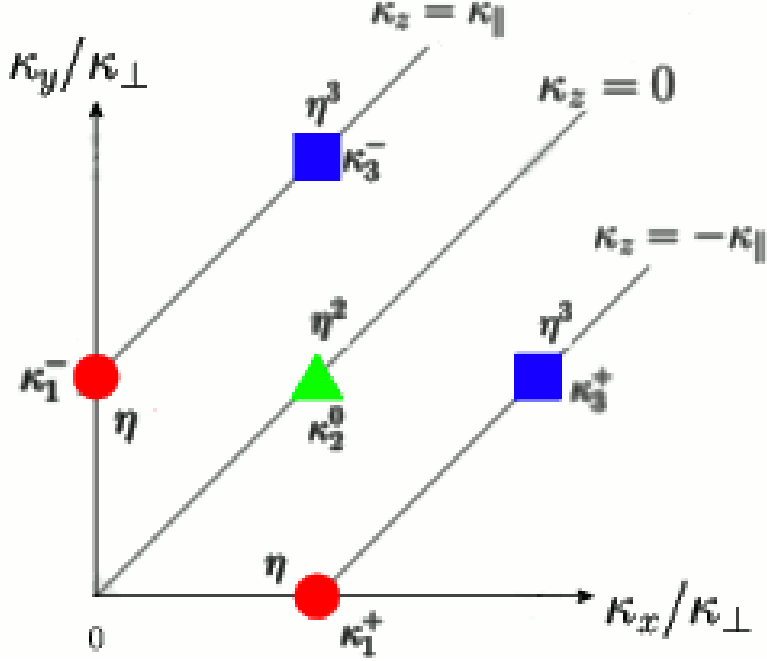

Figure A.5: Adapted from Howes (2013), the figure represents nonlinear interaction between the counter propagating Alfvén waves.

The tertiary modes $\boldsymbol{\kappa}_3^\pm$ denoted by filled squares in figure (A.5), each have linear Alfvén frequency $\omega_0$ and and propagates in a direction of the primary modes (i.e. along the z direction). This sequence of interaction guided by the nonlinear terms in equation (A.28) therefore leads to the secular transfer of energy from low wavenumbers ($\boldsymbol{\kappa}_1^\pm$) to higher wavenumber, e.g. $\boldsymbol{\kappa}_3^\pm$. It is clear from the Figure (A.5) and above discussion that the nonlinear transfer of energy or the turbulent cascade proceeds only along the contours of constant $\kappa_\parallel$ with increasing $\kappa_\perp$. Thus we can easily demonstrate the anisotropic cascade implied by (A.28). We have seen parallel components of counter-propagating Alfvén waves $\kappa_{\parallel 1}$ and $\kappa_{\parallel 2}$, must be opposite in sign. We can see therefore in case of cascade to tertiary mode, equations (A.35) and (A.36) with $\boldsymbol{\kappa}_{\perp 1} \cdot \boldsymbol{\kappa}_{\perp 2} = \mathbf{0}$, implies that $\kappa_{\parallel 3} \leq \kappa_{\parallel 1}$ and $\kappa_{\parallel 3} \leq \kappa_{\parallel 2}$, whereas $\kappa_{\perp 3} \geq \kappa_{\perp 1}$ and $\kappa_{\perp 3} \geq \kappa_{\perp 2}$ (Goldreich & Sridhar 1995, Howes et al., 2013). In other words the cascade gets stronger in the perpendicular direction giving rise to an anisotropy in a sense $\kappa_\perp \gg \kappa_\parallel$.

Above discussion is based on the three wave interaction (two ounterpropagating wave modes leads to secondary wave mode), which in general is valid for weak turbulence (A.33), many of the general qualitative properties of the process persists in strong turbulence.

## A.3 Kinetic Alfvén waves

Starting from incompressible MHD equation (A.28) we have seen how Alfvén waves undergo turbulent cascade, preferentially in the direction perpendicular to the ambient magnetic field and transfer energy to higher wavenumbers i.e., to smaller scales. At high wavenumbers a typical astrophysical plasma is characterized by ion mean free path ($\lambda_i$), ion gyroradius ($\rho_i$) and electron gyroradius ($\rho_i$). These scales are ordered as $\lambda_i > \rho_i > \rho_e$; thus the ion mean free path is usually reached first. The length scale $\lambda_i$ characterizes collisionality in the direction of ambient magnetic field and thus the inequality $\kappa_\parallel \lambda_i \geq 1$ marks the transition from collisional to non-collisional dynamics. Using critical balance condition (Goldreich & Sridhar 1995) the equivalent transition to the collisionless dynamics in perpendicular wavenumber occurs for the scales $\kappa_\perp \rho_i \geq 1$. As the $\rho_i$ reached in the turbulent cascade the ions decouples from the turbulent electromagnetic fluctuations. This leads to the transition of non-dispersive Alfvén waves to dispersive kinetic Alfvén waves (KAW). This also marks the break down of MHD description as well as the one fluid model, we used for the description of MHD wave modes.

To obtain the physical picture of the dispersive KAW in the range of high wavenumber ($\kappa_\perp$) it is convenient to explore the nonlinear term in (A.28) through the turbulent electric and magnetic field fluctuations. The definition of Elssäser variables (A.27) with the assumption $|\delta \mathbf{B}| \ll B_0$ implies that the term of the form $\delta \mathbf{v} \cdot \boldsymbol{\nabla}$ contributes significantly to the nonlinearity. Following ohms law for ideal MHD we can write :

$$\mathbf{E} + \frac{(\delta \mathbf{v} \times \mathbf{B})}{c} = 0 \tag{A.37}$$

Thus the term $\delta \mathbf{v} \cdot \boldsymbol{\nabla}$ in the fluid equation transforms to the form of non-linearity, represented by $\mathbf{E}_\perp \times \mathbf{B}_\perp$. This suggests that the lowest order contribution to the plasma velocity fluctuations is given by :

$$\delta \mathbf{v} = \frac{c \mathbf{E} \times \mathbf{B}}{B_0^2} \tag{A.38}$$

The $\mathbf{E}_\perp \times \mathbf{B}_\perp$ nonlinearity is the dominant nonlinear mechanism underlying the turbulent cascade in MHD as well as kinetic regimes.

### A.3.1 KAW : Hollweg 1999

We briefly review the dispersion relation for the KAW, derived by Hollweg 1999 using the two fluid model. Alternate derivations of the KAW dispersion relation that incorporate slightly different physics can be found in (Hasegawa & Chen, 1974; Lysak & Lotko, 1996). As we have seen for wavenumber $\kappa_\perp \rho_i \geq 1$, kinetic effects dominates and the single fluid MHD description breaks down. However the MHD description can be applied to demonstrate some important properties of the Alfvén waves in a kinetic regime.

We consider the right hand coordinate system as shown in the figure(A.6), where $\mathbf{B_0}$, the ambient magnetic field is in the direction along the $z$ axis. Consider the Alfvén wave with frequency $\omega$ propagates in the $x-z$ plane as $\exp[i(\kappa_x + \kappa_z - \omega t)]$. The magnetic field fluctuations, $\delta B_y$ in the $y$ direction gives rise to the velocity fluctuations $\delta v_y$, which can be interpreted as the lowest order contribution due to the nonlinearity $\mathbf{E} \times \mathbf{B}$ to the plasma velocity

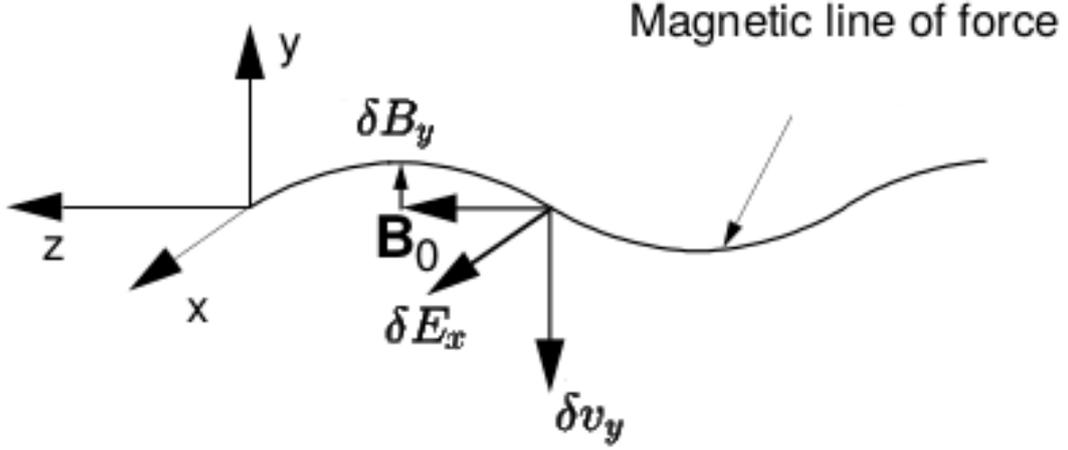

Figure A.6: Geometry for the Alfvén wave propagation and the perturbed components associated with the wave propagating along the $B_0$

given by :

$$\delta v_y = c \frac{\delta E_x}{B_0} \tag{A.39}$$

This expression can alternately be obtained by using the Alfvén wave properties, (A.24) and (A.25). The presence of $\delta E_x$ suggests that there is a current in the x direction carried mainly via the polarization drift of the ions. For $\omega \ll \omega_{ci}$ we write :

$$\delta v_{px} = \frac{1}{\omega_{ci} B_0} \frac{\partial}{\partial t}(\delta E_x) = \frac{q_i}{m_i} \frac{\omega}{\omega_{ci}^2} \delta E_x \tag{A.40}$$

Where $q_i$ and $m_i$ are the ion charge and mass, $\omega_{ci} = q_i B_0/m_i c$ is the ion cyclotron frequency. The polarization drift involves the charge and the mass of the particle. As the polarization effect becomes significant local charge separation occurs and single fluid theory breaks down. We therefore need to consider the effects of charged particles, constituting plasma, separately. Since the polarization drift involves charge and the mass of the particle it gives rise to the compressional effects on the ions and therefor causes density perturbations $\delta n_i$. With $\delta \rho_i = m_i \delta n_i$ where $m_i$ is the ion mass and $n_{0i}$ is the ion number density, we can write the linearized continuity equation (A.7),

$$\frac{\partial}{\partial t}\delta n_i = -n_{0i}\boldsymbol{\nabla}\delta\mathbf{v}. \tag{A.41}$$

From which we get the x-component of the ion drifting $\delta n_i/n_{0i} = \kappa_x \delta v_x/\omega$. Using (A.39) and (A.40) with the x-component of (A.41) we can write,

$$i\frac{\delta n_i}{n_{0i}} = \frac{\kappa_y}{\omega_{ci}}\delta v_y \tag{A.42}$$

Note that the ion inertial length is given by $d_i = v_A/\omega_{ci}$, thus we can write equation (A.42) as,

$$i\frac{\delta n_i}{n_{0i}} = \kappa_y d_i \frac{\delta v_y}{v_A} \tag{A.43}$$

Where $\delta v_y/v_A$ is interpreted as the measure of the first order amplitude of the Alfvén wave. Starting from the dominant nonlinearity $\mathbf{E} \times \mathbf{B}$ in the MHD equations we arrived at an approximate estimate of the condition where compressible effects becomes important. We see from (A.43) that for higher wavenumbers where $\kappa_y^{-1}$ is comparable with the ion inertial length or in other words for $\kappa_y d_i \sim 1$, Alfvén wave exhibits compressibility. The dispersion relation, specifying the properties of the wave departs from that of the incompressible MHD Alfvén wave. In this region $\kappa_y d_i \geq 1$ the Alfvén wave is known as Kinetic Alfvén wave.

We have seen that the electron $\mathbf{E} \times \mathbf{B}$ drift coupled with the ion polarization drift of the charged particles gives rise to a compressible effects for ions at higher wavenumber (A.43). Similarly coupling of ion $\mathbf{E} \times \mathbf{B}$ drift with electron polarization drift gives rise to electron compressibility, when the electron inertia is significant. This improves (A.42), and consequently the condition for the transition to KAW.

A fluid approach to the collision-less plasma constitute two dominant forces on charge particles that transfer momentum. First − charged particles responds to electric and magnetic field via Lorenz force, and second they also respond to the pressure gradient force.

With m and n denoting the mass and number density of particles respectively, we can write the fluid momentum equation for any species in the two fluid case as :

$$mn\frac{d\mathbf{v}}{dt} = mn\left[\frac{\partial \mathbf{v}}{\partial t} + (\mathbf{v}\cdot\boldsymbol{\nabla})\mathbf{v}\right] = qn(\mathbf{E} + \mathbf{v}\times\mathbf{B_0}) - \boldsymbol{\nabla}p. \qquad (A.44)$$

The linearized momentum equation in the direction of ambient magnetic field is therefore given by :

$$m\omega\delta v_z = -iq(\delta E_z) + \kappa_z \gamma k_\beta T_0 \frac{\delta n}{n_0} \qquad (A.45)$$

Here we make use of the plane wave ansatz with wave propagating in $x-z$ direction. Where pressure fluctuations $\delta p = \gamma k_\beta T_0 \delta n$, with $\gamma$ is the adiabatic index, $k_\beta$ is the Boltzmann constant, $T_0$ temperature of plasma species.

The electrons moving parallel to the ambient magnetic field constitute the electric field fluctuations along $z$-axis. Assume quasi-neutrality $\delta n_i / n_{0i} \sim \delta n_e / n_{0e}$ and using (A.42), the $z$ component of (A.45) for electrons is given by :

$$e\delta E_z = i\, m_e\omega\delta v_{ze} - (\gamma_e k_\beta T_{0e})\frac{\kappa_x \kappa_z}{\omega_c}\delta v_y \qquad (A.46)$$

Using (A.41) and (A.42) we can write, $\delta v_{ze} = (\omega/\kappa_z)\delta n_e/n_{0e} = (\omega\,\kappa_x/i\,\kappa_z\,\omega_c)\delta v_y$, which implies :

$$\delta E_z = \left(\frac{m_e\,\omega^2}{e\,\kappa_z^2} - \frac{\gamma_e\,k_\beta T_{0e}}{e}\right)\frac{\kappa_x\,\kappa_z}{\omega_c}\delta v_y \qquad (A.47)$$

The presence of $\delta E_z$, field aligned electric field, which is an important property of the Alfvén wave, suggests that the Faraday's law can be written

as :

$$\kappa_x \delta E_z - \kappa_z \delta E_x = \frac{\omega}{c} \delta B_y \tag{A.48}$$

Consequently the polarization drift (A.40) gets modified to :

$$\delta v_{px} = -\frac{i\, q\, \omega}{m\, \omega_c^2} \delta E_x \left(1 + \gamma \kappa_x^2 \frac{k_\beta T_0}{m\, \omega_c}\right)^{-1} \tag{A.49}$$

Corresponding density fluctuations are then given by :

$$\frac{\delta n}{n_0} = -i \frac{q\, \kappa_x}{m \omega_c^2} \left(1 + \gamma \kappa_x^2 \frac{k_\beta T_0}{m\, \omega_c}\right)^{-1} \delta E_x \tag{A.50}$$

Now using $\mathbf{E} \times \mathbf{B}$ drift with the definition $\omega_c = q\, B_0/mc$ we can write :

$$\delta E_x = \delta v_y \frac{m \omega_c}{q} \tag{A.51}$$

which gives :

$$i \frac{\delta n}{n_0} = \frac{\kappa_x}{1 + \kappa_x^2 L^2} \frac{\delta v_y}{\omega_c} \; . \tag{A.52}$$

Where, $L = k_\beta T_0/(m\, \omega_c^2)$. The first order amplitude of the Alfvén wave $\delta v_y/v_A$ is therefore can be given by :

$$i \frac{\delta n}{n_0} = \frac{\kappa_\perp d_i}{1 + \gamma \kappa_\perp^2 L^2} \frac{\delta v_\perp}{v_A}, \tag{A.53}$$

where we make use of the definition of inertial length $d = v_A/\omega_c$. In writing (A.53) we adopted the notation where the components in the transverse direction, defined with respect to the ambient magnetic field $B_0$, are written as $\kappa_x, \kappa_y \to \kappa_\perp$ and $\delta v_x, \delta v_y \to \delta v_\perp$.

For the special case of warm plasma where ion dynamics dominates we can write (A.53) as follows :

$$i\frac{\delta n_i}{n_{0i}} = \frac{\kappa_\perp d_i}{1+\gamma_i \kappa_\perp^2 \rho_i^2}\frac{\delta v_\perp}{v_A} \qquad (A.54)$$

Using (A.50) we can modify the expression (A.47) and then eliminate $\delta E_x, \delta E_z$ using (A.48) to yield :

$$\omega^2 = \frac{1+\kappa_\perp L^2}{1+\kappa_\perp^2 c^2/\omega_{pe}^2}\kappa_\parallel^2 v_A^2 \qquad (A.55)$$

Here $\omega_{pe}$ is the electron plasma frequency. The equation (A.55) is known as the dispersion relation for the KAWs in the cold plasma described by the two fluid model. The dispersion relation (A.55) contains the thermal effect of the ions but additionally takes into account the electron inertia of the background plasma. If we look at the dispersion relation (A.55) it is immediately clear that for $\kappa_\perp L \ll 1$ we recover the dispersion relation for MHD Alfvén wave (A.17). Usually the quantity $L$ is considered as $L = \rho_i$, ion gyroradius and therefor the more familiar condition for transition to KAW can be written as $\kappa_\perp \rho_i \sim 1$. This implies the transition from incompressible MHD Alfvén waves to compressible kinetic Alfvén waves takes place well before the ion inertial length is reached, however the wave becomes completely dispersive for wavenumbers $\kappa_\perp d_i \geq 1$.